\documentclass[traditabstract]{aa} 
\usepackage{graphicx}
\usepackage{txfonts}
\usepackage{natbib}
\newcommand{\kms}{{\hbox {km\thinspace s$^{-1}$}}}
\newcommand{\Lsun}{{\hbox {L$_\odot$}}}
\newcommand{\Msun}{{\hbox {M$_\odot$}}}

\newcommand{\cmt}{{\hbox {cm$^{-3}$}}}
\newcommand{\cmd}{{\hbox {cm$^{-2}$}}}
\newcommand{\hdo}{{\hbox {H$_{2}$O}}}

\newcommand{\chot}{{\hbox {$C_{\mathrm{hot}}$}}}
\newcommand{\ccore}{{\hbox {$C_{\mathrm{core}}$}}}
\newcommand{\cwarm}{{\hbox {$C_{\mathrm{warm}}$}}}
\newcommand{\cext}{{\hbox {$C_{\mathrm{extended}}$}}}
\newcommand{\cwest}{{\hbox {$C_{\mathrm{west}}$}}}
\newcommand{\ceast}{{\hbox {$C_{\mathrm{east}}$}}}
\newcommand{\chalo}{{\hbox {$C_{\mathrm{halo}}$}}}
\def\t#1#2#3#4#5#6{{\hbox {$#1_{#2#3}\!\rightarrow\!#4_{#5#6}$}}}

\def\13co{$^{13}$CO}
\def\c18o{C$^{18}$O}

\begin{document}
   \title{Herschel\thanks{Herschel is an 
       ESA space observatory with science 
       instruments provided by European-led Principal Investigator consortia
       and with important participation from NASA.}/PACS 
     spectroscopy of NGC~4418 and Arp 220: \\ H$_2$O, H$_2^{18}$O, OH, $^{18}$OH,
     O {\sc i}, HCN and NH$_3$}

   \author{E. Gonz\'alez-Alfonso \inst{1}, 
J. Fischer \inst{2}, J. Graci\'a-Carpio \inst{3}, E. Sturm \inst{3}, 
S. Hailey-Dunsheath \inst{3}, D. Lutz \inst{3}, A. Poglitsch \inst{3},
A. Contursi \inst{3}, H. Feuchtgruber \inst{3}, S. Veilleux \inst{4,5},
H. W. W. Spoon \inst{6}, A. Verma \inst{7}, N. Christopher \inst{7},
R. Davies \inst{3}, A. Sternberg \inst{8}, 
R. Genzel \inst{3}, L. Tacconi \inst{3}
          }

   \institute{Universidad de Alcal\'a de Henares, 
Departamento de F\'{\i}sica, Campus Universitario, E-28871 Alcal\'a de
Henares, Madrid, Spain  
   \email{eduardo.gonzalez@uah.es}
         \and
Naval Research Laboratory, Remote Sensing Division, 4555
     Overlook Ave SW, Washington, DC 20375, USA
         \and
Max-Planck-Institute for Extraterrestrial Physics (MPE), Giessenbachstra{\ss}e
1, 85748 Garching, Germany 
         \and
Department of Astronomy, University of Maryland, College Park, MD 20742, USA
         \and
Astroparticle Physics Laboratory, NASA Goddard Space
       Flight Center, Code 661, Greenbelt, MD 20771 USA
         \and
Cornell University, Astronomy Department, Ithaca, NY 14853, USA
         \and
University of Oxford, Oxford Astrophysics, Denys Wilkinson Building, Keble
Road, Oxford, OX1 3RH, UK 
         \and
Sackler School of Physics \& Astronomy, Tel Aviv University, Ramat Aviv 69978,
Israel
}

   \authorrunning{Gonz\'alez-Alfonso et al.}
   \titlerunning{Herschel/PACS spectroscopy of NGC~4418 and Arp~220}

  \abstract
   {
Full range Herschel/PACS spectroscopy of the (ultra)luminous infrared galaxies
NGC~4418 and Arp~220, observed as part of the SHINING key programme, reveals
high excitation in H$_2$O, OH, HCN, and NH$_3$.   
In NGC~4418, absorption lines were detected with $E_{\mathrm{lower}}$$>$800 K 
(H$_2$O), 600 K (OH), 1075 K (HCN), and 600 K (NH$_3$), while in Arp~220 the
excitation is somewhat lower. 
While outflow signatures in moderate excitation lines are seen in
Arp~220 as have been seen in 
previous studies, in NGC~4418 the lines tracing its outer regions are
redshifted relative to the nucleus, suggesting an inflow with 
$\dot{M}$$\lesssim$12 \Msun\ yr$^{-1}$. 
Both galaxies have compact and warm ($T_{\mathrm{dust}}$$\gtrsim$100 K) nuclear
continuum components, together with a more extended and colder component that
is much more prominent and massive in Arp~220.
A chemical dichotomy is found in both sources: on the one hand, 
the nuclear regions have high H$_2$O abundances,
$\sim$$10^{-5}$, and high HCN/H$_2$O and HCN/NH$_3$ column
density ratios of 0.1$-$0.4 and 2$-$5, respectively, indicating a
chemistry typical of evolved hot cores where grain mantle evaporation
has occurred. On the other hand, the high OH abundance,
with OH/H$_2$O  ratios of $\sim$0.5, indicates the effects
of X-rays and/or cosmic rays. The nuclear media have high surface
  brightnesses ($\gtrsim$$10^{13}$ \Lsun/kpc$^2$) and are estimated to be very
  thick ($N_{\mathrm{H}}$$\gtrsim$$10^{25}$ cm$^{-2}$).
While NGC~4418 shows weak absorption in H$_2^{18}$O and
$^{18}$OH, with a $^{16}$O-to-$^{18}$O ratio of $\gtrsim$250$-$500, the
relatively strong absorption of the rare isotopologues in Arp~220 indicates 
$^{18}$O enhancement, with $^{16}$O-to-$^{18}$O of 70$-$130.
Further away from the nuclear regions, the H$_2$O abundance decreases to
$\lesssim$$10^{-7}$ and the OH/H$_2$O ratio is reversed relative to the
nuclear region to 2.5$-$10.  
Despite the different scales and morphologies of NGC~4418, Arp~220, and Mrk
231, preliminary evidence is found for an evolutionary sequence from infall,
hot-core like chemistry, and solar oxygen isotope 
ratio to high velocity outflow, disruption of the hot core chemistry and
cumulative high mass stellar processing of $^{18}$O. 
}


   \maketitle
%

\section{Introduction}

   From the first CO observations of IRAS galaxies \citep{you84}, much
   effort has been devoted to understanding spectroscopic
   observations of molecules in (ultra)luminous infrared galaxies (hereafter,
   (U)LIRGs), and how they are related to the conditions traced by the
   continuum and line observations of atoms and ions. The launch of the
   Infrared Space Observatory (ISO) and, more recently, the 
   Herschel Space Observatory, have opened a new window to study 
   extragalactic sources in the far-infrared (hereafter, far-IR) domain, where
   the bulk of the (U)LIRG luminosity is emitted. The relevance of far-IR
   molecular spectroscopy of (U)LIRGs is well illustrated with the 
   initial ISO result of \cite{fis99}, that the fine-structure
   lines of atoms and ions decrease in strength, as absorption in
   molecular lines, specifically \hdo\ and OH, increase in depth. 

   Water (\hdo) and hydroxyl (OH) are indeed key molecular species that trace
   relevant physical and chemical properties of the interstellar medium in
   luminous infrared extragalactic sources. Powerful OH mega-masers are
     common in ULIRGs, and \hdo\ masers are found in AGN accretion disks and
     in post-shocked gas associated with nuclear jets \citep[see][for a
     review]{lo05}. With many transitions lying at
   far-IR wavelengths, where the bulk of the luminosity is emitted,
   \hdo\ and OH are mainly excited through absorption of far-IR photons
   \citep[][hereafter G-A04]{gon04}, and thus their excitation helps
     probe the general properties of 
   the underlying far-IR continuum sources \citep{gon08}. Chemically, \hdo\ is
   expected to trace an undepleted chemistry where grain mantles are
   evaporated; both \hdo\ and OH are also boosted in shocks and, especially
     OH, in X-ray and Cosmic Ray Dominated Regions (XDRs and
     CRDRs). Furthermore, OH  
   has turned out to be a unique tracer of massive molecular outflows
   \citep{fis10,stu11}. Recently, \hdo\ has been detected in high-$z$ sources
   \citep{imp08,omo11,lis11,wer11,bra11}, and thus understanding its
   role in the local Universe will be crucial for future routine observations
   of \hdo\ with ALMA in the far Universe. HCN is another key molecule widely
   studied in local (U)LIRGs at (sub)millimeter wavelengths
   \citep[e.g.,][]{aal95}. 
   NH$_3$, first detected in extragalactic sources by
     \cite{mar79}, is another important tracer; in Arp~220, it was detected
   at far-IR wavelengths with ISO 
     (G-A04) and at centimeter wavelengths by \cite{tak05} 
      and recently by \cite{ott11}.
   With its high sensitivity and spectral resolution at far-IR wavelengths,
   the {\em Herschel}/PACS instrument \citep{pil10,pog10} 
   provides new opportunities for using the far-IR transitions of these
     molecules to probe the kinematic, chemical, and radiative activity in IR
     luminous galaxies.

   We report in this work 
   Herschel/PACS spectroscopic observations of \hdo, OH, their $^{18}$O
   isotopologues, NH$_3$, and the surprising detection of highly excited
   HCN, in NGC~4418 and Arp~220. NGC~4418 is a peculiar, single
     nucleus galaxy, with a moderate luminosity ($\sim10^{11}$ \Lsun) but
   with other properties similar 
   to warm ULIRGs: a high $L_{\mathrm{FIR}}/M_{\mathrm{H_2}}$ ratio and
   an extreme [C {\sc ii}] deficit \citep[e.g.][]{gra11}, a high 
   continuum surface brightness, an extremely compact luminosity
   source \citep{eva03}, and warm infrared colours  
   ($S_{25\mathrm{\mu \,m}}/S_{60\mathrm{\mu m}}=0.23$, $S_{60\mathrm{\mu
       \,m}}/S_{100\mathrm{\mu m}}=1.37$). We adopt a distance 
   to the source of 29 Mpc. Arp~220 is the prototypical and nearest ULIRG,
   showing a double nucleus and a luminosity of 
   $\approx1.3\times10^{12}$ \Lsun. We adopt a distance to the source of 72
   Mpc. The observations are described in \S\ref{sec:obser}; models are
     shown in \S\ref{sec:models}; model results are discussed in
     \S\ref{sec:discussion}, and our main conclusions are summarized in
     \S\ref{sec:conclusions}.


\section{Observations and results}
\label{sec:obser}

\subsection{Observations}

   The full ($52.3-98$, $104.6-196$ $\mu$m), high resolution, PACS
   spectra of NGC~4418 and Arp~220, taken as 
   part of the guaranteed-time key programme SHINING, were observed on July
   27th and February 27th (2010), respectively. 
The spectra were taken in high spectral sampling density mode using first and
second orders of the grating. 
The velocity resolution of PACS in first order ranges from $\approx320$ \kms\
at 105 $\mu$m to $\approx180$ \kms\ at 190 $\mu$m, and in second order from
$\approx210$ \kms\ at 52 $\mu$m to $\approx110$ \kms\ at 98 microns.  For NGC
4418, we also present a scan around the [O {\sc i}] 63 $\mu$m line taken in
third order with velocity resolution of $\approx85$ \kms. 
The data reduction was mostly done using the standard PACS reduction and
calibration pipeline (ipipe) included in HIPE 5.0 975. The reduction of
Arp~220 required, however, two additional steps in order to correct for
pointing errors that resulted in significant variations of the 
continuum level of the central spaxel\footnote{PACS has $5\times5$
    spatial pixels (spaxels), each with $9''\times9''$ resolution and covering
    a field of $47''\times47''$ \citep{pog10}.} as well as in wavelength
  offsets generated by pointing errors along the dispersion direction of the
  slit \citep{pog10}. These effects were produced by the 
movement of the source between the central spaxel and a
neighbouring spaxel. In order to correct for them, $(i)$ the emission
  from the two spaxels with highest signal were added together in an
intermediate step in the pipeline, which significantly improved the
continuum baseline of the combined spectrum but increased the noise by a 
factor\footnote{ 
    Since the neighbouring spaxel has a signal much lower than the central
    one, the subsequent final calibration leaves the factor $\sqrt{2}$ nearly
    unchanged.} $\approx\sqrt{2}$; 
  $(ii)$ the time evolution of the pointing shifts was
  reconstructed from the star tracker frames, and wavelength calibration was
  corrected according to Fig.~8 by \cite{pog10}. Pointing errors
  across the slit were found from 0 to $\sim3''$, resulting in velocity shifts
  of $\lesssim60$ \kms. We estimate that the associated uncertainties are below
  30 \kms, though they may be higher for some individual lines. 

   \begin{figure}
   \centering
   \includegraphics[width=8.0cm]{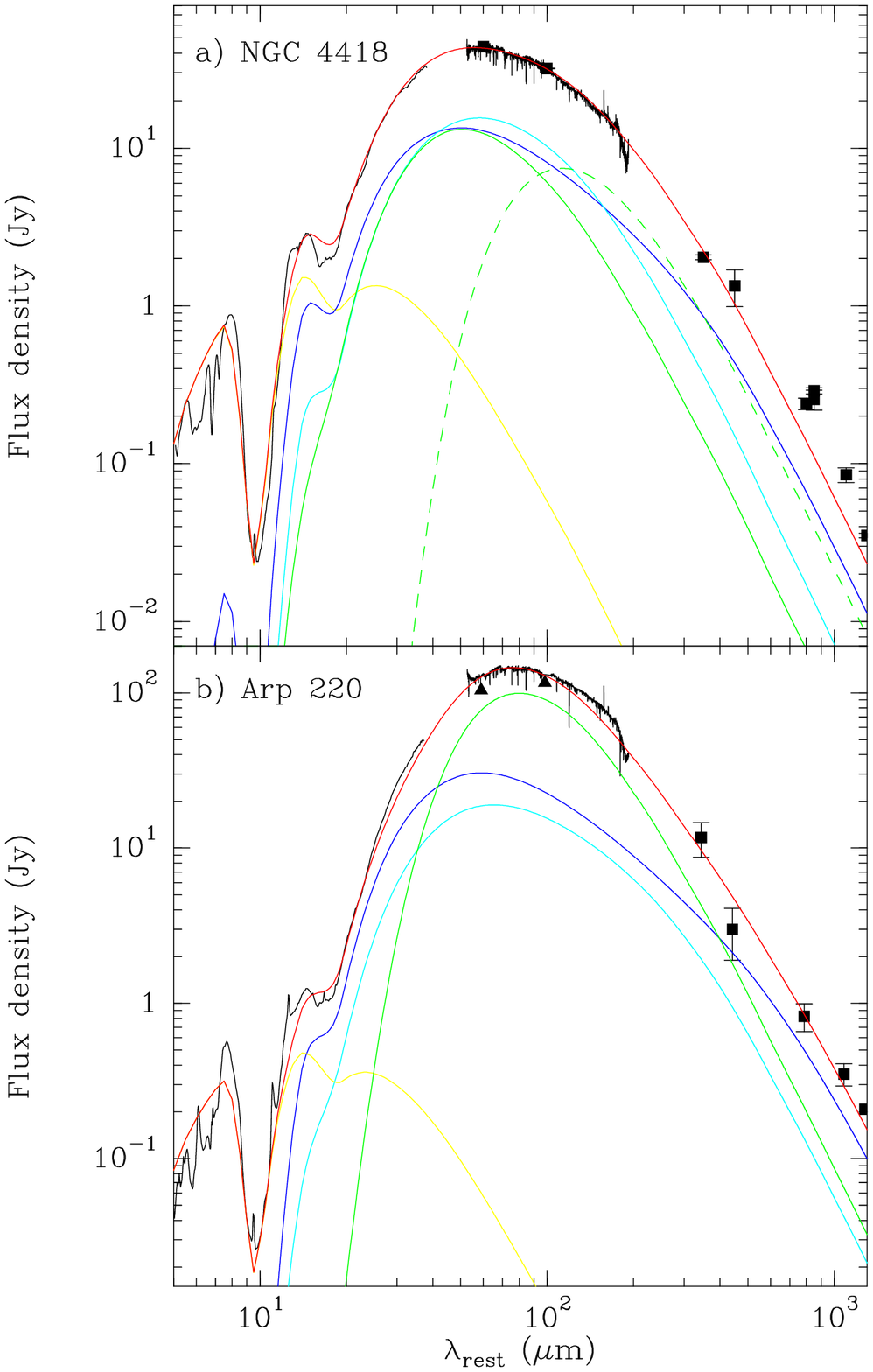}
   \caption{Spectral energy distribution of NGC~4418 and Arp~220 from mid-IR
     to millimeter wavelengths. Both the Herschel/PACS and the Spitzer/IRS
     spectra are shown. Data points at (sub)millimeter wavelengths for NGC
     4418 are from \cite[350 $\mu$m]{yan07}, \cite[450, 800, and 
     1100 $\mu$m]{roc93}, \cite[850 $\mu$m]{dun00,lis00}, and
     \cite[1300 $\mu$m]{alb07}. For Arp~220, they are taken from   
     \cite[450 $\mu$m]{eales89}; \cite[350, 800, and 1100 $\mu$m]{rigo96};
     and \cite[1300 $\mu$m]{saka99,sak08}. Models discussed in
     \S\ref{sec:models} are included. For NGC~4418, the yellow, blue,
     light-blue, and solid-green curves show the models for the hot, core,
     warm, and extended components, respectively, with
     parameters given in Table~\ref{tab:cont}. An additional cold component,
     shown with the dashed-green curve, is included to better fit the SED at
     long wavelengths. For Arp~220, the yellow, blue, light-blue, and
     solid-green curves show the models for the hot, western nucleus,
     eastern nucleus, and extended components, respectively
     (Table~\ref{tab:cont}).  
}   
   \label{continua}
    \end{figure}

   \begin{figure*}
   \centering
   \includegraphics[width=16.0cm]{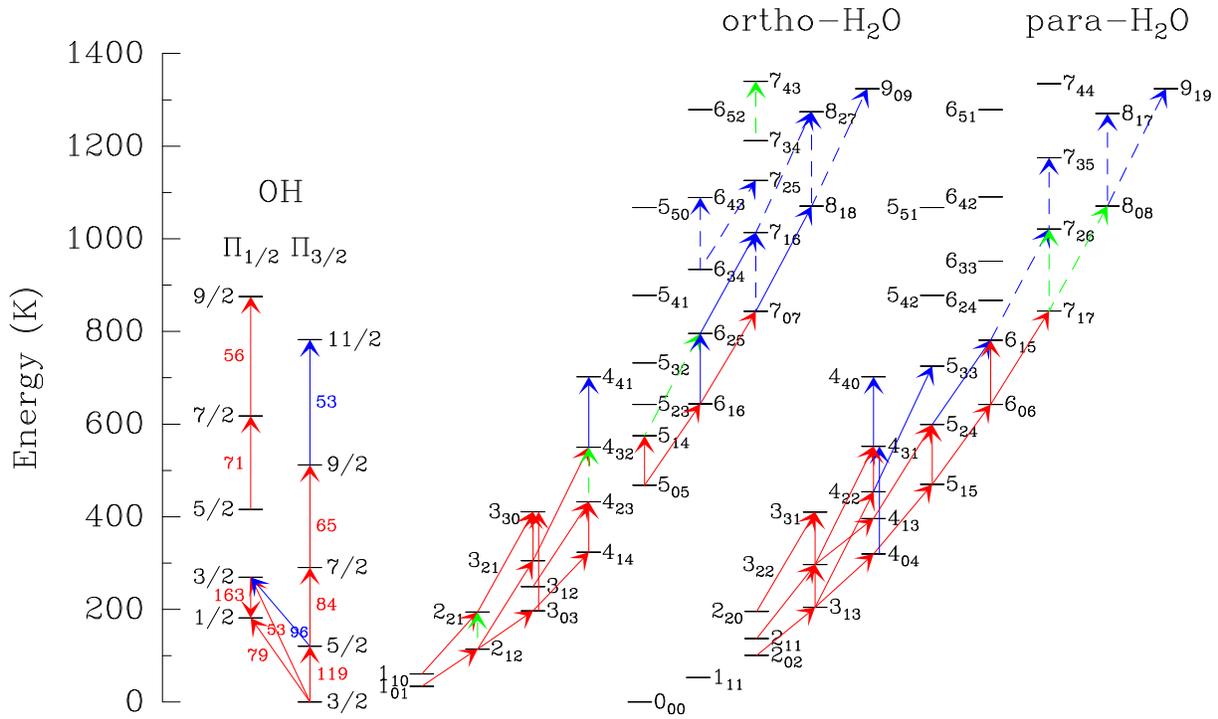}
   \caption{Energy level diagrams of OH and \hdo\ (ortho and para). Red arrows
     indicate lines detected in both sources; blue arrows denote lines
     detected only in NGC~4418, green-dashed arrows mark transitions
     contaminated (blended) by lines of other species, and blue-dashed arrows
     denote marginal detections in NGC 
     4418. The small numbers in the OH diagram indicate rounded
     wavelengths in $\mu$m.}  
    \label{enerdiag}
    \end{figure*}

For the final calibration, the spectrum was normalized to
the telescope flux and recalibrated with a reference telescope spectrum
obtained from observations of Neptune.  
The flux density of the spectrum in the central spaxel(s) was then
rescaled to the value of the total continuum of the source obtained by the
combination of the 25 spaxels of the PACS spectrometer. This method assumes
that the distribution of the IR continuum and the lines are similar for the
spatial resolution of the observations. Indeed the line to continuum
  ratio in the central and the combined 25 spaxels was found to be the same
  for all strong lines except for the ground-state OH 119 and 79 $\mu$m in Arp
  220, which display 9\% and 3\% deeper absorption in the central
  spaxels. The latter OH lines thus appear to show low-level reemission
  (probably through scattering) over kpc spatial scales. In general,
  the comparison indicates that both sources are at far-IR wavelengths
point-sources compared with the PACS $9''\times9''$ spatial resolution
($\sim1.3\times1.3$ kpc$^2$ for NGC~4418 and $\sim3.1\times3.1$ kpc$^2$ for
Arp~220).

   The continua of both sources are shown in Fig~\ref{continua}, including the
   Spitzer IRS spectra \citep{spo07}; the models are discussed in
   \S\ref{sec:models}. Both galaxies show strong silicate absorption at
   9.7 $\mu$m, with PAH emission that is weak in Arp 220 and only
     weakly detected at 11.2 $\mu$m in
   NGC~4418 \citep{spo07}. Besides the obvious differences in luminosity, a
   striking contrast between both sources is the
   $S_{60\mu\mathrm{m}}/S_{25\mu\mathrm{m}}$ ratio, which is $\approx4$ in NGC
   4418 and approaches $\approx13$ in Arp~220. The higher 
   value in Arp~220 indicates that larger amounts of dust surround the
   nuclei in an extended region (ER in G-A04, denoted here as \cext),
     both extinguishing the nuclear emission at 25 $\mu$m and
   contributing to the far-IR continuum. This emission can be
   interpreted as heating of dust by the radiation emanating from the nuclei
   \citep[][G-A04]{soi99}. In NGC~4418, the warmer spectral energy
   distribution (SED) suggests that the amount of dust in \cext\ is 
   lower, but still probably significant (see \S\ref{sec:models}). On the
   other hand, the (sub)millimeter emission in Arp~220 is dominated by the
   nuclei as revealed by interferometric (sub)millimeter observations
   \citep{dow07,sak08}, indicating extreme opacities in, and high luminosities
   from the nuclei. No interferometric observations are available for NGC
   4418.

\begin{figure*}
   \centering
   \includegraphics[width=14.0cm]{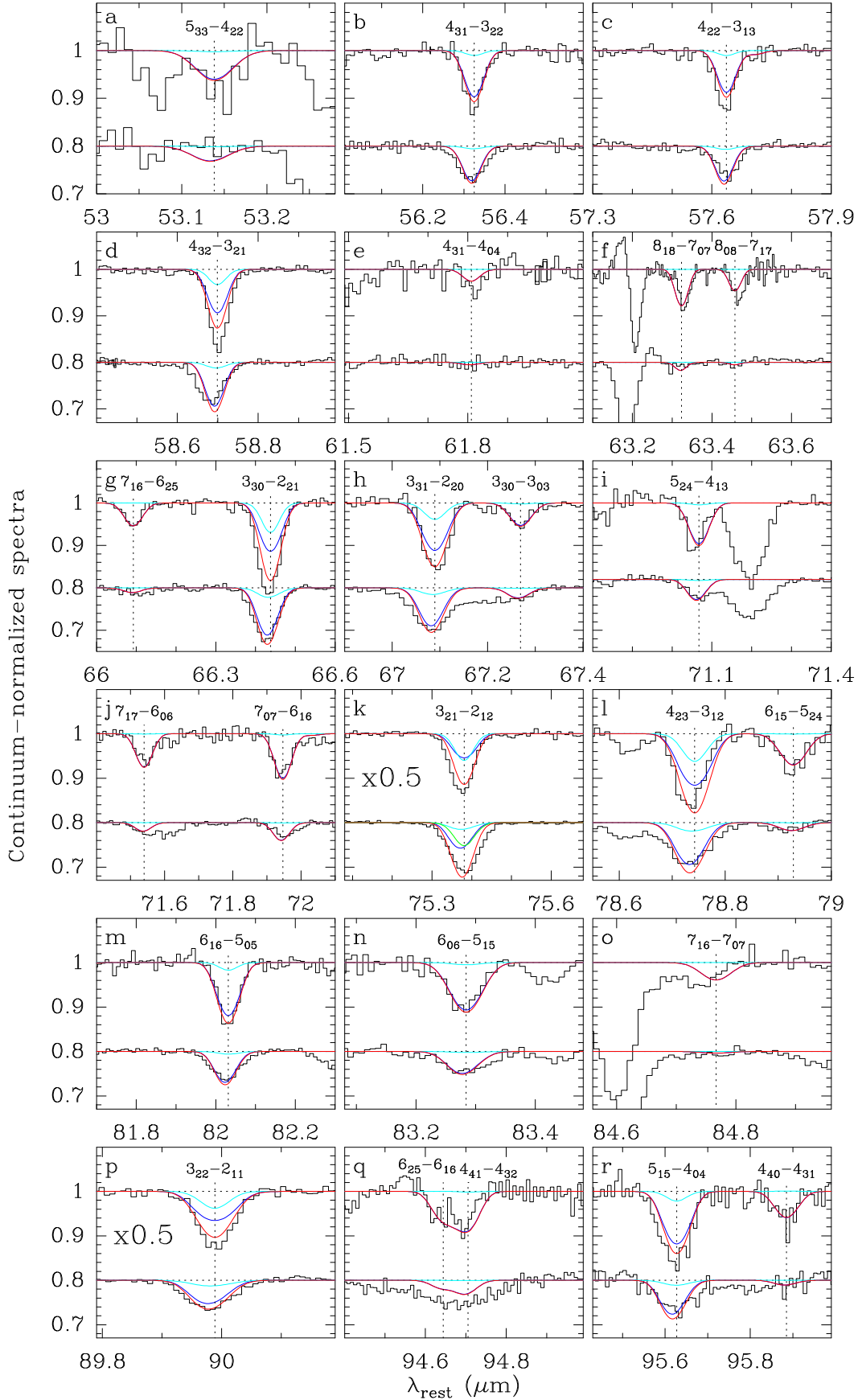}
   \caption{\hdo\ lines in the blue and green bands ($\lambda<100$ $\mu$m)
     detected in NGC~4418 (upper spectra in each panel) and Arp~220
     (lower spectra). The spectra, histograms with black solid
       lines, have been 
     scaled by a factor of $0.5$ in panels k and p. Vertical dotted lines
     indicate the rest wavelengths of the lines relative to $z=0.00705$ and
     $0.0181$ for NGC~4418 and Arp~220, respectively.
     Model results are also shown. For NGC~4418, the blue and
     light-blue curves show the models for the \ccore\ and \cwarm\
     components, respectively; for Arp~220, the blue and
     light-blue lines show the estimated contributions from \cwest\ and
     \ceast, and the green curve 
     (in panel k) shows the contribution by \cext.
     Red is total. Model parameters are given in Tables~\ref{tab:cont} and
     \ref{tab:lines}. 
}
    \label{h2o}
    \end{figure*}

\begin{figure*}
   \centering
   \includegraphics[width=14.0cm]{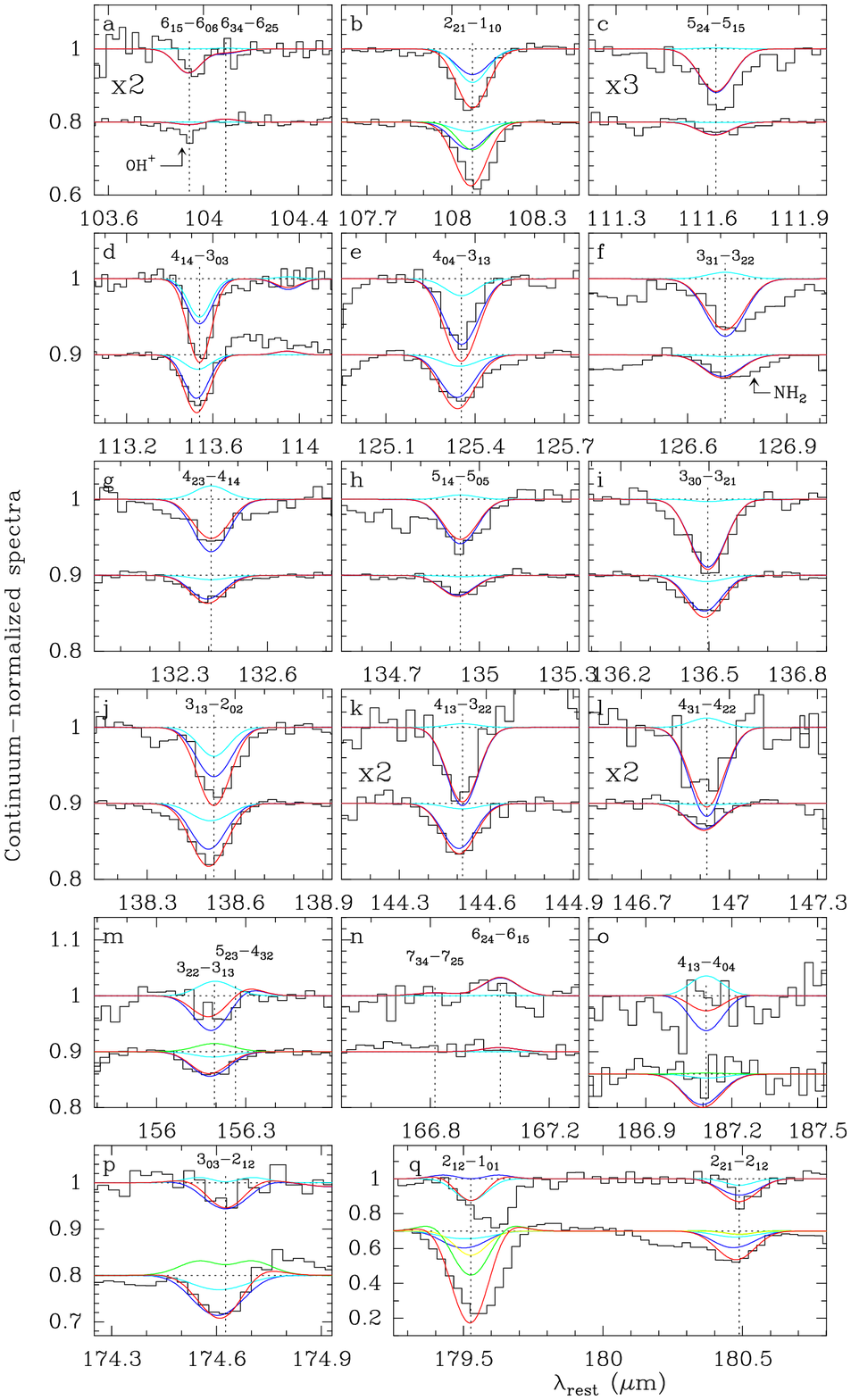}
   \caption{\hdo\ lines in the red band ($\lambda>100$ $\mu$m) detected in NGC
     4418 (upper spectra in each panel) and Arp~220 (lower 
       spectra). The spectra have been scaled by a 
     factor of $2$ in panels a,k and l, and by 3 in panel c. Model results are
     also shown. Color codes for the contributions of the different
     components are given in Fig.~\ref{h2o}, with the yellow curve in panel q
     showing the additional contribution by the halo component (\chalo) in
     front of the nuclear region of Arp~220.}
    \label{h2o-b}
    \end{figure*}

    Spectroscopic data used for both line identification and radiative
     transfer modeling were taken from the JPL \citep{pic98} and CDMS
     \citep{mul01,mul05} catalogues.
   Figure~\ref{enerdiag} shows the energy level diagrams of OH and \hdo.
   The spectra around the \hdo\ lines are
   shown in Fig.~\ref{h2o} (blue and green bands, $\lambda<100$ $\mu$m) and
   Fig.~\ref{h2o-b} (red band, $\lambda>100$ $\mu$m), while high-lying \hdo\
   lines tentatively detected in NGC~4418 
   (blue-dashed arrows in Fig~\ref{enerdiag}) are shown in
   Fig.~\ref{h2o-c}. The OH line profiles are presented in
   Fig.~\ref{oh}. Figs.~\ref{h218o} and \ref{18oh} show excerpts of the
   PACS spectra around the H$_2^{18}$O and $^{18}$OH lines, respectively. 
   The HCN and NH$_3$ lines are displayed in Figs.~\ref{hcn} and
     \ref{nh3}, respectively. Figure~\ref{inoutflow}
   compares the line profiles of the [O {\sc i}] 63 and 145 $\mu$m 
     transitions with those of some OH lines. Figure~\ref{saka}
     compares the line shapes of two HCO$^+$ transitions observed with high
     spatial resolution by \cite{sak09} in Arp 220, with selected lines
     observed with Herschel/PACS. Results of the continuum and line modeling are
     summarized in Tables~\ref{tab:cont} and \ref{tab:lines}, and the
     line equivalent widths of the \hdo, [O {\sc i}] and OH, HCN, NH$_3$,
     H$_2^{18}$O, and $^{18}$OH lines are listed in
     Tables~\ref{tab:fluxesh2o} to \ref{tab:fluxes18oh}.

\subsection{Systemic redshifts}
\label{sec:redshifts}

   Line parameters were derived by subtracting baselines and fitting Gaussian
   curves to all spectral features.
   One relevant question in our study is the zero-velocity adopted for the 
   high-excitation molecular gas in the nuclear region of the galaxies.
   In NGC~4418, the excited H$_2$O lines do not show any evidence
    of systematic shifts indicative of outflowing or inflowing
   gas. Fig~\ref{h2o-fits} indicates that the only \hdo\ line 
   clearly redshifted relative to all others 
   is the ground-state $\t212101$ at 179.5 $\mu$m,
   which is expected to be contaminated by CH$^+$ (2-1) (see below). From all
   other \hdo\ lines, the average redshift is $z=0.007035\pm0.000070$ 
   ($cz=(2110\pm20)$ \kms), where the error denotes the standard
     deviation of the mean. The inferred redshift is fully consistent
     with observations 
   of CO, CN, HNC, and other molecular species observed at millimeter
   wavelengths \citep[e.g.][]{aal07}. The Sloan Digital Sky Survey (SDSS)
   spectrum of NGC~4418 also gives a consistent redshift for the optical
     absorption/emission lines of $z=0.0070$ and $0.0072$, respectively
     (D. Rupke, private communication). Finally, this 
    is also consistent with the redshift measured for the high-lying OH lines
    (see below), which also trace the nuclear component, and with that of
   the [C {\sc ii}] emission line at $157.741$ $\mu$m, $z=0.007050$, which we
   adopt as the reference value.
   In Arp~220, \cite{sak09} detected P-Cygni profiles in excited submillimeter
   lines of HCO$^+$ toward both nuclei, indicating dense outflowing gas in
   the nuclei (see also Fig.~\ref{saka} and \S\ref{sec:kinematics}). The
   transition from absorption to emission occurs at a velocity 
   of $\approx5420$ \kms\ toward both nuclei, which we adopt as the systemic
   redshift ($z=0.0181$). The foreground absorption traced by CO (3-2) toward
   the western nucleus lies at a lower velocity, $\approx5350$ \kms.
   The line velocity shifts ($V_{\mathrm{shift}}$) relative to the above
   systemic redshifts, the line widths, and the equivalent widths
   ($W_{\mathrm{eq}}$) are plotted in Fig.~\ref{h2o-fits} for the \hdo\ lines
   and in Fig.~\ref{oh-fits} for the OH lines.

\subsection{\hdo\ and OH}
\label{sec:obser_h2o_oh}

   Based on the \hdo\ spectra and the molecular line positions shown in
   Figs.~\ref{h2o} and \ref{h2o-b}, most \hdo\ lines are free of contamination
   from lines of other species, with the following exceptions: 
   as mentioned above, the \hdo\  \t212101\ line
   in Fig.~\ref{h2o-b}q at 179.5 $\mu$m is blended with CH$^+$; the \t808717\
   line in Fig.~\ref{h2o}f is blended with highly excited NH$_3$ (see
   Fig.~\ref{nh3}a); the \t524413\ (Fig.~\ref{h2o}i) and the \t221212\ lines 
   (Fig.~\ref{h2o-b}q) may have some contribution from CH; finally, the
   spectral feature at 104 $\mu$m associated in Fig.~\ref{h2o-b}a with the
   \t615606\ line is probably contaminated by OH$^+$. Other contaminated lines
   marked with dashed-green arrows in Fig.~\ref{enerdiag} are not 
   shown. The OH lines in Fig.~\ref{oh}
   have relatively weak contamination by NH$_3$ in the red $\Lambda-$component
   of the 84 $\mu$m doublet (panel e, see Fig.~\ref{nh3}c), and by \hdo\ in
   the blue $\Lambda-$component of the 65 $\mu$m doublet (panel f). 

   The detections summarized in Fig.~\ref{enerdiag} indicate extreme
   excitation in the nuclei of both galaxies. In Arp~220, lines with lower
   level energy up to $E_{\mathrm{lower}}\approx600$ K in both \hdo\ 
   (\t707616\ and \t717606, Fig.~\ref{h2o}j, the latter contaminated by
   excited NH$_3$) and OH ($\Pi_{1/2} 9/2\rightarrow7/2$, Fig. \ref{oh}i)
   are detected. NGC~4418 shows even higher-lying lines,
   with detected \hdo\ lines as high as \t818707\ and
   \t716625\ (Fig.~\ref{h2o}f,g; $E_{\mathrm{lower}}\gtrsim800$ K). These are
   the highest-lying \hdo\ lines detected in an extragalactic source to
   date.   

   The relative strengths of the \hdo\ and OH lines in NGC~4418
   and Arp~220 show a striking dependence on $E_{\mathrm{lower}}$. The
   lowest-lying ($E_{\mathrm{lower}}\lesssim200$ K) \hdo\ lines, especially
   the \t212101\ transition at $179.5$ $\mu$m (Fig.~\ref{h2o-b}q), are
   stronger in Arp~220 than in NGC~4418, but higher-lying lines are stronger
   in NGC~4418 (see 
   Fig.~\ref{h2o-fits}). The effect is even more clearly seen in OH 
   (Fig.~\ref{oh-fits}), where the ground-state lines (upper row in
   Fig.~\ref{oh}) are much stronger in Arp~220, the intermediate $\Pi_{3/2}
   5/2\rightarrow3/2$ transitions at 84 $\mu$m (Fig. \ref{oh}e) 
     have similar strengths, and higher excited OH lines
   (Fig. \ref{oh}f,g,h,i) are stronger 
   in NGC~4418. This is partially a consequence of the \cext\ component
   in Arp~220: since $W_{\mathrm{eq}}$ are calculated relative to the observed
   (nuclear+\cext) continuum, the high-lying lines arising from the nuclei are
   reduced due to the contribution by \cext\ to the
   continuum, which is more prominent in Arp~220. 
   In other words, the high-lying lines
   are more continuum-diluted in Arp~220 than in the relatively ``naked'' NGC
   4418, which is consistent with the continuum models shown in
   \S\ref{sec:models}. On the other hand, \cext\ in Arp~220 produces
   strong absorption in the low-lying lines, thus increasing their
   $W_{\mathrm{eq}}$ relative to that in NGC~4418.   

   Nevertheless, dilution in the continuum cannot explain the higher
   excitation indicated by the very high-lying lines of \hdo\ and OH 
   in NGC~4418. This is best seen in Fig.~\ref{h2o-fits}gh, where the line
   fluxes are normalized relative to that of the excited \t330221\ \hdo\
   line at 66.4
   $\mu$m. The latter line (Fig.~\ref{h2o}g) is well detected in both sources
   and, with $E_{\mathrm{lower}}=160$ K, is not expected to be significantly
   contaminated by absorption of extended low-excitation \hdo\ in Arp~220. 
   Lines with $E_{\mathrm{lower}}$ above 500 K are shown to produce, relative
   to \t330221, more absorption in NGC~4418 than in Arp~220.
   Also, as mentioned above, the \hdo\ \t818707\ and
   \t716625\ transitions (Fig.~\ref{h2o}f,g) are only detected in
   NGC~4418 (the \t808717\ 
   line in  Fig.~\ref{h2o}f is contaminated by NH$_3$, and the
   feature is not detected in Arp~220). Two more para-\hdo\ transitions
   detected only in NGC~4418 are the \t533422\ and \t440431\ lines
   (Fig.~\ref{h2o}a,r); the lower levels of these lines are non-backbone and
   thus require high columns to be significantly populated.  
   Other \hdo\ lines detected only in NGC~4418 are the \t625616\ and \t441432\
   at $94.7$ $\mu$m, but the Arp~220 spectrum at this wavelength shows a very
   broad feature, possibly due to pointing or instrumental effects
   (Fig.~\ref{h2o}q). 

   \begin{figure}
   \centering
   \includegraphics[width=9.0cm]{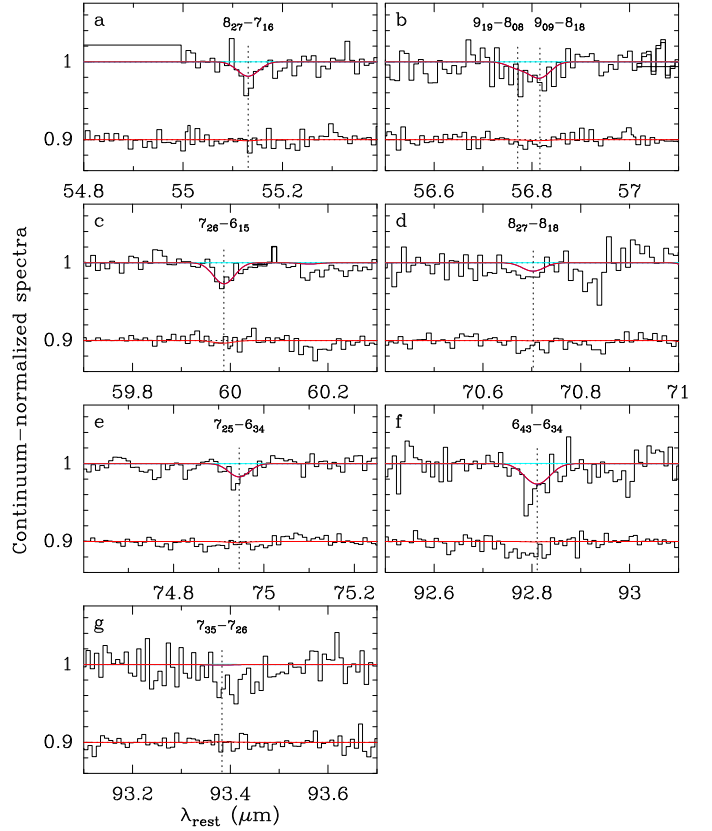}
   \caption{Marginally detected \hdo\ lines in NGC~4418
     (upper spectra in each panel); the corresponding lines from
     Arp~220 are also shown (lower spectra). Model predictions
     for these lines (for \ccore\ in NGC~4418 and for \cwest\ in
     Arp~220) are also shown (red curves).} 
    \label{h2o-c}
    \end{figure}

   \begin{figure*}
   \centering
   \includegraphics[width=12.0cm]{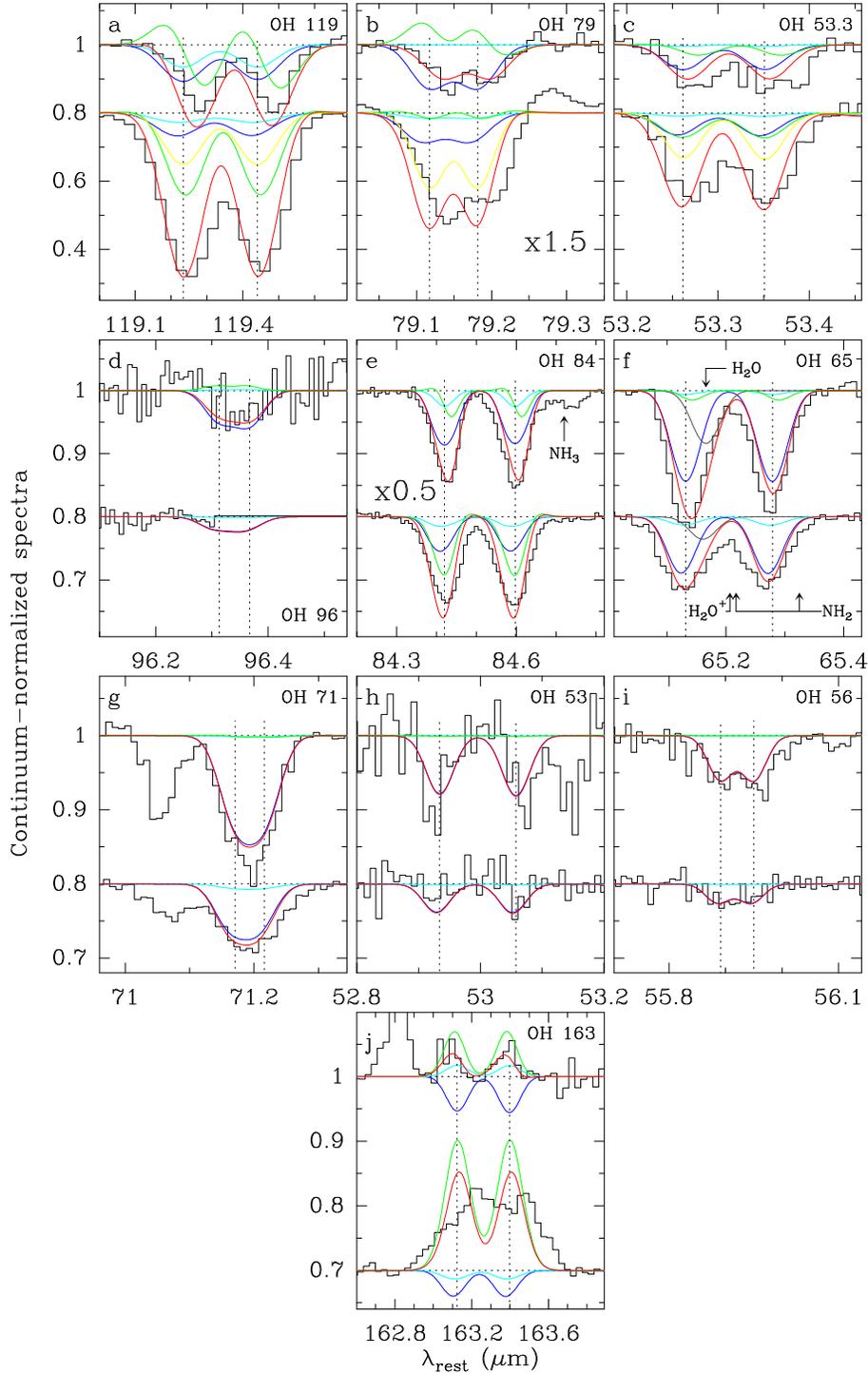}
   \caption{OH lines in NGC~4418 (upper spectra) and Arp~220
     (lower spectra). $\Lambda-$doubling is resolved in most
     transitions, except for 
     the two narrowly separated 71 $\mu$m components that are blended into a
     single spectral feature. Model results are also shown. For NGC~4418, 
     the blue, light-blue, and green curves show the models for the 
     \ccore, \cwarm, and \cext\ components, respectively; for Arp~220, the
     blue, light-blue, green, and yellow curves show the models for 
     \cwest, \ceast, \cext, and \chalo, respectively. The gray curves in
       panel f show the expected contribution of the \hdo\ \t625514\ line to
       the spectra around $65.2$ $\mu$m . The red curves show the total from
       all components.} 
    \label{oh}
    \end{figure*}

   The spectrum of NGC~4418
   clearly shows two intrinsically weak \hdo\ lines with $\Delta
   K_{\pm1}=\pm3$, the \t431404\ line at $61.8$ $\mu$m and the  \t330303\ line
   at $67.3$ $\mu$m (Fig.~\ref{h2o}e,h). 
   All other detected \hdo\ lines have higher transition
   probabilities and, with the exception of the very high-lying
     lines, are expected to be strongly saturated. 
   While the ortho-\hdo\ $67.3$ $\mu$m line
   is also detected in Arp~220 within a wing-like feature that probably has
   some contribution from H$_2^{18}$O \t330221, the para-\hdo\ $61.8$ $\mu$m
   line is not detected in Arp~220. The high columns in NGC~4418 are also
   confirmed by comparing the absorption in the ortho/para pair \t441432\ and 
   \t440431\ (Fig.~\ref{h2o}q,r), which have similar lower level energies
   ($\approx550$ K), wavelengths, and transition probabilities ($A\approx0.15$
   s$^{-1}$). For an ortho-to-para ratio of 3, one would expect the same ratio
   of 3 for $W_{\mathrm{eq}}$ in the optically thin regime, but this is as low
   as $1.15$ indicating that both lines are still well saturated.

   Even higher-excitation lines
   of \hdo, with $E_{\mathrm{lower}}\approx900-1100$ K, may be present in
   the spectrum of NGC~4418 (blue-dashed arrows in Fig~\ref{enerdiag}). 
   Among these features, shown in Fig~\ref{h2o-c}, the
   most likely detection is the \t726615\ line in Fig~\ref{h2o-c}c, all
   other features being marginal. In particular, the $92.8$ $\mu$m feature
   in (f) has a non-Gaussian shape. Nevertheless, all features together
   appear to indicate more extreme excitation in the nucleus of NGC~4418. This
   is plausible, as HCN rotational lines with $E_{\mathrm{lower}}\gtrsim1075$
   K are detected in NGC~4418 (see below).

   As discussed earlier, dilution in the continuum also lowers the
   $W_{\mathrm{eq}}$ of the 
   high-lying OH lines in Arp~220 relative to the values in NGC~4418;
   however, the most excited OH transitions, the $\Pi_{1/2}\,
   9/2\rightarrow7/2$ and $\Pi_{1/2}\, 11/2\rightarrow9/2$ lines at 56 and 53 
   $\mu$m, respectively, are intrinsically 
   stronger in NGC~4418 according to our decomposition of the continuum. 
   Another effect of \cext\ is also worth noting: the only OH transition
   observed in emission above the continuum, 
   the $\Pi_{1/2}\, 3/2\rightarrow1/2$ doublet at 163 $\mu$m, is excited
   through absorption of continuum photons in the 35 and 53.3 $\mu$m doublets
   followed by a cascade down to the ground OH state \citep[see App.~II
   by][]{gen85}, and is therefore expected to arise in the \cext\ 
   component (G-A04). The
   strong OH 163 $\mu$m emission in Arp~220 reflects the massive envelope
   around its nuclei. In NGC~4418, the doublet is weaker but still
   in emission, also hinting at the existence of a \cext\ component
     around its nucleus.

\subsection{H$_2^{18}$O and $^{18}$OH}

\begin{figure}
   \centering
   \includegraphics[width=9.0cm]{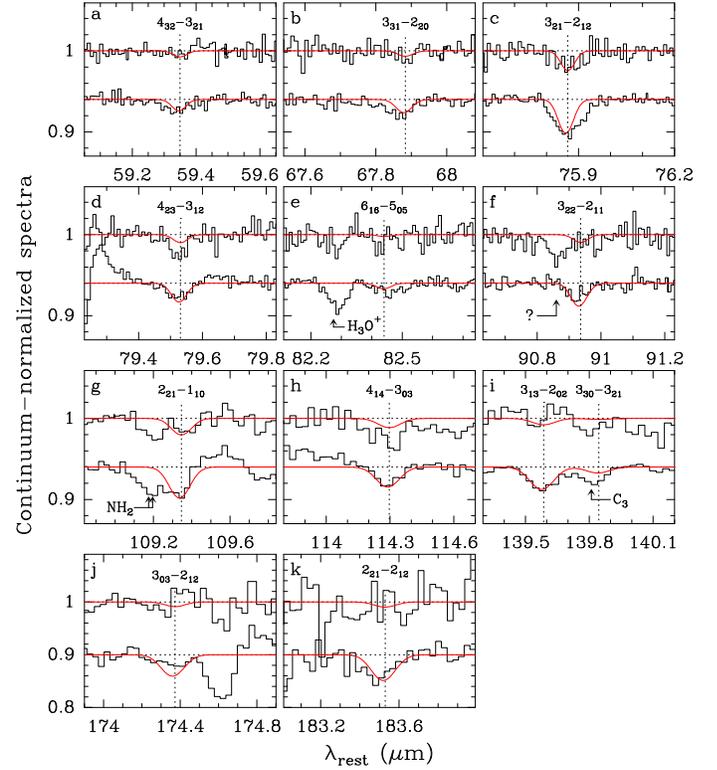}
   \caption{Spectra around the wavelengths of relevant H$_2^{18}$O lines
     in NGC 4418 (upper profiles) and Arp~220 (lower 
       profiles). Model predictions 
     for NGC~4418 (\ccore) with $\mathrm{H_2O/H_2^{18}O}=500$, and for Arp~220
     (\cwest) with $\mathrm{H_2O/H_2^{18}O}=70$ are shown in red.}    
    \label{h218o}
    \end{figure}

\begin{figure}
   \centering
   \includegraphics[width=8.5cm]{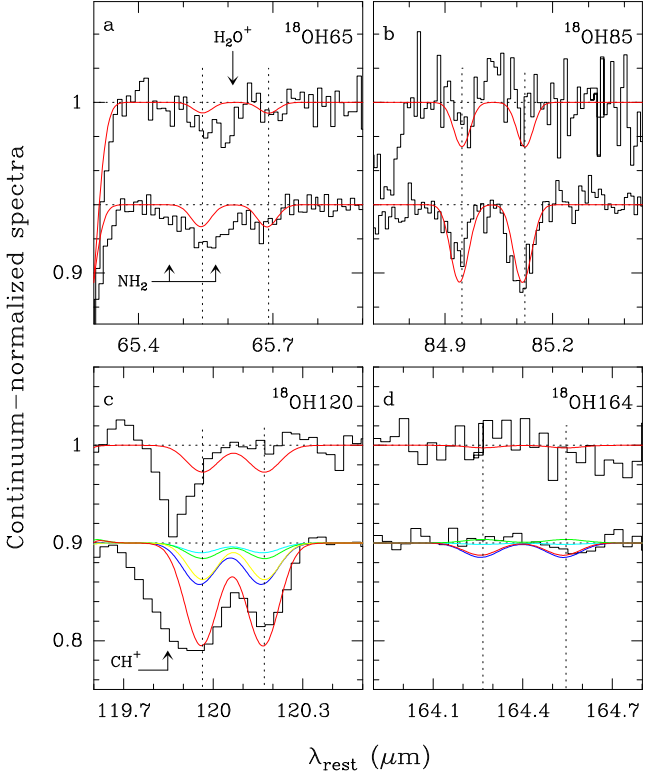}
   \caption{Spectra around the wavelengths of relevant $^{18}$OH lines
     in NGC 4418 (upper profiles) and Arp~220 (lower 
       profiles). Model predictions 
     for NGC~4418 (\ccore) with $\mathrm{H_2O/H_2^{18}O}=500$, and for Arp~220
     with $\mathrm{H_2O/H_2^{18}O}=70$ are also shown. For the $^{18}$OH
     120 $\mu$m doublet, the contributions by the different components 
     in Arp 220 (see Fig.~\ref{oh}) are indicated; red denotes the
       total of all components.}
    \label{18oh}
    \end{figure}

The spectra around relevant H$_2^{18}$O and $^{18}$OH
transitions in NGC~4418 and Arp~220 are displayed in
Figs.~\ref{h218o} and \ref{18oh}. Close spectral features due to H$_3$O$^+$,
NH$_2$, C$_3$, and CH$^+$ are also indicated. H$_2^{18}$O is detected in both
sources, though in relatively low-lying lines. The highest-lying detected line
is the \t432321\ transition with $E_{\mathrm{lower}}\approx270$ K in
Fig.~\ref{h218o}a, as detection of the \t616505\ transition in
panel e is rather marginal. The most striking feature in Fig.~\ref{h218o} is
the relative amount of absorption observed in NGC~4418 and 
Arp~220. While the absorption in the main isotopologues is stronger
in NGC~4418 
than in Arp~220 (Figs.~\ref{h2o} and \ref{h2o-b}), the opposite behavior is
generally found for the absorption in the rare isotopologues. This is best
seen in the \t331220, \t321212, \t221110\ 
(showing redshifted emission in Arp~220), \t313202, and \t221212\ lines. 
There are, however, two spectral features nearly coincident with the
\t423312\ and \t414303\ H$_2^{18}$O lines at $79.5$ and $114.3$ $\mu$m
(panels d and h, respectively), with absorption in NGC~4418 apparently
stronger than in Arp~220, but the latter is subject to an uncertain
  baseline.

The $^{18}$OH lines in Fig.~\ref{18oh}, specifically the $^{18}$OH 85
and 120 $\mu$m doublets, confirm the $^{18}$O enhancement in Arp~220. The
blueshifted $\Lambda-$component of the $^{18}$OH 120 $\mu$m doublet is
strongly contaminated by CH$^+$ $J=3\rightarrow2$ in both sources, but the
uncontaminated redshifted $\Lambda-$component, undetected in NGC~4418, is
strong in Arp~220.  It is also worth noting that the two $^{18}$OH 85
  $\mu$m $\Lambda-$components show rather different absorption depths in
  Arp~220. This $\Lambda-$asymmetry is not expected to arise from radiative
  pumping or opacity effects, because the $\Lambda-$components of any
  transition have the same radiative transition probabilities, and the
  corresponding lower levels tend to be equally populated. Indeed, the
  absorption in the corresponding $\Lambda-$components of the main
  isotopologue at 84 $\mu$m are very similar (Fig.~\ref{oh}e). The $^{18}$OH
  85 $\mu$m $\Lambda-$asymmetry in Arp 220 is further discussed in
  \S\ref{sec:arp220-18o-models}. 
The $^{18}$OH $\Lambda-$component at 65.54 $\mu$m in Fig.~\ref{18oh}a is
partially blended with two lines of NH$_2$ and, together with the expected
contribution by H$_2$O$^+$ at 65.61 $\mu$m, form a broad feature with
uncertain baseline. The red $\Lambda-$component at 65.69 $\mu$m
appears to be stronger in Arp~220 than in NGC~4418.

\subsection{HCN}

HCN is a key molecule widely studied in both Galactic and
extragalactic sources. In NGC~4418 and Arp~220, the low-lying lines at
millimeter wavelengths have been observed by \cite{aal07,aal07b} and
\cite{wie04}. The $\nu_2=1^1$ vibration-rotation band at 14 $\mu$m was
detected in both sources by \cite{lah07}, and rotational emission from the
upper vibrational state in NGC 4418 has been detected by
\cite{sak10}. The $\nu_2=1^1$ l-type HCN lines from $J=4-6$ at centimeter
  wavelengths were detected in absorption toward Arp 220 \citep{sal08}.
Among the 15 extragalactic sources detected in the HCN band, the \cite{lah07}
analysis indicated that NGC~4418 has the second highest HCN column. Recently,
\cite{ran11} have reported the detection of several emission/absorption HCN
lines with Herschel/SPIRE in Arp~220. 

   \begin{figure}
   \centering
   \includegraphics[width=8.5cm]{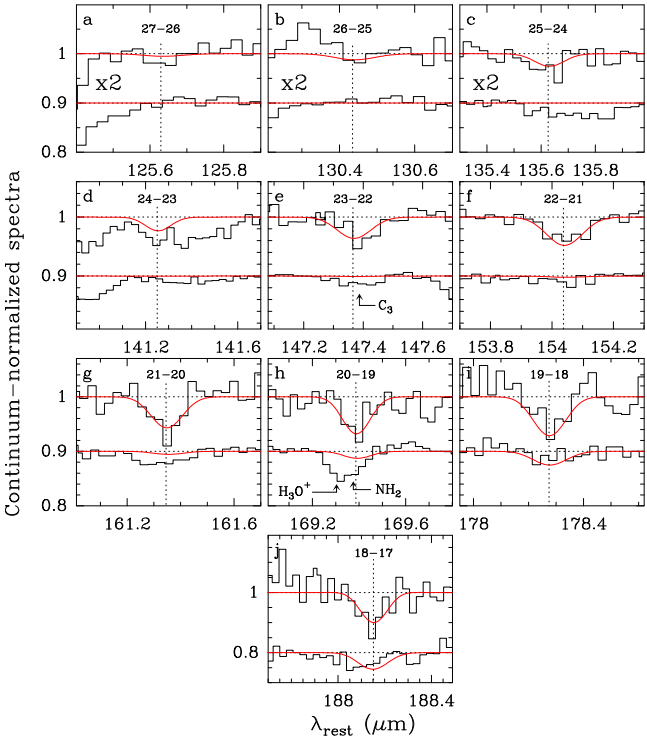}
   \caption{Spectra around the frequencies of HCN lines in NGC~4418
     (upper profiles in each panel) and Arp~220 (lower 
       profiles). Model predictions  
     for NGC~4418 (\ccore) and Arp~220 (\cwest) are included.}    
    \label{hcn}
    \end{figure}

   \begin{figure}
   \centering
   \includegraphics[width=9.0cm]{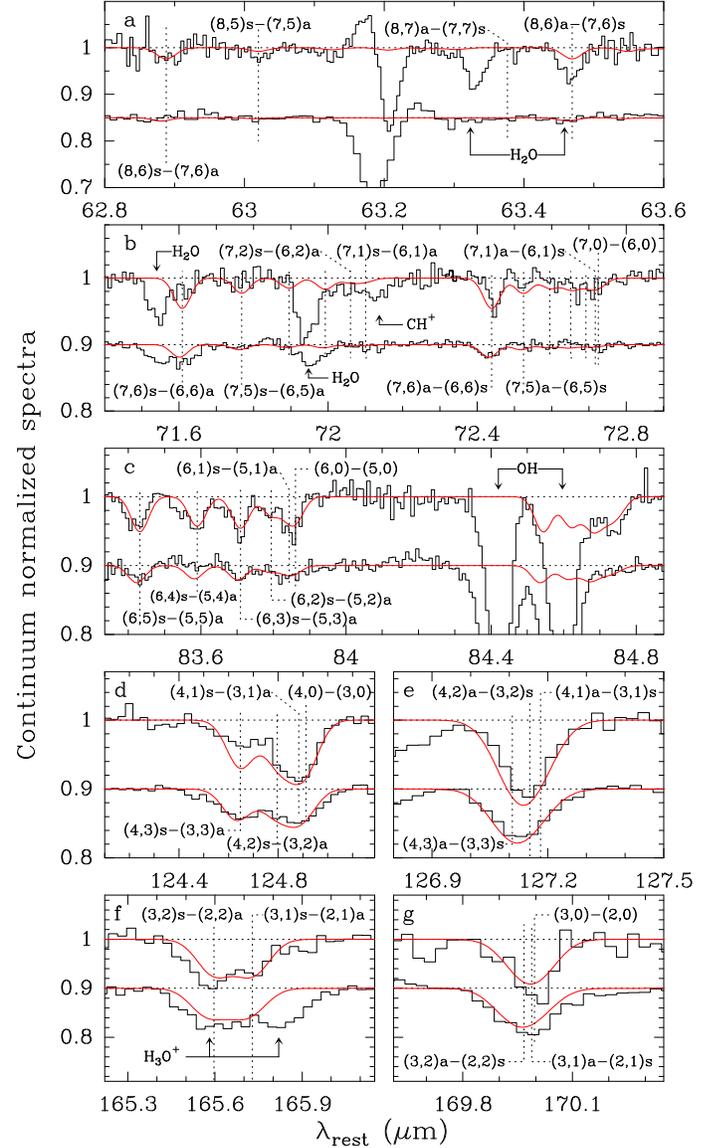}
   \caption{Spectra around the frequencies of relevant NH$_3$ lines in 
     NGC~4418 (upper profiles in each
     panel) and Arp~220 (lower profiles). Model predictions
     for NGC~4418 (\ccore) and Arp~220 (\cwest) are included.}    
    \label{nh3}
    \end{figure}

We report in Fig.~\ref{hcn} the PACS detection of HCN absorption in pure
rotational transitions at far-IR wavelengths in both NGC~4418 and Arp~220. 
In NGC~4418, all transitions from
$J=18\rightarrow17$ ($E_{\mathrm{lower}}=650$ K) to $J=23\rightarrow22$
($E_{\mathrm{lower}}=1075$ K) are clearly 
identified. Furthermore, there are hints of
absorption up to the $J=25\rightarrow24$ transition ($E_{\mathrm{lower}}=1275$
K). In Arp~220, the $J=20\rightarrow19$ line is clearly contaminated
by H$_3$O$^+$ and NH$_2$, and the $J=23\rightarrow22$ one by C$_3$ and
possibly also by NH$_2$. The $J=19\rightarrow18$ line is weak,
suggesting that this transition is tracing the tail of the Spectral Line
Energy Distribution (SLED). However, a
clear feature is found at the wavelength of the $J=21\rightarrow20$ line, 
and the $J=22\rightarrow21$ line could show some marginal absorption 
as well. Nevertheless, we consider the identification of the
$J=21\rightarrow20$ line as questionable given the weakness of the nearby
  rotational HCN lines and the possible contamination by other species.

Fluxes derived from Gaussian fits are shown
in Fig.~\ref{hcn-fits} as a function of $E_{\mathrm{lower}}$, where values
derived from Herschel/Spire observations by \cite{ran11} are included for Arp
220. The HCN excitation in NGC~4418 is extreme, with the SLED apparently
peaking at $J=21\rightarrow20$ ($E_{\mathrm{lower}}\approx900$ K). The
striking characteristic of the far-IR HCN lines is that they are detected in
{\em absorption} against the far-IR continuum, indicating that the dust
temperature and far-IR continuum opacities behind the
observed HCN are high; otherwise the lines would be detected in emission. 
We argue in \S\ref{sec:models} that the \hdo, OH, and HCN lines in NGC~4418
are tracing a high luminosity, compact nuclear source, denoted
as the {\em nuclear core}  (\ccore). In Arp~220, the HCN SLED as seen by
Herschel/Spire \citep{ran11} peaks at $J=16\rightarrow15$
($E_{\mathrm{lower}}\approx500$ K) indicating, like \hdo, more moderate
excitation, but a second, higher-excitation component could be present if the
$161.3$ $\mu$m spectral feature is due to HCN $J=21\rightarrow20$. As shown in
\S\ref{sec:hcnarp220} and \ref{sec:coren4418}, high HCN columns are required
to explain the data.

   \begin{figure}
   \centering
   \includegraphics[width=8.0cm]{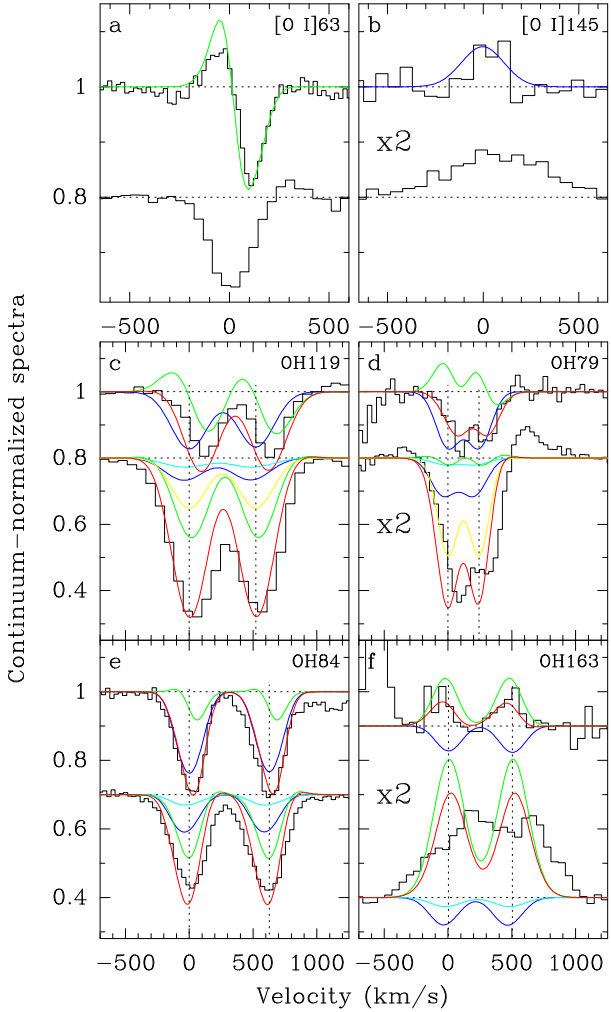}
   \caption{Line profiles of the [O {\sc i}] 63 and 145 $\mu$m (a \& b)
     transitions compared 
     with those of some OH lines (c-f). Upper histograms, NGC~4418; lower
     histograms, Arp~220. Model predictions are also included. For NGC~4418,
     the blue curves show the combined contribution of the \ccore\ and
     \cwarm\ components, and the green curves show the model for
     \cext. For Arp 220, see caption of Fig.~\ref{oh}. Red denotes the
       total of all components.} 
    \label{inoutflow}
    \end{figure}

   \begin{figure}
   \centering
   \includegraphics[width=8.0cm]{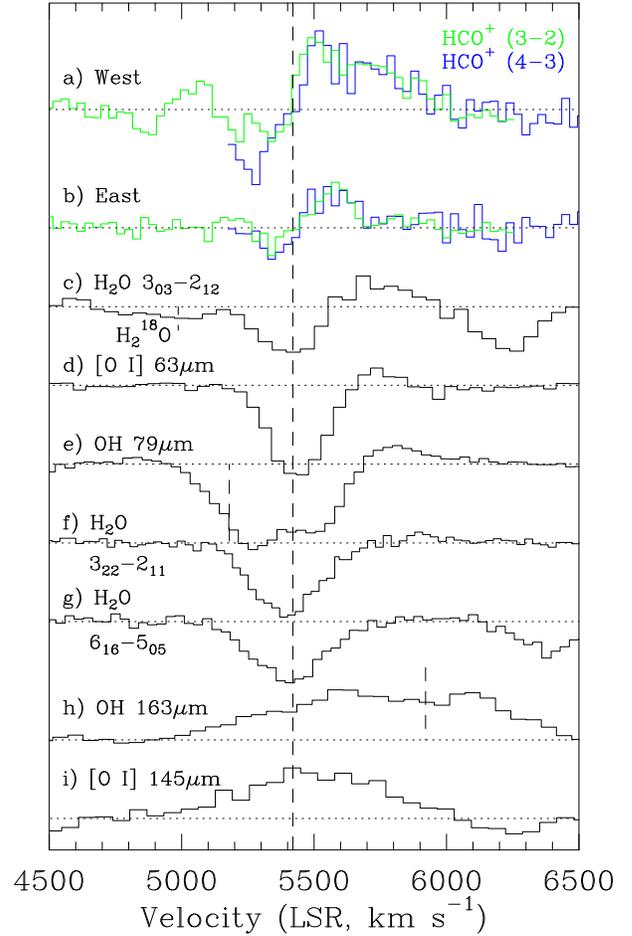}
   \caption{Comparison between the line shapes of the HCO$^+$
     $J=3\rightarrow2$ (green histograms) and $J=4\rightarrow3$ (blue
     histograms) transitions toward (a) the western and (b) eastern nucleus of
     Arp 220 \citep[from][]{sak09}, with selected lines detected with
     Herschel/PACS (c-i). The dashed vertical line indicates our adopted
     systemic velocity (\S\ref{sec:redshifts}). The velocity scale for the OH
     79 (163) $\mu$m doublet is relative to the red (blue)
     $\Lambda-$component, and the systemic velocity for the other
     $\Lambda-$component is also indicated. The velocity of the H$_2^{18}$O
     \t303212\ line in (c) is also indicated.} 
    \label{saka}
    \end{figure}

\subsection{NH$_3$}

Ammonia is another N-bearing species widely observed in Galactic sources
mostly through the pure-inversion transitions from metastable ($J=K$) levels
at centimeter wavelengths \citep[see][for a review]{ho83}. The
rotation-inversion 
transitions lie at far-IR wavelengths, and many of them have been detected
with ISO toward Sgr B2 \citep{cec02,pol07} indicating a hot, low dense
molecular layer interpreted in terms of shock conditions
\citep{cec02}. Detections of far-IR NH$_3$ lines in extragalactic
sources were presented, together with a preliminary analysis,
by G-A04 in Arp~220. In Fig.~\ref{nh3} we show the Herschel/PACS detection
of high-excitation, far-IR NH$_3$ lines in both NGC~4418 and Arp~220. 

NH$_3$ has two species, ortho-NH$_3$ ($K = 3n$) and para-NH$_3$ ($K \neq 3n$),
with an expected ortho-to-para abundance ratio of unity \citep{ume99}. 
All rotational $(J,K)$ levels, except those of the $K=0$ ladder, are split
into 2 inversion doubling sublevels ($a$ for asymmetric and $s$ for
symmetric), which for a given $J$ have decreasing level energy with
increasing $K$.  
Radiative transitions are only allowed within $K-$ladders, thus only
collisions can populate the $K>1$ ladders and the absorption from different
$K-$ladders is then sensitive to the gas temperature. Within a given
$K-$ladder, both 
collisions and absorption of far-IR photons can pump the $J>K$ non-metastable
levels. Since the $a\rightarrow s$ ($s\rightarrow a$) $J+1\rightarrow J$
transitions of different $K-$ladders have similar wavelengths, the lines are
crowded in wavelength and overlapping in some cases, with severe blending
between the $K=0$ and $K=1$ ladders. The PACS domain covers the range
from the $(3,K)a\rightarrow(2,K)s$ ($\approx170$ $\mu$m, Fig.~\ref{nh3}g)
and the $(3,K)s\rightarrow(2,K)a$ ($\approx165.7$ $\mu$m, Fig.~\ref{nh3}f)
lines, up to $(10,K)\rightarrow(9,K)$ at $50-51$ $\mu$m.

Figure~\ref{nh3} also indicates (partial) blending with lines of other
species, as \hdo\ (panels a and b), CH$^+$ (panel b), and H$_3$O$^+$ (panel
f). Most $(6,K)a\rightarrow(5,K)s$ lines are strongly blended with the OH 84
$\mu$m doublet. From the comparison with the modeling described in
\S\ref{sec:models}, the $(6,K)s\rightarrow(5,K)a$ and especially the
$(7,K)a\rightarrow(6,K)s$ lines are so blended that a pseudo-continuum is
expected to be formed, with the consequent uncertainty in the baseline
subtraction. As also found for \hdo, OH, and HCN, NGC~4418 shows in NH$_3$
clear indications of higher excitation than Arp~220, and the absorption in
non-metastable levels relative to metastable ones is also stronger in NGC~4418
(e.g., the $(4,K)s\rightarrow(3,K)a$ group in panel d). 
The ortho-NH$_3$ $(8,6)a\rightarrow(7,6)s$ is probably detected in NGC~4418,
as the blended para-\hdo\ \t808717\ line (see also Fig.~\ref{h2o}f) is
probably less strong than the close ortho-\hdo\ \t818707\ line, and also
because the $(8,6)s\rightarrow(7,6)a$ transition is detected. There are
also some hints of absorption in the $(8,5)s\rightarrow(7,5)a$ line.
In summary, NH$_3$ lines are detected up to $(8,K)\rightarrow(7,K)$ in
NGC~4418 ($E_{\mathrm{lower}}\approx600$ K), and up to $(7,6)\rightarrow(6,6)$ in 
Arp~220 ($E_{\mathrm{lower}}\approx400$ K). 

\subsection{Kinematics}
\label{sec:kinematics}

In Fig.~\ref{saka}, we compare the HCO$^+$ $J=3-2$ (green histograms) and
$J=4-3$ (blue histograms) line profiles toward the western (a) and eastern (b) 
nucleus of Arp 220 \citep[from][]{sak09}, with the line profiles of selected
lines detected with Herschel/PACS (c-i). The HCO$^+$ lines toward both nuclei,
observed with high angular ($0.3''$) and spectral (30 \kms) resolution,
exhibit P-Cygni profiles indicative of outflowing gas \citep{sak09}. 
Specifically, the redshifted spectral features observed in emission have a 
large velocity extent of $\sim500$ \kms\ from the systemic velocity.
This prominent redshifted HCO$^+$ emission has its counterpart in
several lines detected with Herschel/PACS: the \hdo\ \t303212\ line at $174.6$
$\mu$m (Fig.~\ref{saka}c), 
the [O {\sc i}] 63 $\mu$m transition (d), and the OH 79 $\mu$m doublet
(e). The latter also shows deeper absorption in the blue $\Lambda-$component
than in the red one, suggesting the occurrence of redshifted emission in the
blue $\Lambda-$component as well. The line emission features are expected to
be formed in gas located at the back/lateral sides (i.e. not in front of the
nuclei where absorption of the continuum dominates the profile), and will
therefore be redshifted, as observed, if the gas is outflowing. 
The velocity of the emission features ($v_{\mathrm{LSR}}\sim5650-6000$ \kms) is
significantly higher than the redshifted velocities measured for the rotating
disks around both nuclei \citep[up to $\sim5700$ \kms,][]{sak08}, indicating
that little of this emission is associated with the rotation motions. Thus we
conclude that moderate excitation lines of \hdo, OH, and O {\sc i} trace the
outflow at redshifted velocities detected in HCO$^+$. The asymmetric shape of
the OH 163 $\mu$m emission doublet, which peaks at redshifted velocities, is
consistent with this scenario, as is the detection of redshifted emission
features in the Herschel/SPIRE spectrum of Arp 220 in several species
including \hdo\ \citep{ran11}.

The blueshifted velocity extent of the OH 163 and [O {\sc i}] 145 $\mu$m
lines, observed in emission, reaches $\sim400$ \kms\
from the systemic velocity (Fig.~\ref{saka}h-i), and could also trace
outflowing gas at the most extreme velocities. The full velocity extent in
these lines ($\approx1000$ \kms) is similar to that of CO (3-2) \citep{sak09}.
The full velocity extent of the high-lying \hdo\ lines is significantly
lower, $\approx550$ \kms. This is specifically illustrated in 
  Fig.~\ref{saka}f-g for the \t322211\ and \t616505\ lines, which lie in the
  range $90-82$ $\mu$m where PACS has relatively high velocity resolution
  ($130-150$ \kms). In contrast with the OH 163 and the [O {\sc i}] 145 $\mu$m
  lines observed in emission, which probably trace spatially extended gas,
  the formation of the high-lying absorption \hdo\ lines is restricted to
  regions with high far-IR radiation densities that are optically thick at
  far-IR wavelengths, obscuring the emitting gas behind the nuclei. The
  absorption in these \hdo\ lines peaks at 
  around central velocities, indicating that the lines mainly trace gas
  rotating on the surface of the nuclei. However, due to uncertainties in the
  velocity correction due to pointing shifts (see \S\ref{sec:obser}), and the
  similarity in the blueshifted velocity extent of the HCO$^+$ and \hdo\ lines,
  we cannot rule out a significant contribution to the absorption by 
  outflowing gas with velocities up to $\sim150$ \kms. Nor can we rule out
  spatially extended gas inflow, similar to what is inferred and discussed
  below for NGC~4418, since the peak absorption in the OH 79 $\mu$m doublet is
  redshifted by $\approx90$ \kms.

\begin{figure}
   \centering
   \includegraphics[width=8.0cm]{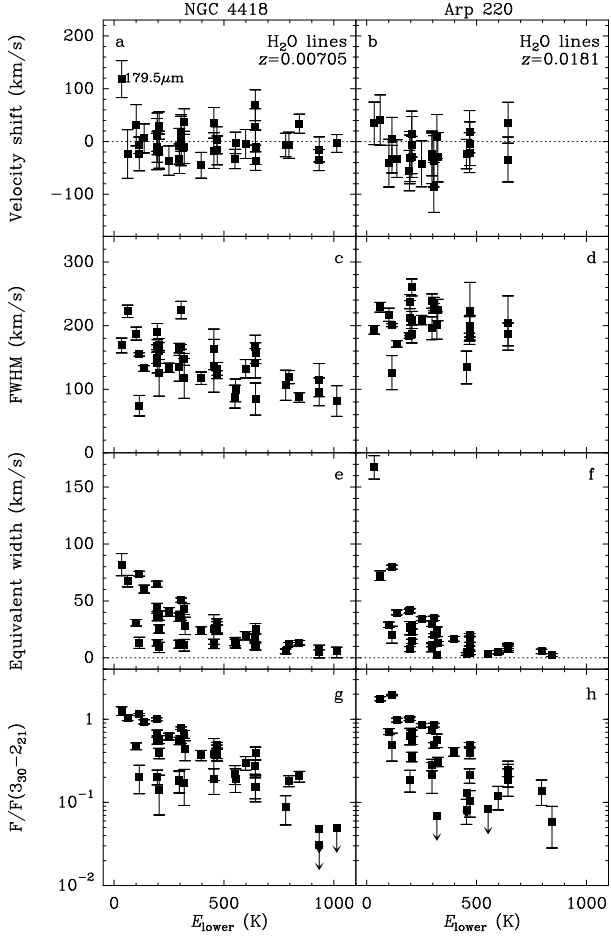}
   \caption{Velocity shifts (a-b), line widths (c-d), and equivalent widths
     (e-f) of the \hdo\ lines in NGC~4418 (left) and Arp~220 (right). Panels
     g-h show the line fluxes normalized to that of the \t330221\ 
       transition at 66.4 $\mu$m (see \S\ref{sec:obser_h2o_oh} 
     for details). Error bars are 1-$\sigma$ uncertainties from Gaussian fits
     to the lines.}      
    \label{h2o-fits}
    \end{figure}

\begin{figure}
   \centering
   \includegraphics[width=8.0cm]{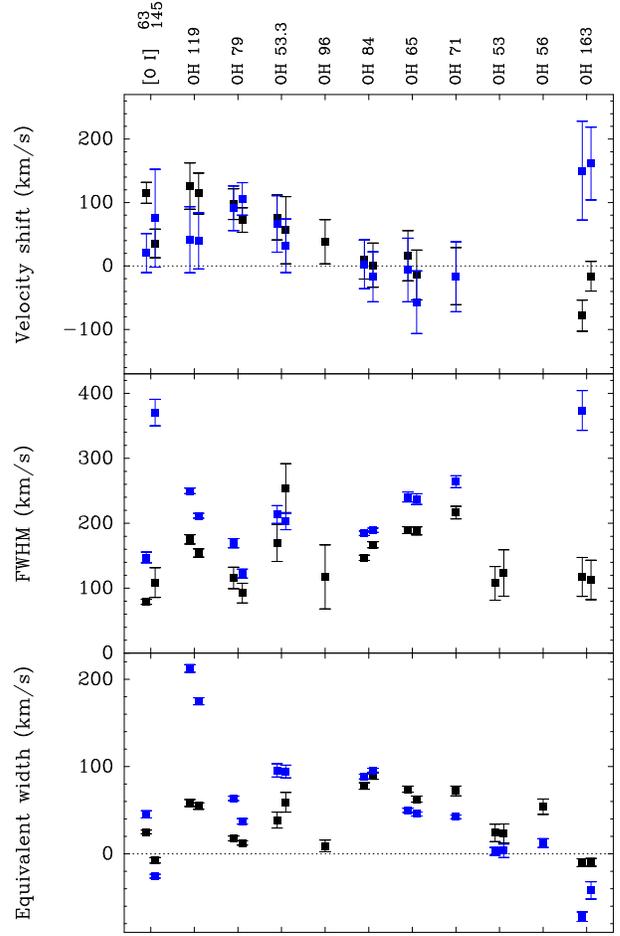}
   \caption{Velocity shifts, line widths, and equivalent widths of the 
     [O {\sc i}] and OH lines in NGC~4418 (black symbols) and Arp~220 (blue
     symbols). Error bars are 1-$\sigma$ uncertainties from Gaussian fits
     to the lines.}     
    \label{oh-fits}
    \end{figure}

   In NGC~4418, the adopted redshift ($z=0.00705$) is derived from the excited
   \hdo\ and OH lines, coinciding within $\approx15$ km
     s$^{-1}$ with both the centroid of the [C {\sc ii}]
   158 $\mu$m line and with the redshift inferred from the 
   stellar absorption at optical wavelengths (\S\ref{sec:redshifts}). No
   indication of outflowing gas in the nuclear region is found. The excited
   \hdo\ and OH lines in NGC~4418 are indeed significantly narrower than in
   Arp~220 (Figs.~\ref{h2o-fits} and \ref{oh-fits}), further indicating the
   relatively quiescent state of the NGC~4418 nucleus. On the other
   hand, the line shapes of the lowest-lying \hdo, OH, and [O {\sc i}] lines
   show an 
   intringuing behavior, opposite to the outflow signatures in Arp~220: the
   absorption features in the ground OH and [O {\sc i}] 63 $\mu$m lines are
   systematically redshifted by $\sim100$ \kms, and the [O {\sc i}] 63 
   $\mu$m line shows a {\em blueshifted} spectral component in emission,
   peaking at $\approx-55$ \kms, i.e. an inverse P-Cygni profile
   (Fig.~\ref{inoutflow}a). The lowest-lying \hdo\ line at 179.5 $\mu$m 
   is also redshifted, but contamination by CH$^+$ $(2\rightarrow1)$ at
   $179.594$ $\mu$m makes the case more uncertain. Additional clues 
   about this component come from the line shapes of the OH 84 and 163 $\mu$m
   doublets: the blue $\Lambda-$component of the 84 $\mu$m doublet
   ($E_{\mathrm{lower}}=121$ K) shows an asymmetric shape with some
   redshifted excess (Fig.~\ref{oh}e and \ref{inoutflow}e; the red
     $\Lambda-$component is probably contaminated by 
   excited NH$_3$, see Fig.~\ref{nh3}c), suggesting 
   that some redshifted gas at $\sim50$ \kms\ is significantly excited.
   The kinematics implied by the OH 163 $\mu$m emission doublet are unclear
     due to the low signal-to-noise ratio (SNR), as one of the
     components is apparently 
   {\em blue}shifted, whereas the other one peaks at central velocities
   (Figs.~\ref{oh}j, \ref{inoutflow}f, and \ref{oh-fits}). In summary,
   the ground-state lines of OH and [O {\sc i}], together with the OH 84 and
   163 $\mu$m doublets, indicate the 
   presence of a component different from the nuclear one in both excitation
   and kinematics, most probably extended in comparison with the nucleus but
   much less massive than the \cext\ component of Arp~220, and the
   velocity shifts suggest that this component may be {\em inflowing} onto the
   quiescent, nuclear region. We further investigate this scenario in
   \S\ref{sec:n4418low}. 

   \begin{figure}
   \centering
   \includegraphics[width=8.0cm]{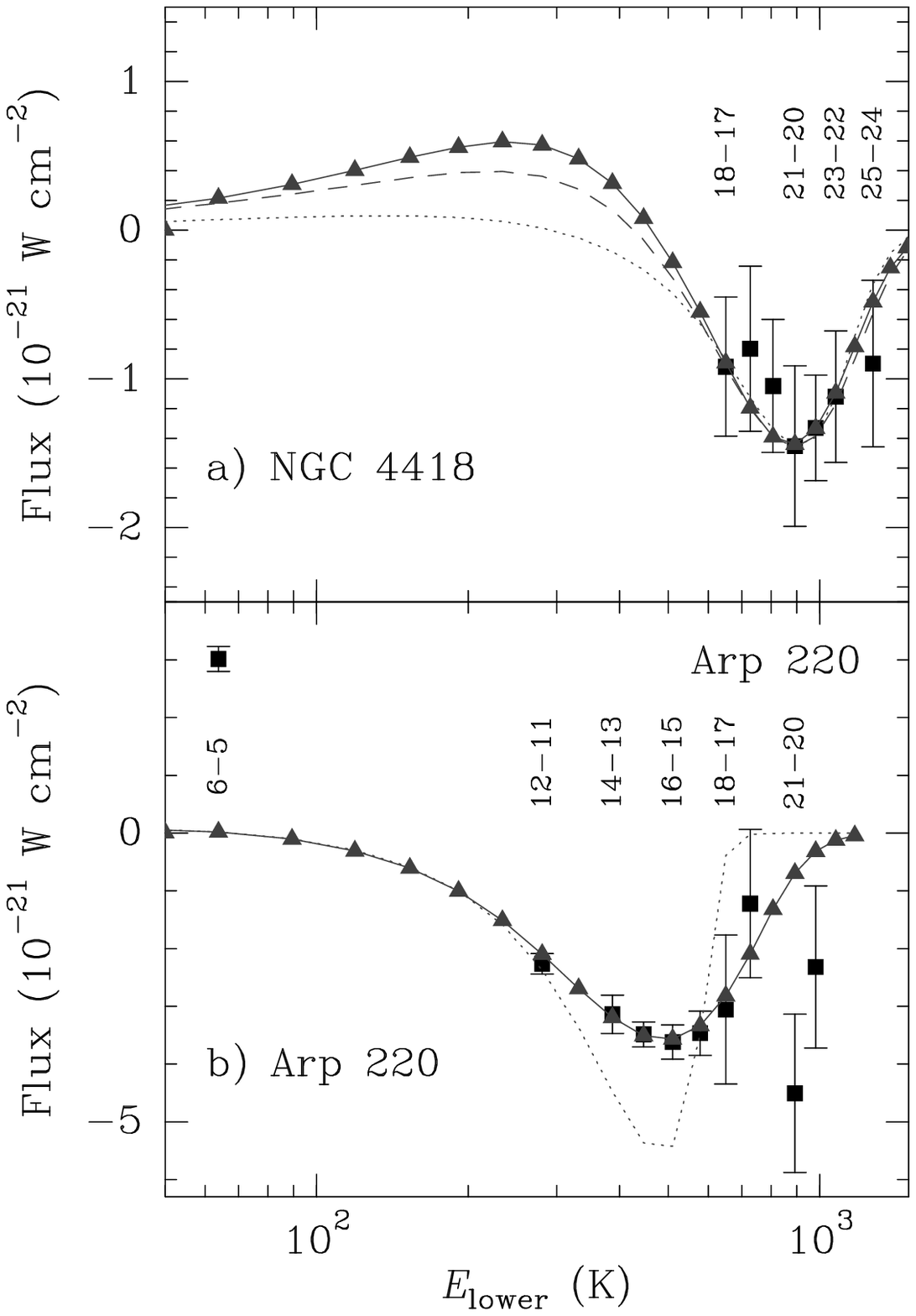}
   \caption{HCN line fluxes in NGC~4418 (a) and Arp~220
     (b). In Arp~220, the fluxes of lines with 
     $E_{\mathrm{lower}}$ up to 570 K have been taken from SPIRE-FTS data
     \citep{ran11}. The curves and triangles show model predictions for
     the \ccore\ component of NGC~4418 and the \cwest\ component of
     Arp~220. The dotted curve in (b) shows results obtained when only
     collisional excitation is included in the model. The solid curve in (a)
     shows the best fit model obtained with $\tau_{200}\approx1$, while the
     dashed and dotted curves correspond to $\tau_{200}=4$ with and without
     gas-dust mixing, respectively.}
    \label{hcn-fits}
    \end{figure}

\section{Models}
\label{sec:models}

\subsection{Overview}

As the data presented in previous sections have shown, the far-IR
spectra of NGC~4418 and Arp~220 are dominated by molecular absorption, 
with emission in only some lines and profiles.  
These rich line spectra and their
associated dust emission cannot be described by a single set of ISM
parameters, but different lines have different excitation requirements
and are thus formed in different regions of the galaxies.
Our approach for both galaxies is to fit these different regions
and conditions, even though they are not spatially resolved with Herschel,
with the smallest possible number of parameterized components.   

   \begin{figure}[ht]
   \centering
   \includegraphics[width=9cm]{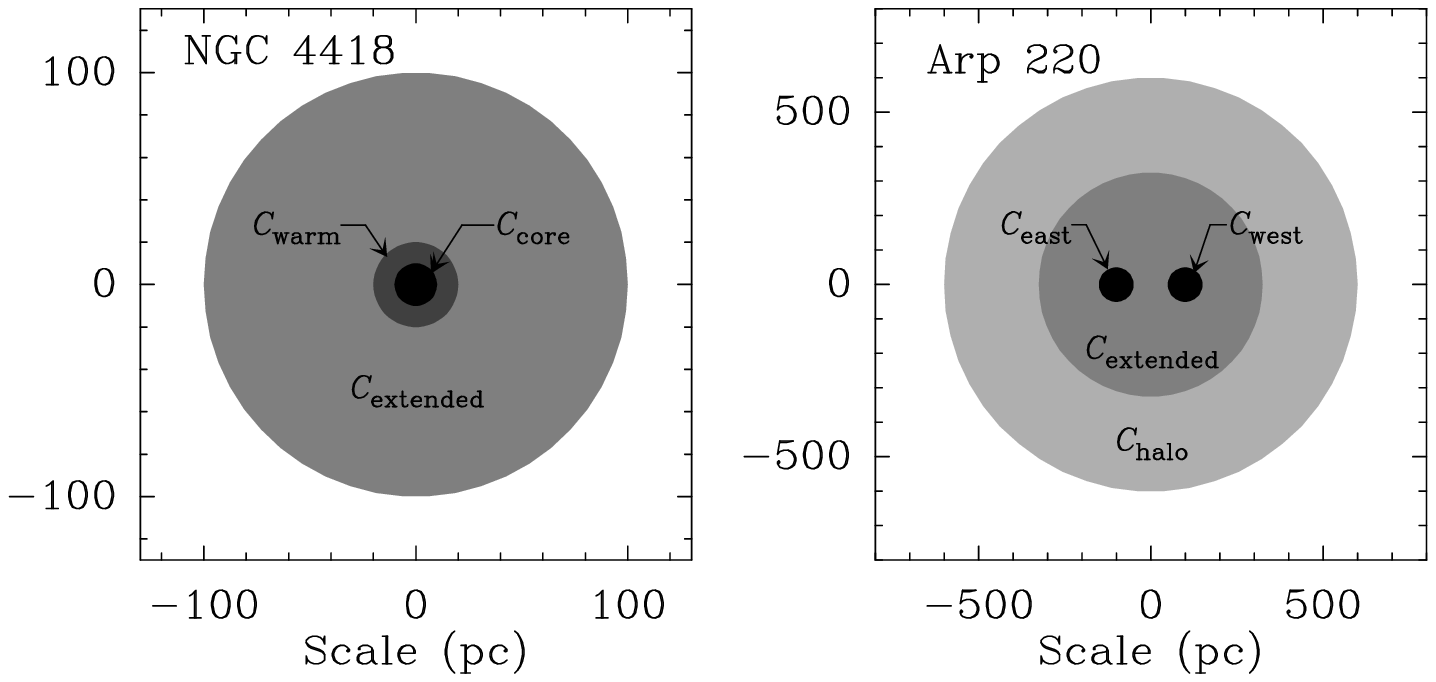}
   \caption{Schematic representation of the modeled sources, showing the
     approximate spatial scales of the different far-IR components (see
     Fig.~\ref{continua} and Tables~\ref{tab:cont} and \ref{tab:lines}). The
     \chot\ components in both galaxies (that account for the mid-IR spectra)
     are most probably associated with the nuclear regions and are not
     included here. The \chalo\ component in Arp~220 has no
     associated continuum, and is responsible for the absorption of the 
       nuclear continuum (\ceast\ and \cwest) in the
     ground-state lines of \hdo, OH, and O {\sc i}. The plot is an
     oversimplification of the actual models, where the 
     different components are modeled separately to account approximately for
     non-spherical symmetry.} 
    \label{cartoon}
    \end{figure}

For NGC~4418, we find that we need (see Fig.~\ref{continua}a) $(i)$ a hot
component (yellow curve, hereafter \chot) that accounts for the mid-IR
continuum; $(ii)$ a nuclear core (blue curve, \ccore) that provides absorption
in the high-lying lines of \hdo, OH, HCN, and NH$_3$; $(iii)$ a warm component
(light-blue, \cwarm), that provides absorption in moderately excited lines of
\hdo\ and OH as well as a significant fraction of the far-IR continuum
emission; and $(iv)$ an extended component (solid green, \cext), which
accounts for the low-lying redshifted lines of OH and O {\sc i}.

For Arp~220, we need (Fig.~\ref{continua}b) $(i)$ a single\footnote{Even if
  the mid-IR arises from the two nuclei, the continuum is simulated with just
  a single component with an effective diameter as given in
  Table~\ref{tab:cont}.} 
hot component (yellow curve, \chot) that accounts for the mid-IR continuum
emission; $(ii)$ the western and eastern nuclear components (\cwest\ and
\ceast\ in blue and light blue, respectively), where the high-lying molecular
lines are formed; $(iii)$ the extended component (\cext, green), which
provides a significant fraction of the far-IR continuum emission and 
which is likely associated with
moderate and low-excitation lines of \hdo\ and OH; and $(iv)$ an additional
absorbing ``halo'' component (\chalo), with no associated intrinsic
continuum but located in front of the nuclei, is required to fit the
absorption in the ground-state lines of \hdo, OH, and O {\sc i}.  

We use single dust temperatures for every component listed above, except for
the \cext\ component in both galaxies where the dust temperature profile
is calculated from the balance of heating (by the inner components) and
cooling \citep{gon99}. In our models, each single dust temperature component
is attenuated by a foreground, screen-like shell, which is parameterized by
its dust opacity at 25 $\mu$m, $\tau_{\mathrm{25,fgr}}$. These screen-like
shells are responsible for the silicate absorption features imprinted on the
various modeled components in Fig.~\ref{continua}. The dust temperature
profile in \cext\ is calculated by assuming spherical symmetry and a
single nuclear heating component located at the center of the modeled source. 

Parameters of the continuum models, and the inferred
molecular parameters, are listed in Tables~\ref{tab:cont} and 
\ref{tab:lines}, respectively. A sketch of the modeled
sources, showing the approximate spatial scales of the different far-IR
components (excluding the \chot\ component of both galaxies), is shown
in Fig.~\ref{cartoon}.  
Each component, however, can be interpreted in terms of 
a single source, as implicitly assumed, or alternatively applied
to each one of an ensemble of $N_C$ smaller clouds of
radius $R_C$ that do not spatially overlap along the line of sight.
The scaling between these two approaches is discussed in G-A04.

The models for Arp~220 are discussed in sections \ref{sec:arp220high}
(high-lying lines) and \ref{sec:arp220low} (low-lying lines), while those for
NGC~4418 are developed in sections \ref{sec:n4418high} (high-lying lines),
\ref{sec:warmn4418} (mid-excitation lines), and \ref{sec:n4418low} (low-lying
lines). The continuum of NGC~4418 is further discussed in
\ref{sec:n4418cont}.

\subsection{The high-lying lines in Arp~220: the nuclear region (\cwest\ and
  \ceast)} 
\label{sec:arp220high}

\subsubsection{H$_2$O}
\label{sec:arp220highh2o}

Our models for Arp~220 are similar to those described in G-A04, and generated
on the 
basis of the interplay between the continuum emission and the molecular line
absorption. Since the molecular excitation in Arp~220 is high and
collisional excitation alone cannot account for it, the excitation
mechanism is expected to be dominated by absorption of dust-emitted photons 
in the nuclear region of the galaxy  \citep[G-A04,][]{gon08,gon10}. The
molecular excitation is thus a function of the dust temperature, $T_d$, and of
$N_{\mathrm{H_2O}}/\Delta v$, and in combination with the observed continuum
provides clues about the general properties of the far-IR continuum
source and its asociated chemistry. If the observed \hdo\ excitation cannot be
reproduced by 
assuming that the dust in the nuclear component has a temperature of ``only''
$T_d=85$ K (derived from the {\em observed} 25 and 1300 $\mu$m emission),
foreground extinction in the far-IR is included to attenuate also the 25
$\mu$m emission from the nuclei in such a way that the far-IR is reproduced
with $T_d>85$ K. This is Scenario 2, S$_2$, in G-A04, based on the continuum
models by \cite{soi99}. Evidence for such foreground extinction in the mid-IR
comes from the strong silicate absorption at 9.7 and 18 $\mu$m, in the
millimeter by self-absorption in CO $(2\rightarrow1)$ \citep{dow07},
and in the far-IR from the [O {\sc i}] 63 $\mu$m line that is observed in
absorption, in contrast to most extragalactic sources in which it is
  observed in emission.

In the present models, we have simulated the dust emission by using a mixture
of silicate and amorphous carbon grains with optical constants from
\cite{dra85} and \cite{pre93}; the mass-absorption coefficient is 550, 
150, and 12.3 cm$^2$/g of dust at 25, 50, and 200 $\mu$m, 
respectively, and the spectral index is
$\beta\approx2$ in the far-IR. We have also attempted to disentangle the
emission from the eastern and western nuclei in Arp~220. We
have modeled the eastern nucleus (\ceast) as a sphere of diameter
$d_{\mathrm{east}}=0.31"$ \citep[][hereafter Sa08]{sak08}, radial opacity 
(i.e. the opacity along a radial path) at
200 $\mu$m $\tau_{\mathrm{200}}=2.7$, and $T_{d,\mathrm{east}}=87$ K. The
emission from \ceast\ is attenuated by foreground dust with
$\tau_{\mathrm{25,fgr}}=1$, yielding $2.5-0.02$ Jy at $25-1300$ $\mu$m
--consistent with measurements by \cite{soi99} and \cite{dow07}. The
corresponding SED of \ceast\ is shown in Fig.~\ref{continua}b in
light-blue, and accounts for a luminosity of 
$L_{\mathrm{east}}=3\times10^{11}$ \Lsun. For the western nucleus (\cwest) we
also use as a first approach the size as derived by Sa08 in the ``disk''
approximation, i.e. a sphere with diameter $d_{\mathrm{west}}=80$ pc ($0.23''$
at 72 Mpc), but allow it to vary to match the observed continuum, and leave as
free parameters the dust temperature $T_{d,\mathrm{west}}$ and foreground
extinction $\tau_{\mathrm{25,fgr}}$. 

   \begin{figure}[ht]
   \centering
   \includegraphics[width=8.4cm]{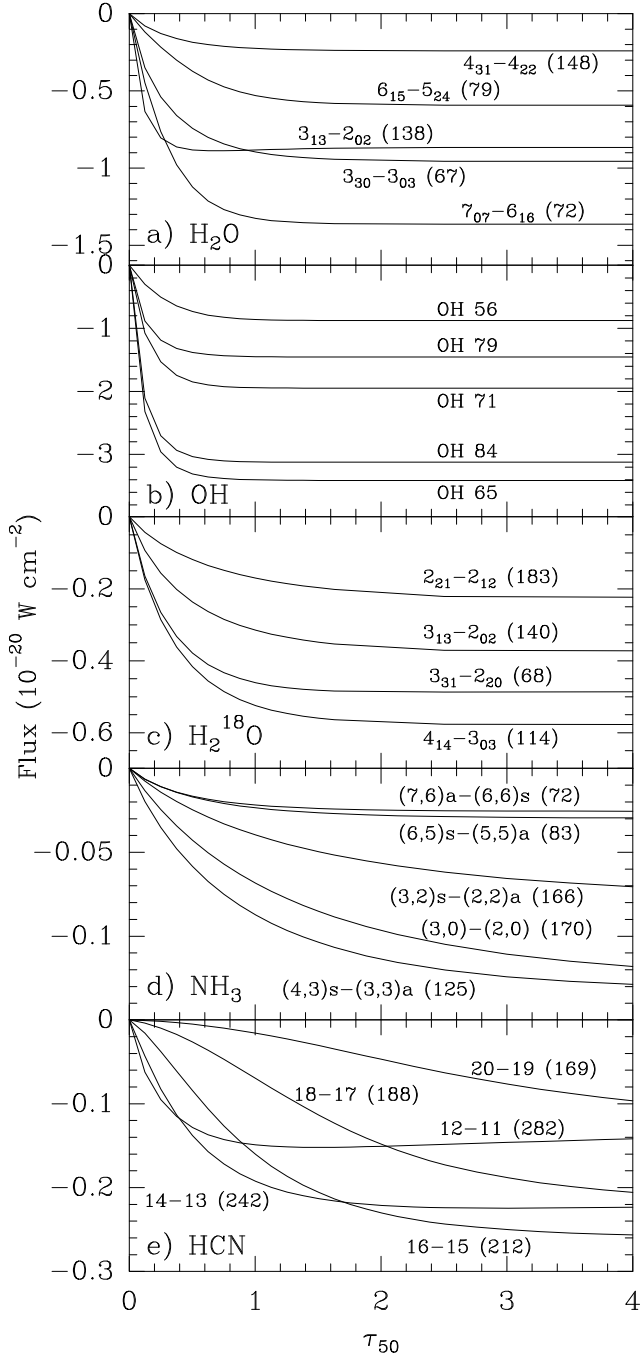}
   \caption{Modeled absorbing flux in Arp 220 of several a) \hdo, b) OH,
     c) H$_2^{18}$O, d) NH$_3$, and e) HCN lines as a
     function of the depth of the molecular shell measured from the
     surface of the continuum source, as parameterized by
     the dust opacity at 50 $\mu$m  ($\tau_{50}$). Numbers in parenthesis
     indicate rounded wavelengths in $\mu$m. The total column of
     molecules is proportional to $\tau_{50}$, and we have adopted
     $N_{\mathrm{H_2O}}=2\times10^{18}\times\tau_{50}$ cm$^{-2}$, 
     and abundance ratios of $\mathrm{OH/H_2O}=0.5$,
     $\mathrm{H_2O/H_2^{18}O}=100$, $\mathrm{HCN/H_2O}=0.3$, and
     $\mathrm{HCN/NH_3}=6$. The dust temperature is 
     110 K.}
    \label{molabs}
    \end{figure}

Calculations for the lines were carried out in spherical symmetry using
the code described in \cite{gon99}. 
The models indicate that most high-lying absorption lines are formed in
  the outermost shell surrounding the far-IR continuum source. This is
  illustrated in Fig.~\ref{molabs}, where the absorbing flux of several lines
  of \hdo, OH, H$_2^{18}$O, NH$_3$, and HCN, 
  is plotted as a function of the depth of the shell where the
  molecules are located, which is parameterized in terms of the continuum
  opacity at 50 $\mu$m ($\tau_{50}$) measured from the 
  surface of the far-IR source. Since the molecular abundances relative to the
  density of dust are uniform in
  these models, the column densities are proportional to $\tau_{50}$. Results
  indicate that the \hdo\ and OH absorption is produced in a thin shell with
  $\tau_{50}\approx1$, as extinction, thermalization by dust emission, as well
  as line opacity effects make results insensitive to the presence of molecules
  deeper into the far-IR source. The optically thinner H$_2^{18}$O lines are
  mostly formed in the same shell, though the lines at long wavelengths
  (e.g. the \t221212\ at 183 $\mu$m) still have significant contribution
  ($\lesssim25$\%) from the $1<\tau_{50}<2$ region. The case for NH$_3$ and HCN
  is somewhat different, as some lines probe deeper regions. The
  high-lying lines of NH$_3$ lie at $\lambda<100$ 
  $\mu$m and are also formed in the $\tau_{50}\approx1$ outermost shell, but
  the low-lying lines at $\lambda>100$ $\mu$m have significant contribution
  from $1<\tau_{50}<3$. The high-excitation HCN lines sample regions deeper than
  the other species ($\tau_{50}\lesssim4$) as their excitation is sensitive to
  the far-IR radiation density at $\lambda\gtrsim170$ $\mu$m (see
  \S\ref{sec:hcnarp220}). 
  Based on these results, the models for the nuclei shown below use a
  ``screen'' approach, i.e. the molecules are mixed with the dust but located
  within a thin shell surrounding the nuclei extended up to a depth of
  $\tau_{50}=2$ for \hdo, OH, and their $^{18}$O isotopologues, and up to
  $\tau_{50}=4$ for NH$_3$ and HCN. The values of the column densities of all
  species below, however, are given for the $\tau_{50}=1$ outermost shell that
  dominates the absorption of \hdo\ and OH, and are lower limits to the true
  columns through the nuclear sources. In order to estimate the abundances
  $\chi_{\mathrm{X}}$ relative to H nuclei, we have normalized these
  column densities per unit of $\tau_{50}$ to a given value of
  $N_{\mathrm{H,norm}}$. For a mass-absorption coefficient of 150 cm$^2$/g of
  dust at 50 $\mu$m and a gas-to-dust mass ratio of 100, $\tau_{50}=1$
  corresponds to a column density of $N_{\mathrm{H,norm}}=4\times10^{23}$
  cm$^{-2}$, where H refers to hydrogen nuclei in both atomic and molecular
  forms. This value of $N_{\mathrm{H,norm}}$ has been applied to all species
  observed in the far-IR, and define the far-IR ``photosphere'' where  most
  molecular absorption is produced.

Line broadening is caused by microturbulence with $v_{\mathrm{tur}}=60$ \kms\
and a velocity shift of 130 \kms\ through the absorbing shell. The 
  latter simulates either the presence of outflowing gas or, more generally,
  gas velocity gradients across the nuclear regions, allowing us to nearly
  match the observed linewidths. Rates for collisional
  excitation are taken from \cite{fau07}; with the adopted $n_{\mathrm{H_2}}=
5\times10^{5}$ cm$^{-3}$ and gas temperature $T_g=150$ K (see
  \S\ref{sec:hcnarp220}),  collisional excitation has little 
effect on the calculated fluxes. The line models have three free parameters:
the dust temperature, $T_{d}$, the \hdo\ column, $N_{\mathrm{H_2O}}$,
and the covering factor of the continuum source, $f_C$.

The bulk of the \hdo\ absorption
is produced in a luminous, compact region with high \hdo\ columns, which
we identify with the double nucleus of Arp 220. A number of previous
  observational studies with high angular resolution have
revealed that the western nucleus is brighter than the eastern one in the
near-IR \citep{arm95}, mid-IR \citep[by a factor of $\sim3$ at 25
$\mu$m;][]{soi99}, and (sub)millimeter wavelengths \citep{dow07,sak08}. 
Given that the excitation of \hdo\ requires a high
brightness continuum source, the observed high-lying \hdo\ absorption is
tentatively attributed to \cwest. 
Nevertheless, the column densities derived below are
independent of whether \cwest\ alone is responsible for the absorption in
the high-lying lines, or \ceast\ has a significant contribution as
well, and the properties we derive for \cwest\ and \ceast\ can be more
generally interpreted as shared by both nuclei. 

\begin{figure}[ht]
   \centering
   \includegraphics[width=8.0cm]{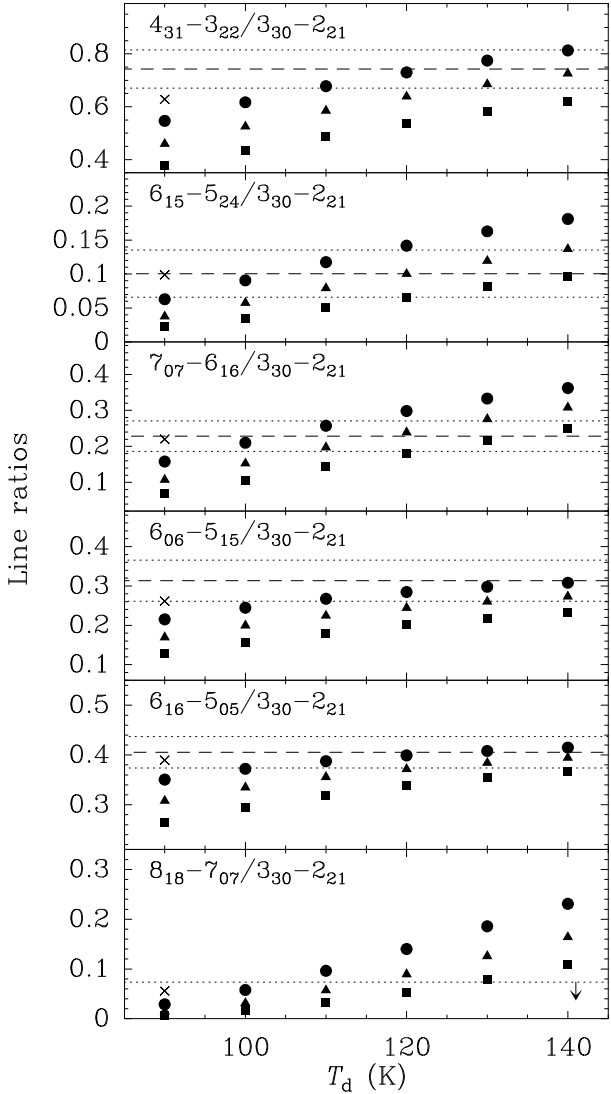}
   \caption{Modeled \hdo\ line ratios in Arp~220 as a function of the dust
     temperature in the western nucleus. The dashed lines indicate the
     observed values, and the dotted lines the estimated upper and lower
     limits. In the lowest panel, the dotted line is the $3\sigma$ upper
       limit for the $8_{18}-7_{07}/3_{30}-2_{21}$ flux ratio. Squares,
       triangles, circles, and crosses (only for 
     $T_{d,\mathrm{west}}=90$ K) show model results for ``screen'' \hdo\
     columns of $N_{\mathrm{H_2O}}= 7.5\times10^{17}$,  
     $1.5\times10^{18}$, $3\times10^{18}$, and $6\times10^{18}$ cm$^{-2}$,
     respectively.}          
    \label{lineratios_arp220}
    \end{figure}

In the nuclear region where the \hdo\ lines are formed, the models shown
  below indicate that most of these lines are strongly saturated; however,   
there are still some high-lying lines that are sensitive to both
$N_{\mathrm{H_2O}}$ and $T_{d}$. For these critical lines, 
Figure~\ref{lineratios_arp220} shows the observed and modeled fluxes relative
to that in the \t330221\ line at 66.4 $\mu$m (Fig.~\ref{h2o}g). As mentioned
in \S\ref{sec:obser}, the \t330221\ is chosen as the normalization line
because it is detected with a high SNR and, with $E_{\mathrm{lower}}=160$ K, 
is not expected to be significantly contaminated by absorption of extended
low-excitation \hdo\ in Arp~220. On the other hand, the
  \t330221\ transition is still low in energy as compared with
the high-lying lines, so that the line flux ratios are sensitive to both
$N_{\mathrm{H_2O}}$ and $T_{d}$. Furthermore, the line flux ratios
do not depend on either the size of the continuum source, or on $f_C$. These
ratios are plotted as a function of $T_{d}$ for different
``screen-like'' \hdo\ columns, ranging from $N_{\mathrm{H_2O}}=
7.5\times10^{17}$ to $6\times10^{18}$ cm$^{-2}$, with dashed lines indicating
the measured values, and dotted lines the estimated upper and lower limits. We
also include in the lowest panel the upper limit derived for the
\t818707\ line. For each $T_{d}$, the foreground extinction is
determined by imposing that the continuum flux density at 25 $\mu$m matches
the observed 7.5 Jy from the western nucleus \citep{soi99}, but this has
little effect on the line flux ratios as the far-IR foreground opacities are
low and, in any case, the considered lines have similar wavelengths. The
minimum $T_{d}$ is 90 K, which is the optically thick, lower limit value
derived for the western nucleus from the 860 $\mu$m submillimeter
continuum by Sa08.

Results of Fig.~\ref{lineratios_arp220} show that the line ratios scale
  linearly with $T_{d}$. The 
highest sensitivity to $T_{d}$ is found for the \t615524\ line, 
since its lower level is non-backbone (Fig.~\ref{enerdiag}). Among the
detected lines, the line ratios can be almost equally well reproduced with
$T_{d}=90$ K and the highest $N_{\mathrm{H_2O}}= 6\times10^{18}$
cm$^{-2}$, or with $T_{d}=140$ K and $N_{\mathrm{H_2O}}=
7.5\times10^{17}$ cm$^{-2}$. However, the model
$T_{d}=140$ K and $N_{\mathrm{H_2O}}=7.5\times10^{17}$ cm$^{-2}$
yields too much absorption in the undetected \t818707\ line, slightly favoring
$T_{d}\lesssim130$ K. Concerning the continuum models, if we
assume that the observed high-lying \hdo\ absorption is dominated by
  \cwest, and that the photosphere at 25 $\mu$m has the same size as the
(sub)millimeter source, a reasonable match to the $20-35$ $\mu$m SED with
$T_{d,\mathrm{west}}\le110$ K is not possible; for example,  
the model with $T_{d,\mathrm{west}}=90$ K yields $\approx4$ Jy at 25 $\mu$m with
no foreground extinction and the size derived from the submillimeter
(Sa08), insufficient to account for the observed $7.5$ Jy at 25 $\mu$m
\citep{soi99}. 
Furthermore, any reasonable fit to the continuum requires foreground
absorption at $20-35$ $\mu$m, and the absolute \hdo\ fluxes also require
  a continuum source larger than the submillimeter one for
  $T_{d,\mathrm{west}}\lesssim120$ K. However, the photosphere at 25 $\mu$m  
(i.e. the nuclear region with $\tau_{25}\ge1$) may be
larger and colder than the submillimeter source
if heated by the central core (Sa08), because the optically thinner
  submillimeter emission samples warm regions that are obscured at 25
  $\mu$m. For $T_{d,\mathrm{west}}=90$ K, the required source diameter is
$d_{\mathrm{west}}=150$ pc, yielding $L_{\mathrm{west}}=6.5\times10^{11}$
\Lsun, while for $T_{d,\mathrm{west}}=130$ K, $d_{\mathrm{west}}=80$ pc and
$L_{\mathrm{west}}=8.2\times10^{11}$ \Lsun. 

In summary, a range of
$T_{d,\mathrm{west}}=90-130$ K and a corresponding ``screen'' 
column per unit of $\tau_{50}$ of
$N_{\mathrm{H_2O}}=(60-7.5)\times10^{17}$ cm$^{-2}$ are derived,
with the colder ($<120$ K) sources yielding diameters above the
(sub)millimeter observed values. The corresponding luminosities are in the
relatively narrow range $L_{\mathrm{west}}=(6.5-8.2)\times10^{11}$ \Lsun.
The corresponding estimated \hdo\ abundances are
$\chi_{\mathrm{H_2O}}\sim(15-2)\times10^{-6}$ relative to H nuclei.

As the reference model shown in Figs.~\ref{h2o}-\ref{h2o-c} for detailed
comparison with data, we use the combination $T_{d,\mathrm{west}}=110$ K and
$N_{\mathrm{H_2O}}=2\times10^{18}$ cm$^{-2}$, with complete coverage of
the source ($f_C=1$).
The corresponding continuum is shown with a blue curve in
Fig.~\ref{continua}b, with the diameter of the source (106 pc)
increased by $\sim30$\% relative to the submillimeter source, and with a
foreground opacity $\tau_{\mathrm{25,fgr}}=1.4$. 
The data also suggests the presence of a lower excitation component to
  attain a better fit to the \t322211, \t221110, and \t303212\ lines at 90,
  108, and 174 $\mu$m, respectively. Tentatively associating this component with
  the eastern nucleus, we derive $N_{\mathrm{H_2O}}= 3\times10^{17}$ cm$^{-2}$
and $f_C=0.4$. Its contribution to the
\hdo\ SLED, shown with light-blue curves in Figs.~\ref{h2o} and \ref{h2o-b},
is expected to be significant only for low-lying \hdo\ lines. 
Continuum and line parameters of the models are listed in
  Tables~\ref{tab:cont} and \ref{tab:lines}.

\subsubsection{OH}

The observed high-lying OH doublets in Arp~220, namely the 
$\Pi_{3/2}$ $9/2\rightarrow7/2$ and $11/2\rightarrow9/2$ at 65 and 53 $\mu$m, 
and the $\Pi_{1/2}$ $7/2\rightarrow5/2$ and $9/2\rightarrow7/2$ at 71 and 56
$\mu$m, are compared in Fig.~\ref{oh} (lower spectra) with the models for the
western and eastern nuclei (blue and light-blue curves). Collisional rates
between OH and H$_2$ were taken from \cite{off94}. As for \hdo, the absorption
in these lines is dominated by a source with high OH column, which we
tentatively identify with \cwest. In the model of Fig.~\ref{oh}, we have
used the same parameters as for \hdo, and varied the OH column
  density to match the observed absorption in the high-lying lines. We derive
  a OH/\hdo\ abundance ratio of $0.5-1$; higher OH columns
  overpredict the absorption in the OH 53 $\mu$m doublet
  (Fig.~\ref{oh}h). The fit to the high-lying OH lines is satisfactory
except for the underprediction of the width of the $\Pi_{3/2}$
$9/2\rightarrow7/2$ 65 $\mu$m doublet, which is probably contaminated by
  relatively weak lines of \hdo\ (gray curve in Fig.~\ref{oh}f), NH$_2$, and
  H$_2$O$^+$. Using the same source sizes as for \hdo, we find a similar
  $\mathrm{OH/H_2O}\sim0.5-1$ ratio for $T_{d}=90-130$ K. An additional
  extended region is required to fit the low-lying OH lines, which is
discussed in \S\ref{sec:arp220low}.


\subsubsection{H$_2^{18}$O and $^{18}$OH}
\label{sec:arp220-18o-models}

Although the detected H$_2^{18}$O lines arise from low-lying
levels, we find that these absorption features are mainly
produced toward the nuclear region with the 
highest column densities. Fig.~\ref{h218o} shows the best fit for the
H$_2^{18}$O lines using the same reference model as for the main
isotopologue, i.e. $T_{d}=110$ K. The best fit H$_2^{18}$O column density 
is $\approx3\times10^{16}$ \cmd, implying an
H$_2^{16}$O-to-H$_2^{18}$O ratio as low as $\sim70$, and the upper
  limit for this ratio is estimated to be $\approx95$.
The H$_2^{16}$O-to-H$_2^{18}$O ratio slightly depends on the dust
temperature, and a value of $\approx100-150$ is found for the model with
$T_{d}=90$ K, though the overall fit is in this case less
  satisfactory. Thus our data suggest an enhancement of $^{18}$O in
Arp~220 relative to the solar value of $\sim5$. This is confirmed by the
models for $^{18}$OH shown in Fig.~\ref{18oh}, where the same
$^{16}$OH-to-$^{18}$OH ratio of $70$ is used with $T_{d}=110$
  K. The $^{16}$OH-to-$^{18}$OH ratio would also increase up to $120-150$ K
  for $T_{d}=90$ K.  
The $^{18}$OH 65 and 85 $\mu$m lines are nearly reproduced, but not the
asymmetry in the $^{18}$OH 85 $\mu$m doublet (Fig.~\ref{18oh}b). Similar
intensity asymmetries of 
$\approx40$\% have been also observed in most OH $\Lambda-$doublets by
\cite{goi11} toward the Orion bar, where the lines are observed in
emission. As \cite{goi11} indicate, this is probably due to asymmetries
in the collisional rates between OH and para-H$_2$ \citep[see
also][]{off94}. In our absorption lines, the asymmetry would arise from the
collisional excitation of $^{18}$OH in the ground $\Pi_{3/2} \, J=5/2
\rightarrow 3/2$ transition, which favors the $J=5/2^+$ sublevel where the
$85.1$ $\mu$m component arises. Since collisions with ortho-H$_2$ quench the
asymmetry, the gas is not very warm but should be dense enough so that
  collisional excitation with para-H$_2$ competes efficiently with the
  excitation induced by the radiation field. 
The asymmetry is not observed in the main isotopologue
(Fig.~\ref{oh}e), indicating that either the higher column of OH efficiently
mixes the populations 
of the sublevels, or that OH is tracing warmer gas with increased
contribution by collisions with ortho-H$_2$. However, 
our model with $n_{\mathrm{H_2}}=5\times10^{5}$ cm$^{-3}$ and $T_g=150$ K does
not reproduce it, and several tests with higher density and lower $T_g$ were
also unsuccesful. Further, since the ground-state $^{18}$OH 120 $\mu$m
doublet is not reproduced with the warm component alone, additional
absorption by the \cext\ and \chalo\ components described below, with the
same $^{16}$OH-to-$^{18}$OH ratio of $70$, is included (Fig.~\ref{18oh}c).

\subsubsection{HCN}
 \label{sec:hcnarp220}

Our models for the western nucleus of Arp~220 can also reproduce the fluxes of
the HCN lines reported by \cite{ran11} and the $J=18\rightarrow17$ and
$J=19\rightarrow18$ lines
detected with PACS (Figs.~\ref{hcn} and \ref{hcn-fits}). For HCN, both
collisional excitation and absorption of
far-IR photons in the highest-lying lines are required to reproduce the
observations. At long wavelengths (i.e. relatively low$-J$ lines), collisional
excitation in a warm and dense region is the primary excitation mechanism.
For moderate dust opacities, these lines would be detected in emission above
the continuum. If collisions are able to excite the molecules up to the
rotational level where the dust emission becomes optically thick and the
far-IR radiation becomes strong, absorption of far-IR photons in high-$J$
lines continues to excite the molecule to higher-lying levels and the
corresponding HCN lines are observed in absorption. Thus the transition from
emission to absorption lines is a measure of the dust opacity and temperature
of this component, and the high-lying HCN lines detected in absorption are
tracing high far-IR radiation densities as well as a warm-dense environment
with high HCN column densities. 

A grid of models were generated by varying $n_{\mathrm{H_2}}$ in the
  range $10^5-3\times10^6$ cm$^{-3}$ and $T_g$ between 110 and 250 K. For each
  value of $n_{\mathrm{H_2}}-T_g$, both $N_{\mathrm{HCN}}$ and $f_C$ were
  varied to fit the observed SLED. In all these models,
  $T_{d}=110$ K was adopted. As discussed in
  \S\ref{sec:arp220highh2o}, the high excitation HCN lines can be formed deeper
  into the dusty, continuum source than lines from other species
  (Fig.~\ref{molabs}), and we 
  extend our HCN shell up to a depth of $\tau_{50}=4$.
  Rates for HCN collisions with He
  were taken from \cite{dum10}, and scaled for the reduced mass of the 
  HCN-H$_2$ system to obtain the expected collisional rates with H$_2$. Our
  most plausible results were found for $n_{\mathrm{H_2}}=(4-8)\times10^5$
  cm$^{-3}$ and $T_g\sim150$ K, for which
  $N_{\mathrm{HCN}}=(6-4)\times10^{17}$ cm$^{-2}$,
  respectively. $N_{\mathrm{HCN}}$ is twice these values for $T_g=110$  
  K. Decreasing $n_{\mathrm{H_2}}$ below $2\times10^5$ cm$^{-3}$ would involve
  HCN columns above $10^{18}$ cm$^{-2}$, implying extreme HCN abundances
  above $2.5\times10^{-6}$ that we consider unlikely. The increase of 
  $n_{\mathrm{H_2}}$ above $10^6$ cm$^{-3}$ involves lower HCN columns, but
  the relative absorption in the $J=12\rightarrow11$ line becomes somewhat
  underpredicted and the value of $f_C$ (or the size of the continuum source)
  is above that derived for \hdo. Results for $T_g=150$ K and
  $n_{\mathrm{H_2}}=4\times10^5$ cm$^{-3}$, with the resulting
  $N_{\mathrm{HCN}}=5.6\times10^{17}$ cm$^{-2}$, 
  are shown in Figs.~\ref{hcn} and \ref{hcn-fits}. The involved far-IR
  continuum source has the same size as for \hdo\ and OH. Similar values of
  $N_{\mathrm{HCN}}=(3-6)\times10^{17}$ cm$^{-2}$ are obtained for
  $T_d=130-90$ K, though the sizes of
  the continuum and HCN sources are in these cases higher and lower than for
  $T_d=110$ K, respectively, and similar to those found for \hdo\ and OH. 

In our most plausible models, the HCN/\hdo\ ratio is 0.1-0.4
(Table~\ref{tab:lines}), and $\chi_{\mathrm{HCN}}=(1-2)\times10^{-6}$. In 
Fig.~\ref{hcn-fits}, the calculated HCN SLED is 
compared with results for a model that neglects the radiative excitation by
dust (dotted curve), showing that the dust has a flattening effect on the
predicted SLED. Less than 10\% of the total
column is stored in levels associated with the observed lines; the highest
populated levels are $J=5,6$. Nevertheless, owing
to the high continuum opacity and dust temperature in the nuclear
region, as well as to self-absorption in the outermost shells of the
 HCN region, all HCN lines above $J=6\rightarrow5$ are predicted in
 absorption. The emission in the $J=6\rightarrow5$ line \citep{ran11} is not 
reproduced, probably indicating that the line is formed in a colder and
more extended region associated with relatively weak continuum. On the
other hand, the model in Fig.~\ref{hcn-fits} accounts for the observed
absorption up to $J=19\rightarrow18$ but fails to explain the absorption
  coincident with the $J=21\rightarrow20$ transition at 161.3 $\mu$m. If
this feature is due to HCN, a region as warm as that
proposed for NGC~4418 (\S\ref{sec:coren4418}) would be involved. 

Our models also include the HCN excitation through the
pumping of the vibrationally excited $\nu_2=1^1$ state, 
but we find the effect
negligible for the ground-vibrational state lines. Evidently, this is a
consequence of the low $T_{d}$ used in our modeling. \cite{ran11}
have proposed that the pumping of the $\nu_2=1^1$ state dominates the HCN
excitation, requiring a dust temperature of $T_{d}>350$ K. If not
completely extincted, this hot component would show up at mid-IR wavelengths,
and there is indeed a hot component (\chot) in our continuum modeling with
$T_{d}=400$ K (yellow curve in Fig.~\ref{continua}b) attenuated by
$A_{\mathrm{V}}\approx60$ mag, which yields an unattenuated flux at 14 $\mu$m
of $\approx2$ Jy -twice as observed. However, \chot\ has an effective
diameter of only $\approx5$ pc, while the observed HCN absorption must be
produced over spatial scales of $\sim100$ pc because, otherwise, the far-IR HCN
absorption would be negligible. Thus, if the HCN molecules are exposed to the
radiation of \chot, this radiation will be diluted and it remains
unclear if it can significantly affect the populations. As \cite{ran11}
suggest, the alternative is that there is a hot component much more prominent,
but very extincted, in such a way that it does not show up in the observed
SED. The occurrence of this component is plausible, but the HCN
molecules responsible for the absorption are in front of, 
and thus apparently exposed to the non-diluted, unextincted 
radiation field characterized by a ``moderate'' $T_{d}$ in the range
$90-130$ K for \cwest, as the continuum models indicate. The associated
dust is very likely optically thick at 14 $\mu$m and thus would obscure
the emission from the above buried 
hot component. Our models for the nuclear region are, however, structureless,
and more refined models are required to clarify the role of the $\nu_2=1^1$
pumping for these lines. Finally, we note that our derived HCN column is a
factor of 20 higher than that derived by \cite{lah07} from the 14 $\mu$m
absorption band; as the authors state, however, their analysis is based on a
covering factor of unity of the mid-IR continuum and their columns are
then lower limits. 

\subsubsection{NH$_3$}

Our model for NH$_3$ is shown with red lines in Fig.~\ref{nh3} and uses the
same parameters as for HCN: $n_{\mathrm{H_2}}=4\times10^{5}$ cm$^{-3}$, 
$T_g=150$ K, $T_{d}=110$ K, and the same source size. 
The model treats line overlaps, and uses the collisional rates from
\cite{dan88}. The derived column density per unit of $\tau_{50}$ is 
$N_{\mathrm{NH_3}}\approx1.5\times10^{17}$ cm$^{-2}$, i.e. 
$\mathrm{HCN/NH_3}\sim4$ and $\chi_{\mathrm{NH_3}}\approx4\times10^{-7}$.
The best fit for all lines is found for $T_g=150$ K, as for $T_g=250$ K the
lower-lying $(3,K)\rightarrow(2,K)$ lines would be underestimated. There is
still some missing flux in these lines at $T_g=150$ K, indicating possible
low-level absorption in a more extended region.
Within a given $K-$ladder, excitation through absorption of far-IR photons
dominates over collisional excitation.
The data are consistent with an ortho-to-para ratio of 1, indicating a
  formation temperature of at least $\sim30$ K \citep{tak00}. 
  Similar columns are obtained for $T_{d}=90$ and 130 K, with the
  appropiate source sizes as derived for \hdo, OH, and HCN.
Half of the derived NH$_3$ column is stored in the metastable ($J=K$) levels, 
giving $\tau_{50}=1$ metastable columns in agreement with the value inferred
by \cite{ott11} from the inversion lines for $T_{ex}=50$ K
($N_{\mathrm{NH_3}}=0.84\times10^{17}$ cm$^{-2}$).

\subsection{The low-lying lines in Arp~220: the extended and halo
  components (\cext\ and \chalo)}
\label{sec:arp220low}

\subsubsection{H$_2$O}

In Arp~220,
the absorption in the low-lying \t321212\ (Fig.~\ref{h2o}k), \t221110\
(Fig.~\ref{h2o-b}b), \t303212\ (Fig.~\ref{h2o-b}p), and especially the
ground-state \t212101\ (Fig.~\ref{h2o-b}q) lines, 
require a more extended, low-excitation region (\cext, G-A04). This is
consistent with the continuum models, as the nuclei cannot account for the
bulk of the far-IR emission between 50 and 300 $\mu$m, and an extended
component heated in part by radiation from the nuclei, but with a
possible contribution from extended star formation, is invoked (green curve in
Fig.~\ref{continua}b). In our models, $T_d$ in
\cext\ is calculated from the balance between heating (by the 
far-IR emission from the nuclei) and radiative cooling, attaining values
between 40 and 90 K. In the model, 
\cext\ extends up to an uncertain diameter of 
$d_{\mathrm{extended}}=650$ pc, has a column of $N_{\mathrm{H}}\sim8\times10^{23}$
cm$^{-2}$, an average density of $n_{\mathrm{H}}\sim10^{3}$ cm$^{-3}$, and a
mass of $M\sim3\times10^9$ \Msun\ (Table~\ref{tab:cont}).
  Most of its luminosity ($\approx7\times10^{11}$ \Lsun) is
  likely reemission, rather than intrinsically generated.

   \begin{table*}
      \caption[]{Parameters of the continuum models.}
         \label{tab:cont}
          \centering
          \begin{tabular}{lccccccccc}   
            \hline
            \noalign{\smallskip}
            Source  &  C$^{\mathrm{a}}$ & Diameter & $T_d$ & $\tau_d$ &
            $N_{\mathrm{H}}^{\mathrm{b}}$ & $M^{\mathrm{c}}$ &
            $L^{\mathrm{d}}$ & $\tau_{\mathrm{25,fgr}}^{\mathrm{e}}$ &
            $L_{\mathrm{att}}^{\mathrm{f}}$ \\ 
                    &   &  (pc)  &  (K) & at 200 $\mu$m & (cm$^{-2}$) &
                    ($10^8$ \Msun) & (\Lsun) & & (\Lsun) \\ 
            \noalign{\smallskip}
            \hline
NGC~4418 & \chot\ & $5.2$ & $350$ & $0.007$ & $3.5\times10^{22}$ &
$8\times10^{-5}$ & $8.5\times10^{10}$ & $1.3$ & $9.0\times10^{9}$ \\ 
         & \ccore\ & $20$ & $140-150$ & $1-2$ & $(0.5-1)\times10^{25}$
         & $0.16-0.33$ & $(6-8)\times10^{10}$ & $1.5$ & $3.1\times10^{10}$ \\ 
         & \cwarm\    & $31$  & $110$ &  $0.5$ & $2.4\times10^{24}$
         &$0.2$ & $6\times10^{10}$ & $1.2$ & $2.9\times10^{10}$ \\ 
         & \cext\    & $200$ & $90-30$ & $0.03$ &
         $(1.0-1.3)\times10^{23}$ & $0.18$ & $2.4\times10^{10}$ & $0$ &
         $2.4\times10^{10}$ \\ 
Arp~220  & \chot\ & $5.2$ & $400$ & $0.007$ & $3.4\times10^{22}$ &
 $8\times10^{-5}$ & $1.4\times10^{11}$ & $1.1$ & $2.0\times10^{10}$ \\
         & \cwest\ &  $150-80$   & $90-130$  &  $6-12$ &
$(3-6)\times10^{25,\mathrm{g}}$ 
         & $55-30^{\mathrm{g}}$ & $(6.5-8.2)\times10^{11}$ & $1.4$ &
$3.7\times10^{11}$ \\ 
         & \ceast\    &  $108$   & $87$  &  $2.7$ & $1.3\times10^{25}$
         & $13$ & $3\times10^{11}$ & $1.0$ & $2.1\times10^{11}$ \\ 
         & \cext\      &  $650$  & $90-40$ & $0.17$ & $8\times10^{23}$ & $30$
         & $6.9\times10^{11}$ & $0$ & $6.9\times10^{11}$ \\ 
            \noalign{\smallskip}
               \noalign{\smallskip}
            \hline
         \end{tabular} 
\begin{list}{}{}
\item[$^{\mathrm{a}}$] Component: in Arp~220, \cwest\ (\ceast) is the western
  (east) nucleus.  
\item[$^{\mathrm{b}}$] Column density of H nuclei, calculated assuming a
  mass-absorption coefficient of 12.3 cm$^2$/g at 200 $\mu$m and a
  gas-to-dust mass ratio of 100.
\item[$^{\mathrm{c}}$] Estimated mass, assuming spherical symmetry. 
\item[$^{\mathrm{d}}$] Unattenuated luminosity of the component. 
\item[$^{\mathrm{e}}$] Foreground opacity at 25 $\mu$m. 
\item[$^{\mathrm{f}}$] Attenuated luminosity of the component. 
\item[$^{\mathrm{g}}$] The estimated mass of \cwest, based on
  the flux density at 860 $\mu$m reported by \cite{sak08} and on an
  emissivity index of $\beta=2$ in the submillimeter 
  ($\kappa_{860}\approx0.7$ cm$^2$ g$^{-1}$ of dust), is higher than the
  dynamical mass of $\sim10^9$ \Msun\ estimated by several authors
  \citep{dow98,sak08}. This probably indicates that $\beta=1$ is a better
  approach in the submillimeter (yielding $\kappa_{860}\sim3$ cm$^2$ g$^{-1}$ of
  dust), as estimated for some Galactic massive star-forming
  regions \citep[e.g.][]{zha07}. 
\end{list}
   \end{table*}

   \begin{table*}
      \caption[]{Derived \hdo\ column densities and abundances, column density
        ratios, and covering factors of the continua.}
         \label{tab:lines}
          \begin{tabular}{lccccccccc}   
            \hline
            \noalign{\smallskip}
            Source  &  C$^{\mathrm{a}}$ & $N_{\mathrm{H_2O}}^{\mathrm{b}}$ &
            $\chi_{\mathrm{H_2O}}^{\mathrm{c}}$ & OH/\hdo$^{\mathrm{d}}$ & 
            HCN/\hdo$^{\mathrm{d}}$ & HCN/NH$_3^{\mathrm{d}}$ & 
            $f_C$(\hdo)$^{\mathrm{e}}$ & $f_C$(OH)$^{\mathrm{e}}$ &
            $f_C$(HCN,NH$_3$)$^{\mathrm{e}}$  \\
                    &   &  (cm$^{-2}$)  &  &  & & & & &  \\
            \noalign{\smallskip}
            \hline
NGC~4418 & \ccore\  & $(2-6)\times10^{18}$  & $(0.5-1.5)\times10^{-5}$ &
$0.3-0.6$ & $0.1-0.3$ & $2-4$ & $0.85$ & $1$ & $0.7$    \\
         & \cwarm\ & $\sim5\times10^{16}$  & $(0.5-2.5)\times10^{-7}$ &
         $\sim0.4$ & &  & 1 & 1 &     \\
         & \cext\ & $\sim0.2\times10^{16}$  & $\sim2\times10^{-8}$ & $5-10$ &  
  &  & 0.45 & 0.45 &     \\
Arp~220  & \cwest\ & $(0.8-6)\times10^{18}$ & $(0.2-1.5)\times10^{-5}$ &
$0.4-1.0$ & $0.1-0.4$ & $3-5$ & 1 & 1 & 1 \\ 
         & \ceast\ & $3\times10^{17}$ & $0.8\times10^{-6}$ & $\sim1$ &
 &  & $0.4$ & $0.4$ &  \\ 
         & \cext\      &  $3\times10^{16}$  & $4\times10^{-8}$ & $\sim3$  &
          &  & 1 & 1 &  \\  
         & \chalo\      &  $(5-10)\times10^{15}$  & $5\times10^{-8}$ &
         $\sim10$ & &  & 0.7 & 1 &  \\  
            \noalign{\smallskip}
               \noalign{\smallskip}
            \hline
         \end{tabular} 
\begin{list}{}{}
\item[$^{\mathrm{a}}$] Component (see Table~\ref{tab:cont}). In Arp~220,
  \chalo\ is an absorbing, widespread component in front of the nuclear
  region.  
\item[$^{\mathrm{b}}$] Column densities per unit of dust opacity at 50
    $\mu$m, $\tau_{50}$, for \ccore\ and \cwarm\ in NGC
  4418, and for \cwest\ and \ceast\ in Arp~220, with an estimated
    column of H nuclei per $\tau_{50}$ of $4\times10^{23}$
  cm$^{-2}$. For the \cext\ component in both sources, mixed 
  values (i.e., dust and molecules are coexistent throughout the source;
    see text for details).   
\item[$^{\mathrm{c}}$] Estimated \hdo\ abundance relative to H nuclei.
\item[$^{\mathrm{d}}$] Column density ratios.
\item[$^{\mathrm{e}}$] Covering factor of the continuum (same value for HCN
  and NH$_3$).
\end{list}
   \end{table*}

With $N_{\mathrm{H_2O}}=3\times10^{16}$ cm$^{-2}$ and
$\chi_{\mathrm{H_2O}}\approx4\times10^{-8}$,
the \hdo\ line absorption due to \cext\ is shown with green curves in
Figs.~\ref{h2o} and \ref{h2o-b} only for the lines mentioned above. We have
used for this component a ``mixed'' approach (i.e. dust and molecules coexist)
because of the moderate far-IR opacities. 
We expect reemission in the \t303212\ and \t322313\ lines (G-A04); in the
former line reemission is indeed observed but redshifted (\S\ref{sec:obser}). 
On the other hand, a transition zone between the nuclear region and
  \cext\ is traced by the \hdo\ lines in emission at
  $\lambda>200$ $\mu$m detected by \cite{ran11}, as will be discussed in a
  future paper.

The depth of the ground-state \t212101\ \hdo\ line at 179.5 $\mu$m,
however, is not quite 
reproduced even with the absorption by \cext. The line is probably
contaminated by CH$^+$ 
$(2\rightarrow1)$, but its contribution is not expected to completely fill
the observed absorption. This is not surprising, as foreground
low-excitation molecular absorption of the far-IR emission from the nuclei
is not included in the calculations of \S\ref{sec:arp220high}.
Thus we have also included a low-excitation component, \chalo, with no
  intrinsic continuum but located in front of the nuclei, so that it absorbs 
the far-IR emission from the nuclei. 
This was denoted as the ``halo'' in G-A04, and it is probably a
combination of a part of \cext\ located in 
the sight line toward the nuclei, together with more widespread, foreground,
low-excitation gas. The \hdo\ column of this component is uncertain 
due to line saturation, but it is consistent with
$N_{\mathrm{H_2O}}\sim(5-10)\times10^{15}$ cm$^{-2}$ (Table~\ref{tab:lines}). 
The absorption in the [O {\sc i}] 63 $\mu$m line (Fig.~\ref{inoutflow}a) is 
also expected to be produced in low-excitation gas, 
which we identify with the \chalo\ component. From the
observed $W_{\mathrm{eq}}\approx45$ \kms\ (Fig.~\ref{oh-fits}) corrected 
for the estimated fraction of the nuclear 63 
$\mu$m continuum (36\%), we obtain $N_{\mathrm{O^0}}\sim2.5\times10^{19}$
cm$^{-2}$. For an oxygen abundance of  $\chi_{\mathrm{O^0}}=2.5\times10^{-4}$
\citep{car01}, \chalo\ is characterized by $N_{\mathrm{H}}\sim10^{23}$
cm$^{-2}$ and $\tau_{\mathrm{25,fgr}}\sim1$. 
The corresponding \hdo\ abundance is  
$\chi_{\mathrm{H_2O}}\sim0.5\times10^{-7}$ relative to H nuclei, similar
  to the value found in Galactic translucent clouds \citep{plu04}.
The inferred $N_{\mathrm{H}}\sim10^{23}$ cm$^{-2}$ and 25 $\mu$m continuum
opacity of $\sim1$ are close to the values obtained from the fit to the
continuum for the absorbing layer in front of the nuclei and of the hot
component (Table~\ref{tab:cont}). While the column density across the
  \chalo\ component is only accurate within a factor 2, the above
  comparison suggests that \chalo\ is the origin of the foreground
extinction of the mid-IR nuclear emission.

\begin{figure}[ht]
   \centering
   \includegraphics[width=8.0cm]{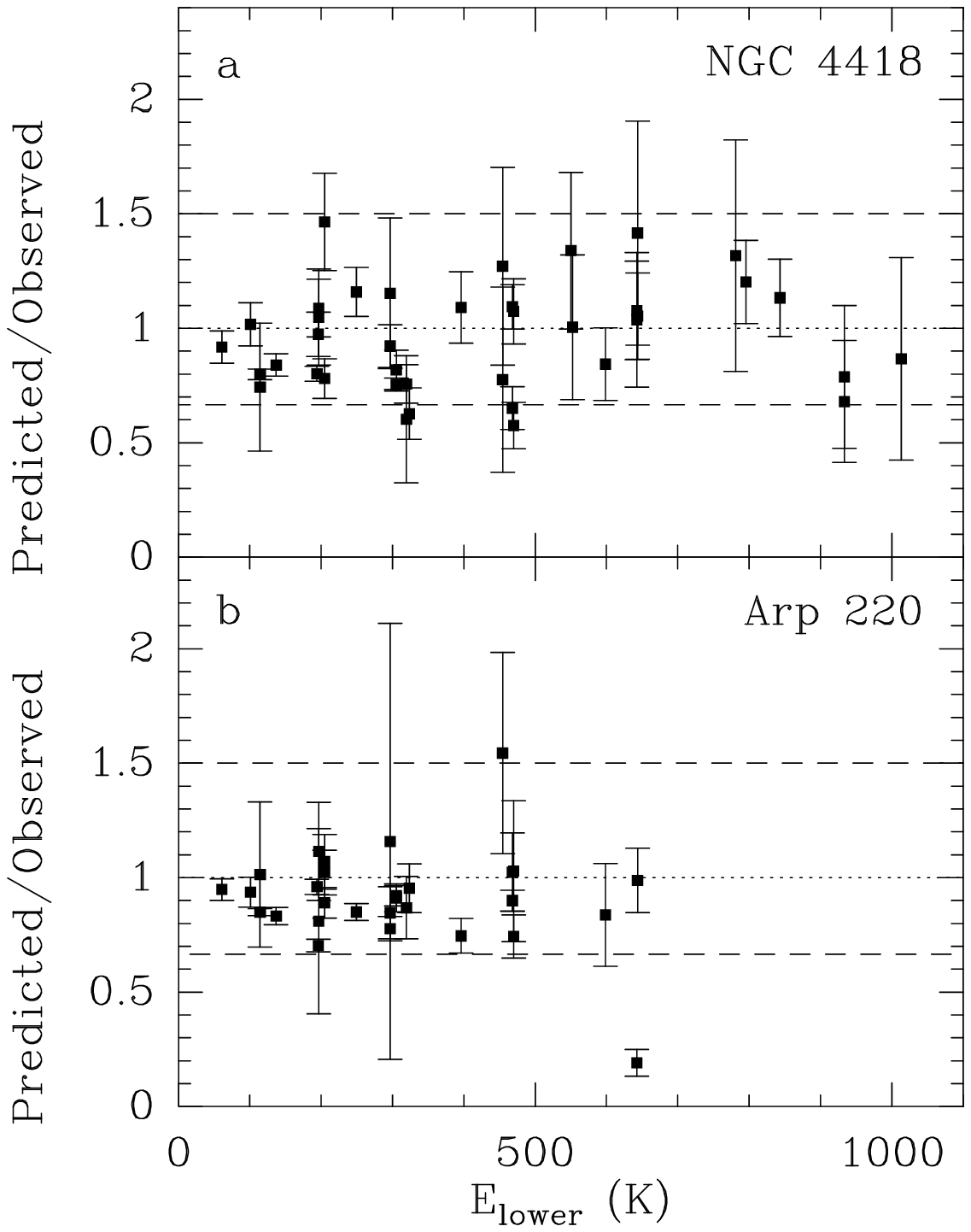}
   \caption{Predicted-to-observed \hdo\ flux ratios in NGC~4418 
     (upper panel) and Arp~220 (lower panel) for the models
     including all components. See \S\ref{sec:models} for details.}    
    \label{h2o_predobs}
    \end{figure}

Figure~\ref{h2o_predobs}b shows the predicted-to-observed flux ratios of the
detected lines for the full reference model including
\cwest, \ceast, and \cext. Values for the contaminated \t717606\
and \t212101\ lines are not plotted. Only two lines, the \t431422\ ($147$
$\mu$m) and \t615606\ ($104$ $\mu$m) show departures larger than 50\%, but the
small predicted-to-observed value of the latter probably indicates that the
feature is contaminated by OH$^+$ (\S\ref{sec:obser}). Among the undetected
lines, the absorption in the \t533422\ and \t413404\ lines at $53.1$
and $187.1$ $\mu$m (Figs.~\ref{h2o}a and \ref{h2o-b}o) are too strong 
in all relevant models.  

\subsubsection{OH}

The far-IR spectrum of \hdo\ has only one ground-state line at $179.5$
$\mu$m. Since this line is strongly saturated, the species is better
suited to describe the region of highest molecular excitation. OH has,
  however, within the PACS wavelength coverage three ground-state
doublets with different strengths at 119, 79, and $53.3$ $\mu$m
(Fig.~\ref{enerdiag}), allowing us to better characterize the \cext\ and
\chalo\ components. 
This is illustrated in Fig.~\ref{oh}a-c, where the highly excited OH in the
nuclei is expected to produce weak absorption 
in these ground-state lines, thus requiring a more extended region where
relatively low-excited OH absorbs a significant fraction of the far-IR
continuum of the galaxy. Two more OH transitions also trace \cext:
the $\Pi_{3/2}$ $7/2\rightarrow5/2$ doublet at 84 $\mu$m (Fig.~\ref{oh}e), and
the  $\Pi_{1/2}$ $3/2\rightarrow1/2$ doublet at 163 $\mu$m. The former line,
with $E_{\mathrm{lower}}=120$ K, is also too strong to be reproduced with the
model for the nuclei, and should be significantly excited in 
\cext. The 163 $\mu$m doublet, the only one observed in emission, is
excited in \cext\ through absorption of continuum photons in the
ground-state $53.3$ and 35 $\mu$m lines, followed by cascade down to the
ground OH state \citep{gen85}, and thus traces large
spatial-scale gas illuminated by the strong far-IR continuum (G-A04). We
indeed expect the 163 $\mu$m doublet to be in absorption toward the nuclei
(Fig.~\ref{oh}j), and thus \cext\ is required to produce emission in the
line. The above pumping mechanism also has observational effects on the
ground-state 79 $\mu$m line: since every 163 $\mu$m photon should be
accompanied, through downward cascade, by a 79 $\mu$m line photon, the
emission in the 163 $\mu$m doublet ensures emission in the 79 $\mu$m
transition as well. This is indeed observed in Arp~220 
(Figs.~\ref{oh}b and \ref{inoutflow}d),
where the 79 $\mu$m doublet shows an emission feature at the most redshifted
velocities of 300 \kms\ with respect to the systemic one; at lower
velocities, one would expect a combination of 
emission and absorption with the latter dominating for suitable geometries
and/or sufficiently large amounts of OH in front of the far-IR continuum
source. With 163 $\mu$m absorption in front of the nuclei, and
emission from more extended regions, the 163 $\mu$m doublet
is expected to be redshifted in outflowing gas, as observed in Arp~220
(Fig.~\ref{oh}j). 

In our specific model for Arp~220, the required OH column and abundance in
\cext\ are $N_{\mathrm{OH}}\approx10^{17}$ cm$^{-2}$ and
$\chi_{\mathrm{OH}}\approx1.3\times10^{-7}$, much lower than towards the nuclear
region and favoring an OH/\hdo\ column ratio of $\sim3$. This ratio reverses
the value found in the nuclear region where OH/\hdo\ is $\sim0.5$. 
The \cext\ component then accounts for the 163 $\mu$m line, though 
the redshift is not reproduced as no outflow is included in this component,
and for the absorption at 84 $\mu$m. Similar to \hdo, however, the full
absorption in the ground-state 119, 79, and $53.3$ $\mu$m doublets is 
not reproduced; in particular, the absorption in the 79 $\mu$m line is 
predicted to be very weak. We again find that the missing
absorption can be attributed to \chalo, 
requiring a column of $N_{\mathrm{OH}}\sim8\times10^{16}$ cm$^{-2}$. The
contribution of this component to the absorption in the ground-state 119, 79,
and $53.3$ $\mu$m doublets is shown with yellow curves in Fig.~\ref{oh}a-c.
The OH/\hdo\ column density ratio in \chalo\ is estimated to be
$\sim10$, but the relative OH and \hdo\ columns in the \cext\ and \chalo\ 
  components are uncertain due to the difficulty of disentangling
 the two components.  

 The ground-state 35 $\mu$m doublet of OH ($\Pi_{1/2}\rightarrow
  \Pi_{3/2} \, 5/2\rightarrow 3/2$) has been observed with ISO by
  \cite{ski97}; correcting their continuum flux density at 35 $\mu$m 
    (62 Jy) to the lower value of 40 Jy indicated by the Spitzer IRS
    spectrum in Fig.~\ref{continua}b, the observed absorbing flux in 
  the 35 $\mu$m doublet is $\approx1.3\times10^{-12}$ erg s$^{-1}$ cm$^{-2}$.
  The joint contribution of \cwest, \ceast, \cext, and \chalo\ predicts a
  comparable absorbing flux of $\approx1.1\times10^{-12}$ erg s$^{-1}$
  cm$^{-2}$, dominated by \chalo\ (G-A04).


\subsection{The high-lying lines in NGC~4418: the nuclear core (\ccore)}
\label{sec:n4418high}

\subsubsection{HCN and \hdo}
\label{sec:coren4418}

Similar to Arp~220, NGC~4418 shows strong mid-IR silicate absorption
\citep[Fig.~\ref{continua}, see also][]{spo01} and absorption in the 
[O {\sc i}] 63 $\mu$m line, indicating important mid-IR extinction.  
We have thus followed a similar analysis as for Arp~220, based on the
combination of the continuum and line absorption. In NGC~4418,
line broadening is purely modeled as microturbulence with 
$v_{\mathrm{tur}}=60$ \kms. 

The extreme excitation of H$_2$O and HCN provides the most stringent
requirements for the dust temperature ($T_{d}$), gas temperature
($T_{\mathrm{g}}$), density ($n_{\mathrm{H_2}}$), and column densities in the
nuclear region. The high-lying lines of these species arise in a compact
component that we denote as the ``nuclear core'' (\ccore). Preliminary
models for HCN in the \ccore\ component of NGC~4418 are shown in
Figs.~\ref{hcn} and \ref{hcn-fits}, with the HCN shell extended up to a
  depth of $\tau_{50}=4$. We generated several models that
  included/excluded excitation through collisions, absorption of far-IR
  radiation, and pumping through the $\nu_2=1^1$ state, and found that 
HCN is primarily excited through collisions, with the 
far-IR emission significantly modifying the distribution of populations
in the highest lines. With the 
SLED apparently peaking at $J=21\rightarrow20$, the \ccore\ model requires 
$n_{\mathrm{H_2}}\sim3\times10^{6}$ cm$^{-3}$ and $T_g$ in the
most probable range of $220-250$ K, significantly higher than in Arp~220. 
Decreasing $n_{\mathrm{H_2}}$ by a factor of 2 would increase $T_g$ to
  $\approx300$ K. 
The high density and $T_g$ values are also required to reduce the otherwise
too strong OH absorption in the ground-state lines,  
to account for the NH$_3$ observations described below, and to
reproduce the far-IR CO emission (discussed in a forthcoming paper).
In spite of the high collisional excitation, the HCN lines are detected in
absorption, indicating that the dust in \ccore\ is warm and at least a
fraction of \ccore\ is optically thick at far-IR wavelengths. 
A reasonable fit to the HCN lines is obtained with $T_{d}=140-150$ K and
$\tau_{\mathrm{200}}\gtrsim1$ for the continuum. With
  $\tau_{\mathrm{200}}\approx1$ (solid curve and triangles in
  Fig.~\ref{hcn-fits}), we expect the rotational lines $J_{\mathrm{upper}}\leq14$
  to be in emission. The strength of the emission lines would decrease if
  $\tau_{\mathrm{200}}\approx4$ (dashed curve), and further weaken with a
  screen model for HCN (dotted curve).
The inferred HCN column per unit of $\tau_{50}$ is
$N_{\mathrm{HCN}}=(6-7.5)\times10^{17}$ cm$^{-2}$, 
i.e. $\chi_{\mathrm{HCN}}=(1.5-2)\times10^{-6}$.  As for Arp~220, we
find that the pumping of the $\nu_2=1^1$ state has little effect on the
excitation and strength of the detected high-$J$ ground-vibrational state
lines, as expected from our moderate value for $T_{d}$ \citep{sak10}.

The minimum diameter of \ccore, $d=16.5$ pc, is obtained
by assuming $f_C=1$. The corresponding luminosity is
$\approx6\times10^{10}$ \Lsun, about 60\% of the 
entire luminosity of the galaxy. 
The \ccore\ component cannot be much bigger
or warmer as its luminosity would exceed the bolometric luminosity of the
galaxy (assuming spherical symmetry for the emitting radiation). 
We also attempted to fit the HCN SLED with lower $T_{d}$ in the range
$110-140$ K. At 130 K, the source size is increased by 18\%
relative to the 150 K case, the \ccore\ luminosity is decreased by 20\%, and
the $J=25\rightarrow24$ line is underestimated; thus our modeling favors
$T_{d}\gtrsim130$ K. With $T_{d}=150$ K, most energy is emitted at wavelengths
shorter than 40 $\mu$m, but the flux density at $150-190$ $\mu$m is still
$4.5-3$ Jy, and this is the continuum absorbed by the HCN. In
Fig.~\ref{continua}a, the SED of \ccore\ (blue curve), with
an adopted $d=20$ pc, is attenuated by foreground dust with
$\tau_{\mathrm{25,fgr}}=1.5$. The corresponding covering factor for HCN is
$f_C\approx0.7$. With $\tau_{\mathrm{200}}=1-2$, the radial column through 
\ccore\ is $N_{\mathrm{H}}\sim(0.5-1)\times10^{25}$ \cmd, sufficient to
strongly attenuate the X-ray emission from an AGN located at the center.

The \ccore\ model was applied to \hdo, and results are shown in
Figs.~\ref{h2o} and \ref{h2o-b} (blue curves). The highest excited \hdo\
  lines are the least saturated ones, which makes them most sensitive
to the column density, and we derive $N_{\mathrm{H_2O}}$ in the
range $(3-6)\times10^{18}$ cm$^{-2}$. The \hdo\ abundance relative to H 
is as high as $\chi_{\mathrm{H_2O}}=(0.5-1.5)\times10^{-5}$, and the corresponding
HCN/\hdo\ ratio in \ccore\ is $0.1-0.3$. The \hdo\ models also confirm
  the very warm ($T_{d}=140-150$ K) source of far-IR emission, though the dust
  mixed in the \hdo\ shell has a somewhat lower $T_{d}=130$ K to avoid strong
  emission in the \t624615\ line at 167 $\mu$m (Fig.~\ref{h2o-b}n).  
\hdo\ excitation through collisions has two main effects: 
first, it increases the
excitation of the low-lying lines among the lowest $\approx8$ levels of ortho
and para-\hdo\ (over that of pure radiative pumping), providing a better
fit to the observations; second, it also significantly affects some
high-lying lines that, although radiatively 
pumped, are populated from the reservoir of \hdo\ in the low-lying
levels. This combination of increased collisional excitation in the low-lying
levels, coupled to the radiative pumping to upper levels, affects mainly the
\t818707\ and \t615524\ transitions, which are strengthened by
$\sim40$\% relative to models where collisional excitation is neglected. 

Our best fit model for \hdo\ uses $\tau_{\mathrm{200}}=1-1.5$;
a higher value (e.g. $\tau_{\mathrm{200}}\approx4$), still compatible with
HCN, would produce too strong absorption in the \hdo\ lines at 
long wavelengths ($\gtrsim150$ $\mu$m, specifically in the \t303212\ 
transition in Fig.~\ref{h2o-b}p). On the other hand,
the covering factor for \hdo\ is slightly
higher than for HCN. These points could indicate that the
very excited HCN is more concentrated than \hdo\ (and mostly OH) toward
the warmest region of \ccore. Figures \ref{h2o}-\ref{h2o-b} and
  \ref{h2o_predobs} indicate that the model for \ccore\ is 
satisfactory for a number of \hdo\ lines, 
and Fig.~\ref{h2o-c} shows that the
model with $N_{\mathrm{H_2O}}=5\times10^{18}$ cm$^{-2}$ predicts 
absorption comparable to that observed in most of the marginally detected
\hdo\ lines.

\subsubsection{OH}

The high-lying OH doublets in the $\Pi_{3/2}$ ladder at 65 and 53 $\mu$m, and
in the $\Pi_{1/2}$ ladder at 71 and 56 $\mu$m, are compared in Fig.~\ref{oh}
(upper spectra) with the model for \ccore\ (blue curves). In this model we
have also adopted the same parameters as for \hdo, and the fit to OH
shown in Fig.~\ref{oh} is obtained with an OH abundance decreased by factor of
$2.5$ relative to that of \hdo. The OH 65 and 71 $\mu$m
  doublets are somewhat underestimated. The 
OH/\hdo\ ratio in \ccore\ of NGC~4418, $\sim0.4$, is similar to that found in
the nucleus of Arp~220. Like \hdo, the model for OH also favors 
$\tau_{\mathrm{200}}\approx1$ to avoid too strong absorption in the 163
  $\mu$m doublet, and the covering factor for OH is 
higher than for HCN and \hdo\ (Table~\ref{tab:lines}). Similar to
Arp~220, all other OH lines 
have important contribution from a more extended region, and are discussed
below in \S\ref{sec:warmn4418} and \ref{sec:n4418low}.

\subsubsection{NH$_3$}

The \ccore\ model for NH$_3$, shown with red lines in Fig.~\ref{nh3}, uses the
same parameters as derived above for HCN:
$n_{\mathrm{H_2}}=3\times10^{6}$ cm$^{-3}$, $T_{d}=150$ K,
and $T_g=200$ and 250 K, but results for these two $T_g$ values were found
to be very similar. The NH$_3$ data are mostly compatible with
  $\tau_{200}\approx1$. The derived column is 
$N_{\mathrm{NH_3}}\approx2.5\times10^{17}$ cm$^{-2}$, i.e. 
$\mathrm{HCN/NH_3}\approx3$ and $\chi_{\mathrm{NH_3}}\approx6\times10^{-7}$.
In comparison with the model for Arp~220, the higher $T_{d}$ in NGC
4418 translates into a higher excitation of non-metastable levels along a
given $K-$ladder, and the higher $T_{\mathrm{g}}$ implies higher excitation
across $K-$ladders. The former effect is responsible for the asymmetry between
the $124.6$ and $124.9$ $\mu$m features in Fig.~\ref{nh3}d, which is not fully
accounted for by the model; nevertheless, the effect is sensitive to the
overlap between the $K=0$ and $K=1$ lines. On the other hand, the model
accounts for the absorption in the $(8,6)s\rightarrow(7,6)a$ line and,
with the contribution of \hdo, for the feature at $63.47$ $\mu$m, though the
marginal absorption in the $(8,5)s\rightarrow(7,5)a$ transition is not
quite reproduced. 
As for Arp~220, the $(3,K)a\rightarrow(2,K)s$ feature (panel g) is not
fully reproduced, suggesting a more extended, lower-excitation component for
the low-lying lines.

\subsubsection{H$_2^{18}$O and $^{18}$OH}

Relative to the main isotopologues, the H$_2^{18}$O and $^{18}$OH lines in NGC
4418 are weaker than in Arp~220. We have modeled these lines in NGC~4418
with the same nuclear \ccore\ model as for \hdo\ and OH, and 
modeled results are compared with observations in Figs.~\ref{h218o} and
\ref{18oh}. For H$_2^{18}$O, the model for the \ccore\ component (red curves
in Fig.~\ref{h218o}) that reproduces the \t331220\ and \t321212\ lines uses a
H$_2$O-to-H$_2^{18}$O ratio of 500. 
While the model predictions are consistent with most observed H$_2^{18}$O
  features, the relatively strong H$_2^{18}$O \t423312\ line at 79.5
    $\mu$m (Fig.~\ref{h218o}d) is 
  underestimated by the model. On the other hand, the $^{18}$OH lines are only
  marginally detected in the spectrum of NGC~4418. Our model (red lines in
  Fig.~\ref{18oh}) also uses an OH-to-$^{18}$OH 
ratio of 500, with the result that the undetected red $120.2$ $\mu$m
$\Lambda-$component and the $^{18}$OH 85 $\mu$m doublet are
  overpredicted. Still, the low SNR $^{18}$OH 
lines at 65 and 85 $\mu$m are affected by an uncertain baseline;
owing to additional uncertainties in the OH column and distribution, we
  estimate a lower limit for the OH-to-$^{18}$OH ratio of 250. In summary, the
$^{16}$O-to-$^{18}$O ratio in NGC~4418 is estimated to be $\gtrsim250$, with a
best fit overall value of $\sim500$.

\subsection{The mid-excitation \hdo\ lines in NGC~4418: the warm component
  (\cwarm)} 
\label{sec:warmn4418}

Figures \ref{h2o} and \ref{h2o-b} show that the model for \ccore\
underestimates the absorption in some mid- and low-lying lines of \hdo\
(\t330221, \t321212, \t322211, \t221110, \t414303, \t313202, and \t212101),
and Fig.~\ref{continua} indicates that the SED of the attenuated \ccore\
cannot account for the bulk of the far-IR emission. Both points indicate the
occurrence of another nuclear component that we denote as the warm component,
\cwarm. 

The parameters for \cwarm\ are more uncertain than those for
\ccore. Assuming that the continuum emission from \cwarm\ is attenuated 
by the surrounding extended component (\cext) described below, the dust
temperature is $\sim85-110$ K and $d\sim30-40$ pc (Table~\ref{tab:cont}). 
The model for the continuum and \hdo\ lines of \cwarm\ is shown with
light-blue curves in Figs.~\ref{continua}, \ref{h2o}, and \ref{h2o-b}. The
\hdo\ lines mentioned above can be reproduced with
$N_{\mathrm{H_2O}}\sim5\times10^{16}$ 
cm$^{-2}$, much lower than toward \ccore. With an estimated
$\tau_{\mathrm{200}}\approx0.5$, the model for \cwarm\ produces emission in
some lines at $\lambda\gtrsim130$ $\mu$m. The \hdo\ abundance is  
$\mathrm{several}\,\times10^{-7}$.

The model for \cwarm\ is applied to OH in Fig.~\ref{oh} (light-blue
curves), with $\mathrm{OH/H_2O}=0.4$. The main effect of 
\cwarm\ is to increase 
the absorption in the 119 and 84 $\mu$m doublets, and to generate reemission
in the 163 $\mu$m doublet. Nevertheless the joint contribution of the \ccore\
and \cwarm\ components predicts the 163 $\mu$m doublet in absorption,
in contradiction to Fig.~\ref{oh}j, thus
indicating the presence of a more extended component described below.

\subsection{The low-lying lines in NGC~4418: the extended component (\cext)}
\label{sec:n4418low}

Figure~\ref{inoutflow} compares the [O {\sc i}] 63 and 145 $\mu$m lines
and the 119, 79, 84 and 163 $\mu$m OH lines in NGC~4418 and Arp~220,
emphasizing the different kinematics found in both sources. 
In Arp~220, the OH 163 $\mu$m doublet, and the emission features in the 
[O {\sc i}] 63 and OH 79 $\mu$m transitions, are redshifted relative to the
systemic velocity, and thus reveal outflowing gas though affected by
foreground absorption. The broad [O {\sc i}] 145 $\mu$m
line supports this kinematic scenario. In NGC~4418, however, the line
profiles are narrower than, and in some aspects opposite to those in
Arp~220. The main 
spectral characteristics of these lines in NGC~4418, as discussed in
\S\ref{sec:kinematics}, are: 
$(i)$ the [O {\sc i}] 63 $\mu$m absorption is {\em red}shifted by
$\approx110$ \kms, and an emission feature is detected at {\em blue}shifted
velocities; $(ii)$ the absorption in the OH 119 and 79 $\mu$m lines is also
redshifted by $\approx100$ \kms, with some indications of higher redshift in
the 119 $\mu$m than in the 79 $\mu$m doublet (Fig.~\ref{oh-fits}); $(iii)$ a
redshifted asymmetry is seen in the red $\Lambda-$component of the 84 $\mu$m
doublet; $(iv)$ the 163 $\mu$m doublet, with relatively low SNR, is
detected in emission with one of the components apparently 
{\em blue}shifted with a velocity comparable to
that of the [O {\sc i}] 63 $\mu$m emission feature, though the other one peaks
at central velocities. The lowest-lying \hdo\ line at 179.5 $\mu$m 
is also redshifted, but it is contaminated by CH$^+$ $(2\rightarrow1)$.

On the other hand, the high-lying \hdo\ and OH lines in NGC~4418 do not show
indications of any outflow, but peak at the rest-frame velocity as derived
from for the optical absorption/emission lines. Only the ground-state 
[O {\sc i}] 
and OH lines are redshifted, together with some indications of a lower
redshift in the excited OH 84 $\mu$m line. This indicates that this kinematic
component is tracing a relatively low-excitation component around the galaxy
nucleus. Near-IR observations \citep{eva03} do not show evidence for a nearby
``contaminating'' companion where these lines could be formed (the
closest galaxy, VV 655, is at $\sim25$ kpc), but a linear,  
compact structure with fainter structures extending radially out to $\sim1''$
($140$ pc) in radius is apparent. This morphology may suggest that the
redshifted spectral features are formed in (the inner region of) these
extended streamers, thereby suggesting the possibility of an envelope
inflowing onto the nucleus of NGC~4418. 

We first constrain the column of the redshifted absorbing component in
front of the nucleus by calculating the column of O {\sc i} from
the $W_{\mathrm{eq}}$ of the 63 $\mu$m absorption feature. 
Assuming that the line is 
optically thin at all velocities, that the shell covers the whole far-IR
continuum source at 63 $\mu$m (43 Jy), and ignoring reemission in the line,
we derive $N_{\mathrm{O}{\mathsc i}}=5.3\times10^{18}$ \cmd. This is a lower
limit, as  
models described below indicate that there is significant reemission from the
envelope (i.e. from \cext) in both the 63 $\mu$m continuum and in
the line; we estimate from the models a more accurate
$N_{\mathrm{O}{\mathsc i}}=(1.4-1.8)\times10^{19}$ \cmd. Assuming a gas-phase
O {\sc i} abundance typical of translucent 
clouds \citep[$\chi_{\mathrm{O}{\mathsc i}}=2.5\times10^{-4}$ relative 
to H,][]{car01,lis01}, $N_{\mathrm{H}}\approx6.5\times10^{22}$ \cmd,
or $A_{\mathrm{v}}\sim 30$ mag. Since oxygen
depletion may be enhanced for those columns \citep{ber00}, the allowed range
is $A_{\mathrm{v}}\approx 30-60$ mag. This extinction is at least half of that
required to account for the silicate absorption at 10 $\mu$m. 

One important parameter for the inflowing scenario is the amount of gas
that is inflowing on the lateral and back sides of the nucleus. In order to
constrain the geometry, calculations in spherical symmetry are performed and
compared with data. The model for this extended component (\cext\ in
Tables~\ref{tab:cont} and \ref{tab:lines}) simulates a spherical envelope
with inner and outer radii of $R_{\mathrm{inner}}\approx20$ and
$R_{\mathrm{outer}}\sim100$ pc ($0.15''$ and $0.7''$) inflowing onto the warm
nucleus. The value of $R_{\mathrm{inner}}$ is chosen to match the
  estimated outer radius of \cwarm. The line profiles are best
  matched with a density profile that varies as
$\sim r^{-2}$, with a highest value of $n_{\mathrm{H}}=2.5\times10^3$ \cmt\ in
the neighborhood of the nucleus. At $R_{\mathrm{outer}}$,
  $n_{\mathrm{H}}\approx10^2$ \cmt, and thus the value for the outer radius
  is not critical. 
While the [O {\sc i}]63 $\mu$m absorption can be reproduced by either a
turbulent 
shell inflowing with uniform $v=-110$ \kms, or by assuming a radial velocity 
gradient across the envelope, the OH shifts in Fig.~\ref{oh-fits} suggest a
progressive gas {\em deceleration} with decreasing distance to the galactic
nucleus, with the thickest 119 $\mu$m line, tracing the outermost
regions, showing the highest $V_{\mathrm{shift}}$, and optically thinner
(79 $\mu$m) or excited (84 $\mu$m) lines peaking closer to the rest velocity. 
We thus have empirically adopted a velocity field with $-30$ \kms\ at 
$R_{\mathrm{inner}}$ and $dv/dr=-2.1$ km s$^{-1}$ pc$^{-1}$ (i.e. $-200$ \kms\
at $R_{\mathrm{outer}}$), roughly simulating gas in nearly free-fall at the
outermost radius, and subsequent deceleration due to radiation
  pressure (see \S\ref{sec:lumin}) and/or interaction with orbiting gas, 
with gas approaching the systemic velocity in the neighborhood of the
nucleus.

Our model for OH, shown in Fig.~\ref{inoutflow}, includes both
the nuclear component (the combined emission of the \ccore\ and \cwarm\
components is shown with blue lines) and the inflowing component (\cext,
green lines). 
An interesting feature of the line profiles is that, at central velocities
where the excited lines show deep absorption (and the OH column densities are
high), the absorption in the ground 119 and 79 $\mu$m lines is relatively
weak. The model for the nucleus (\ccore+\cwarm) predicts too much absorption
in these lines even adopting the high density of $3\times10^6$
cm$^{-3}$ and $T_{\mathrm{g}}=200$ K for \ccore. Nevertheless, a reasonable 
match to the profiles involves partial cancellation between the absorption
toward the continuum source and emission (through direct scattering at 119
$\mu$m, and pumping through 53 and 35 $\mu$m for the 79 $\mu$m transition)
from \cext\ around the nucleus. This process also accounts for
the emission in the 163 $\mu$m doublet, as every 163 $\mu$m photon should be
accompanied, through downward cascade, by a 79 $\mu$m line photon. 
In the model, the 163 $\mu$m components are blueshifted by $-45$ \kms,
  but the observed shift is unclear as discussed in \S\ref{sec:kinematics}. 
On the other hand, the adopted velocity gradient predicts
absorption at $\sim-50$ \kms\ in the 84 $\mu$m excited line, nearly matching
the observed asymmetric 84 $\mu$m line shape. The model for OH is then
consistent with an extended envelope and favors the scenario of an inflow
extended on the lateral sides, though the lack of blueshifted emission
in one of the $\Lambda-$components of the 163 $\mu$m doublet leaves the case
inconclusive. The OH column across \cext\ is
$N_{\mathrm{OH}}\approx1.5\times10^{16}$ \cmd, and
$\chi_{\mathrm{OH}}\sim(1.5-3)\times10^{-7}$. From the absorption in 
the \hdo\ \t212101\ line in Fig.~\ref{h2o-b}q, which is contaminated by
CH$^+$, we estimate an uncertain $N_{\mathrm{H_2O}}\sim2\times10^{15}$ \cmd.

The model for OH has been applied to O {\sc i}. The absorption component
  of the 63 $\mu$m line is reproduced with an O {\sc i}/OH abundance
ratio of $10^3$, close to the value expected for translucent clouds.
We first ignore the possible contribution by the nucleus
to the emission feature of the [O {\sc i}] 63 $\mu$m line. The resulting line
shape is 
compared with the observed profile in Fig.~\ref{inoutflow}a (green curve), and
shows that $(i)$ the velocity field and density profile adopted for OH roughly
match the absorption in the [O {\sc i}] 63 $\mu$m line, though the modeled
absorption is somewhat overestimated at low velocities; $(ii)$ in our
spherically symmetric model, the emission feature is formed in the inflowing
envelope through scattering of 63 $\mu$m photons, overestimating by 70\% the
flux of the observed spectral feature (325 Jy \kms\ between $-200$ and 0
\kms). Our model for O {\sc i} thus supports an inflow with wide opening
angle, but with the amount of gas inflowing on the lateral sides being lower
than in front of it. 

   \begin{figure}
   \centering
   \includegraphics[width=8.0cm]{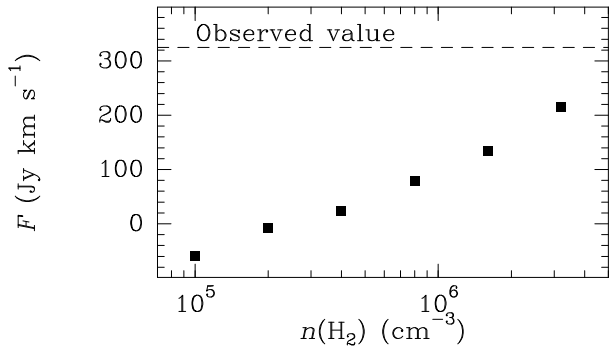}
   \caption{Modeled [O {\sc i}] 63 $\mu$m emission above the continuum, 
     integrated over blueshifted velocities (i.e. between $-200$ and 0
       \kms\ with respect to the systemic velocity), for the nucleus of
     NGC~4418 as a function of the H$_2$ density. The dashed line indicates
     the observed value.}     
    \label{oidens}
    \end{figure}

The above solution for the blueshifted emission in the [O {\sc i}] 63
$\mu$m line, i.e. that the feature is formed in the inflowing envelope, 
 is, however, not unique, because of the possible contribution by the nucleus
 to this feature. Indeed, the [O {\sc i}] 145 $\mu$m 
line, though weak and too narrow as compared with the instrumental spectral
resolution, is detected in the PACS spectrum (Fig.~\ref{inoutflow}b), and the
line should be formed in the nuclear region. Here we attempt to address
the conditions under which the emission in the [O {\sc i}] 63 $\mu$m
  line may be also formed in the nuclear region.
We have then calculated the
expected nuclear emission in the [O {\sc i}] 63 $\mu$m line as a function of
the assumed nuclear gas density. In all models, the nuclear O {\sc i} column
is fixed to match the observed [O {\sc i}] 145 $\mu$m line. We have used a
``mixed'' approach 
as the continuum opacity at 145 $\mu$m is moderate and the inner regions of
the nucleus can contribute to the [O {\sc i}] 145 $\mu$m
line. We have adopted the dust model for the \cwarm\ component with 
$T_{{\mathrm d}}=110$ K, and the gas at $T_{{\mathrm g}}=150$ K. 
Collisional rates between O {\sc i} and H$_2$ are taken from \cite{jaq92}.
Results for the blueshifted (i.e. integrated between $-200$ and 0
\kms) emission are compared with the observed value in Fig.~\ref{oidens}. At
the relatively low density of $10^5$ \cmt, the excitation of the [O {\sc i}]
63 $\mu$m line is below $T_{{\mathrm d}}$, and the line is predicted in weak
absorption. However, the absorption changes to emission for higher
densities, and the blueshifted integrated flux attains a value comparable to
the observed one at $n_{{\mathrm H_2}}\approx 3\times10^6$ \cmt. 
Thus a significant contribution by the compact nucleus to the [O {\sc i}] 63
$\mu$m emission feature is not ruled out from the point of view of the
radiative transfer. For $n_{{\mathrm H_2}}= 3\times10^6$ \cmt, the ``mixed'' 
O {\sc i} column is $N_{{\mathrm O^0}}\approx 1.5\times10^{20}$ \cmd, and 
$\chi_{{\mathrm O^0}}\sim 3\times10^{-5}$. 

The case of the [O {\sc i}] 63 $\mu$m emission appears then somewhat ambiguous
on the basis of these radiative transfer models. Still, the OH models and the
requirement of an extended region to account for the emission in the OH 163
$\mu$m doublet, favor \cext\ as the main contributor to the [O {\sc i}] 63
$\mu$m emission feature, but the relative contribution by the nucleus is hard
to constrain. We further discuss this issue in
\S\ref{sec:disc:inflow}. 

\subsection{The continuum of NGC~4418}
\label{sec:n4418cont}

Our continuum fit to NGC~4418, shown in Fig~\ref{continua}a and with
parameters listed in Table~\ref{tab:cont}, is composed
of: $(i)$ a hot component (\chot, yellow curve) with $T_{d}=350$ K and an
effective diameter of $d\approx5$ pc, attenuated by $A_{\mathrm{V}}=70$ mag.
It accounts for the mid-IR emission including the silicate 10 $\mu$m
absorption.  
$(ii)$ The nuclear core (\ccore, blue curve) with $T_{d}=150$ K and
$d=20$ pc, attenuated by $A_{\mathrm{V}}=81$ mag. It is required to explain the
high-lying molecular absorption observed in \hdo, HCN, OH, and NH$_3$. 
$(iii)$ The warm component (\cwarm, light-blue curve) with $T_{d}=110$
K and $d=31$ pc, attenuated by $A_{\mathrm{V}}=65$ mag. The parameters of this
component are somewhat uncertain and could include regions with lower
$T_{d}\sim85$ K. \cwarm\ mainly accounts for some mid-excitation
lines of \hdo\ and OH. $(iv)$ The extended component (\cext, solid green
curve) with non-uniform $T_{d}$ and $d\approx200$ pc, which mainly accounts
for the redshifted OH absorption and for the emission in the OH 163 $\mu$m
doublet. The emission from the above components still underestimates the 
observed continuum at $\lambda\gtrsim100$ $\mu$m, and therefore we have added 
a more extended, cold component with $d=650$ pc and $T_{d}=50-25$ K
(dashed-green curve in Fig.~\ref{continua}a). This cold component, however,
has no obvious molecular counterpart at far-IR wavelengths and is
included in order to obtain a reasonable fit to the SED at long
wavelengths, though the total emission (red curve in Fig.~\ref{continua}a) still
underestimates the (sub)millimeter emission. Possible causes are the
wavelength dependence of the dust absorption coefficient, and the uncertain
dust opacity and size of the nuclear core at (sub)millimeter wavelengths. 

The unattenuated luminosities of the hot, core, and warm components are
$8.5\times10^{10}$ \Lsun\ (i.e. 85\% of the bolometric luminosity), 
$\sim7\times10^{10}$ \Lsun, and $6\times10^{10}$ \Lsun, respectively. The sum
of the three exceeds the bolometric luminosity, implying that, unless there
are important geometrical effects that make us underestimate the true galaxy
luminosity, most emission of the core and warm components is not intrinsically
generated, but largely reemission due to heating from inside. 
 Nevertheless, clear departures from spherical symmetry are inferred from
  our modeling, as e.g. the extinction affecting \chot\ is much lower than
  the extinction through \ccore, so the latter component can not be
  interpreted as a homogeneous shell surrounding \chot.

\section{Discussion}
\label{sec:discussion}

\subsection{Model uncertainties}

Our schematic models adopt uniform densities and uniform dust temperatures 
for the nuclear components, which is clearly a rough approach as it
ignores their inner structures and clumpiness. The use of spherical symmetry
is also certainly inaccurate, especially in NGC~4418 where a linear structure
for the nuclear region is observed in the near-IR \citep{eva03}. Thus our models
should be considered equivalent spherical/uniform approaches to complex
structures. 

Toward the nuclear region of Arp~220 and NGC~4418,
our column density estimates are dust-limited, in the sense that the line
absorption traces the outermost regions where $\tau_{d}<1$. In this
far-IR photosphere, we estimate that our derived columns per unit of
  $\tau_{50}$ are probably accurate to within $50\%$ for a given
$T_{d}$. Conversion to abundances is more uncertain. We have normalized the
inferred column densities to a total H column of 
$N_{\mathrm{H,norm}}=4\times10^{23}$ \cmd. However, the above H column depends on
the gas-to-dust ratio as well as on the assumed properties of dust grains. 
 Furthermore, the HCN absorption is observed at wavelengths longer than
  most \hdo\ and OH lines, and thus traces higher H columns than the
  latter species.  The molecular abundances are thus reliable only to
  within a factor of $2-3$. Nevertheless, abundance {\em ratios}
    between different molecular species are
  based on the same approach, and are therefore more accurate.

\subsection{The nuclear regions of NGC~4418 and Arp~220}

\subsubsection{A hot core chemistry}

The far-IR Herschel/PACS observations of NGC~4418 and Arp~220 reveal extreme
excitation in the key molecular species \hdo, OH, HCN, and NH$_3$. The
lines are observed mostly in absorption against the continuum, and our
analysis indicates that very high column densities and abundances of these
molecules surround (and are associated with) compact, far-IR thick,
luminous continuum components. 

One of the most intringuing results of our analysis is the high abundance
derived for HCN, $(1-2)\times10^{-6}$ in both Arp~220 and NGC~4418. At
millimeter wavelengths, HCN has 
been found to be enhanced over HCO$^+$ in ULIRGs \citep{gra06} and in
Seyferts over starbursts \citep{ima07}, and this has been proposed to be a
diagnostic of AGN activity. \cite{cos11}, however, argue that for high
densities and columns HCO$^+$ is expected to be stronger than HCN in XDRs
\citep{mei07}. Models by \cite{mei05} predict HCN abundances and cumulative
column densities in XDRs that fall by far too short to explain the values we
derive in both NGC~4418 and Arp~220. Likewise, the peak HCN abundances
predicted by \cite{lep96} for XDRs, $\mathrm{a\, few}\times10^{-8}$, are
about two orders of magnitude lower than the values we derive.

The high HCN abundances in both sources are very likely tracing a hot-core
like chemistry, as has been proposed by previous authors,
  i.e. by \cite{lah07} from the HCN 14 $\mu$m absorption in
  these and other (U)LIRGs, \cite{mar11} for Arp~220 based 
on multispecies observations at 1.3mm, and \cite{aal07} and \cite{cos10}
to explain 
the luminous HC$_3$N emission in NGC~4418. HCN is well known to be enhanced in
Galactic hot cores, as in the Orion hot core where the abundance is
$\sim10^{-6}$ \citep{sch92}, about two orders of magnitude higher than in cold
molecular clouds. Time-dependent chemical models for hot cores by \cite{rod01}
predict a linear increase of the HCN abundance with time for $T_g=300$ K,
attaining values of $10^{-5}$ relative to H$_2$ after $\sim10^6$ yr of
chemical evolution. The authors describe this model result as ``pathological''
as such high abundances are not found in Galactic hot cores, and suggest that
these regions do not survive for more than a million years. However,
extragalactic nuclear regions may survive long enough to develop such high 
HCN abundances.

 The proposed evolved hot core chemistry finds support in the
  $\mathrm{HCN/NH_3}$ ratio of $2-5$ estimated in both Arp~220 and NGC~4418. 
  After $\gtrsim10^6$ yr of chemical evolution, the models by \cite{rod01} 
  predict that a similar value of the ratio can be reached provided that the
  gas temperature is high enough, though HCN is predicted to continue
  increasing with time. The 
  ratio found by \cite{rod01} at such late stages of hot core evolution
  is independent of whether NH$_3$ was injected into the gas phase from mantles
  or formed directly in the gas phase.

The hot-core scenario is also quite consistent with the warm dust temperatures
derived in the nuclear regions of NGC~4418 and Arp~220, and with the
derived high \hdo\ abundances. 
In the Orion hot core, \cite{mel10} derive
\hdo\ abundances of $\sim10^{-5}$ from Herschel/HIFI observations.
Theoretically, an ``undepleted chemistry'' where grain mantles are evaporated
is expected to increase the gas-phase \hdo\ abundance up to values $>10^{-5}$
\citep[e.g.][]{cha97}, i.e. orders of magnitude relative to the values
inferred in cold molecular clouds and PDRs \citep{mel05}, where \hdo\ ice is 
expected to be efficiently formed upon hydrogenation of atomic oxygen depleted
on mantles in cold environments \citep{ber00,hol09}. 
If the gas attains sufficiently high temperatures ($T_g\sim300$ K), \hdo\ will
be further enhanced in the gas phase via neutral-neutral reactions
\citep[e.g.][]{eli78}.  

The current models underpredict the low-$J$ lines of HCN observed
by \cite{aal07} in NGC~4418, who derived densities of $\sim10^5$ \cmt; we
derive much higher densities, $\sim3\times10^6$ \cmt, for the
high-$J$ lines. Likewise, our models for the high-J lines in Arp~220
  underpredict the $J=6\rightarrow5$ measurement of \cite{ran11}. It is
plausible that these sets of lines probe different 
components, with the higher-$J$ lines tracing very dense gas close to the
regions of high far-IR emission with temperatures of $T_d>100$ K, while the
low-$J$ lines trace a surrounding or interclump component with more
moderate densities. No lines of HNC are found in the PACS spectra of
NGC~4418 or Arp~220, while at 
millimeter wavelengths HNC is overluminous relative to HCN \citep{aal07b}.

The high OH abundances in both NGC~4418 and Arp~220 do not easily fit
into the hot core scenario. Observationally, OH observations of Orion BN/KL
by \cite{mel90} indicate that the emission/absorption is mostly produced by
the shocked gas associated with the outflow, with little contribution, if any,
by the hot core. Recently, \cite{wam11} derived a very low
$\mathrm{OH/H_2O}\sim10^{-3}$ abundance ratio in the warm ($T>100$ K),
quiescent component of the Galactic high-mass star-forming region W3 IRS 5. 
Models by \cite{cha97} indeed predict a very low $\mathrm{OH/H_2O}$ ratio in
hot cores as most oxygen at high $T$ is expected to be chanelled into \hdo.

\subsubsection{X- and cosmic rays}
\label{sec:xcrays}

Another component, with chemistry dominated by photoprocesses and/or cosmic
rays, is thus required to account for the OH enhancement in the nuclear region
of both galaxies. Multitransition observations of OH in the Orion bar and
chemical modeling by \cite{goi11} reveal a dense/warm PDR component with 
an OH abundance of $\mathrm{a\, few}\times10^{-6}$. 
However, the OH enhancement is
expected to be restricted to the external $A_{\mathrm{V}}\lesssim1$ layers
where $\mathrm{OH/H_2O}>1$ due to \hdo\ photodissociation by the FUV field
\citep{goi11}. In NGC~4418 and Arp~220, we require widespread OH to account
for the observations and, if PDRs alone are invoked, the surfaces of the
nuclear regions in both sources (with typical sizes of 40 and 100 pc) would
have to be covered with such high-density, low-$A_{\mathrm{V}}$ PDRs with a
covering factor of $\gtrsim0.5$ and, for typical individual PDR columns of
$N_{\mathrm{OH}}\sim\mathrm{a\, few}\times10^{15}$ \cmd\ \citep{goi11},
$\gtrsim100$ of such regions are needed along the line of sight. The
impact of FUV photons on 
large amounts of gas can be enhanced under special geometries, as in the walls
of cavities excarvated by outflows \citep{bru10}; OH is boosted in outflows
where  \hdo\ is photodissociated by strong FUV fields
\citep{mel90,spi00}. However, again a large number of overlapping outflows
would be required across columns of $\mathrm{several}\times10^{23}$
cm$^{-2}$. Furthermore, the outflow signatures in Arp~220 are only
  observed in lines of moderate excitation, and NGC~4418 does not show any
hint of outflowing gas, indicating that high-velocity (as compared with
  Galactic standards) outflows are not responsible for the 
    observed chemistry. Nevertheless, the presence of Orion-like
    outflows with velocities of $\sim30$ \kms\ is not ruled out.

A promising scenario to explain the high OH columns and abundances is
through the effect of cosmic/X-rays, which penetrate in high-$A_{\mathrm{V}}$
regions much more deeply than UV photons. The OH/\hdo\ ratio as predicted by
XDR and CR models of high density molecular regions is $>1$, and
$\chi_{\mathrm{OH}}$ easily exceeds $10^{-6}$ \citep{mal96,mei11}. Large amounts
of OH can be formed through dissociative recombination of H$_3$O$^+$, which in
turn is formed through the sequence
$\mathrm{O^+}\rightarrow\mathrm{OH^+}\rightarrow\mathrm{H_2O^+}
\rightarrow\mathrm{H_3O^+}$ \citep{mal96}; as we will report in a forthcoming
paper, these molecular ions are indeed detected in absorption from
excited levels in both sources.
However, both NGC~4418 and Arp~220 have
strong X-ray deficits relative to the FIR emission \citep{mai03,iwa05}. From
the X-ray flux reported by \cite{mai03} from NGC~4418,
$F_X(0.3-8\,\mathrm{keV})=1.5$ $\mathrm{erg\,s^{-1}\,cm^{-2}}$, and assuming a
point-like X-ray source located at the center of our \ccore\ component, the
X-ray flux at the surface of \ccore\ ($r\sim10$ pc) is $\sim0.1$
$\mathrm{erg\,s^{-1}\,cm^{-2}}$; a similar 
estimate for Arp~220 with the fluxes reported by \cite{iwa01} yields
an X-ray ($0.5-10$ keV) flux at the surface of the western nucleus ($r\sim50$
pc) of $\sim0.5$ $\mathrm{erg\,s^{-1}\,cm^{-2}}$. Since the lines observed with
PACS are formed around far-IR thick components with
$N_{\mathrm{H}}\gtrsim5\times10^{24}$ cm$^{-2}$, any putative strong and
buried X-ray source would be 
severely extincted, and the above distance-corrected estimates are probably
lower limits due to foreground absorption but still relevant within a factor
of 3. Conversion to ionization rates ($\zeta$) and energy deposition
  rates per unit of hydrogen density ($H_X/n$) is very sensitive to the 
adopted X-ray spectrum \citep{war99}; adopting   
$\zeta\,(\mathrm{s^{-1}})\sim2\times10^{-13}
F_X\,(\mathrm{erg\,s^{-1}\,cm^{-2}})$, we infer
$\zeta\sim (0.2-1)\times10^{-13}$ s$^{-1}$ 
and $H_X/n\lesssim (1-5)\times10^{-30}$ erg cm$^3$ s$^{-1}$ \citep{mal96} for
NGC~4418-Arp~220 in the far-IR nuclear photosphere traced by the OH. 
It is unclear if such low energy deposition rates per unit density can 
produce the high nuclear OH abundance we measure at least in NGC~4418, where
very high densities are found. Nevertheless, cosmic rays are an alternative,
and both X- and cosmic rays could
efficiently generate OH from a high reservoir of \hdo\ that is formed by other
means (i.e. by a hot core chemistry or by low-velocity outflows), through
photodissociation of \hdo\ by internally generated UV photons
\citep{gre89,war99} and/or reactions of \hdo\ with He$^+$, H$^+$,
H$_2^+$, and H$_3^+$ to form OH directly or via the H$_2$O$^+$ and H$_3$O$^+$
intermediates; this high \hdo\ reservoir is evident in both sources. Detailed
chemical modeling is required to differentiate between these, and other
alternatives; while \hdo\ may be expected to be abundant in regions where 
HCN-NH$_3$ or OH are enhanced, it remains to be determined whether the
HCN-NH$_3$ and the OH can be cospatial. 

\subsubsection{Sources of luminosity}
\label{sec:lumin}

The presence of high columns of HCN and \hdo\ and the implied hot core
chemistry in the nuclear regions neither supports nor rules out the
presence of a deeply buried AGN. It could simply indicate that,
regardless of the nature of the intrinsic luminosity source(s), frequency
degradation in a very thick medium (as UV radiation is absorbed by dust
  and re-emitted at longer wavelengths by sampling progressively colder
  media), makes far-IR photons more relevant for the chemistry of large
fractions of the nuclear gas (through dust heating and mantle evaporation)
than the quickly absorbed UV photons. 

Taking the Galactic Orion hot core as a template, its size
\citep[$\approx10''$ diameter, or 0.02 pc at 414 pc; e.g.][]{wil00}
is $\sim900$ and $\gtrsim4000$ times smaller (and the square of these numbers
in surface) than those estimated for the \ccore\ and \cwest\ components of NGC
4418 and Arp~220, respectively. If the HCN absorption is interpreted in terms
of hot cores around young stars, then $\sim8\times10^4$ and 
$\sim1.5\times10^7$ cores are inferred, respectively. Likewise, if a
typical hot core harbours a luminous star with $10^5$ \Lsun, $\sim7\times10^5$
and $\sim7\times10^6$ young stars are inferred, respectively. This
interpretation, however, has the drawback that 
Galactic hot cores are short lived, with a time scale of order
$10^5$ yr \citep{wil01}. Even if the high pressure in the nuclear regions
\citep{spo05} allows the hot cores to survive longer, 
a synchronized birth of $\sim10^7$ hot cores in $<1$ Myr
over spatial scales of $\gtrsim80$ pc, involving a propagation
velocity of stellar formation of $>100$ \kms, seems unlikely.
The observed absorption is rather interpreted in terms of large hot-core like
structures (disk/torus/envelope) not necessarily
associated with the very first stages of star formation.

 The inferred hot core chemistry, complemented by the strong effects of
  X- and/or cosmic rays, is found to be physically associated with warm and
  luminous sources of far-IR emission.
The surface brightnesses of the nuclear regions in both sources are
very high, $\sim7\times10^{13}$ \Lsun/kpc$^2$ over 20 pc for 
the \ccore\ component of NGC~4418, 
and $(0.9-4)\times10^{13}$ \Lsun/kpc$^2$ over $150-80$ pc for the
western nucleus of Arp~220. In the latter source, these are lower limits if the
dust is heated from inside \citep{dow07}.    
Sa8 indicates that the luminosity from the Arp~220 western nucleus could be
generated by $200-1000$ super-star clusters.
 An important contribution from star formation to the luminosity of the
  western nucleus is indeed evident \citep[e.g.,][]{smi98,lon06}, 
  and the apparent lack of strong X-ray emission from X-ray binaries
  associated with star formation \citep{lon06} can be interpreted in
  terms of extreme absorption column densities (Sa08, this work).

 On the other hand, the broad 1.6 GHz OH megamaser emission terminal
 velocity of 800 \kms\ in Arp~220 \citep{baa89}, comparable to those of OH in
 the far-IR absorption lines of AGN driven ULIRGs \citep{fis10,stu11}, is much
  higher than the OH velocities observed in the far-IR spectrum of Arp
  220. We interpret this as  
  evidence for both far-IR extinction of the high velocity gas accelerated in
  the innermost nuclear region and deceleration of the outflowing gas with
  increasing involved columns. A very compact source for the outflowing gas
  would also support the presence of an energetically
  significant, strongly buried AGN that would also explain the 
  somewhat different distributions traced by the compact centimeter-wave
  sources and the submillimeter continuum emission (Sa08).
In NGC~4418, the extreme HCN and \hdo\ excitation on 
a scale of 20 pc in diameter, the large fraction of luminosity coming from
such a compact region, and the indications of internal heating, favor the
scenario of a far-IR thick disk/torus/envelope surrounding a nascent
AGN, as has been proposed in previous works \citep[e.g.][]{spo01}, or a
hyper-star cluster $\sim70$ times more luminous than the super-star
  cluster in NGC 5253 \citep{tur00}.

In NGC 4418, the outward
  acceleration due to radiation pressure can be described by 
  $a_{\mathrm{rad}}\sim\kappa\sigma T_d^4/c\sim(2-5)\times10^{-6}$ cm
  s$^{-2}$, where $\kappa\sim2-5$ cm$^{2}$ g$^{-1}$ is the Rosseland mean
  opacity at $100-150$ K \citep{tho05} and $\sigma$ is the Stefan-Boltzmann
  constant. We find that $a_{\mathrm{rad}}$ is
  similar to the gravitational acceleration, 
  $a_{\mathrm{grav}}=GM/R^2\sim(2.5-5)\times10^{-6}$ cm s$^{-2}$, where
  $M=(1.6-3.2)\times10^{7}$ \Msun\ 
  (for $\tau_{200}\sim1-2$) and $R=10$ pc have been estimated
  (Table~\ref{tab:cont}). If the stellar mass is not much higher than the gas
  mass in the \ccore\ component, the above comparison suggests that radiation
  pressure plays an important role in supporting the nuclear structure against
  self-gravity in an Eddington-like limit \citep{sco04}. 
  In Arp~220 (\cwest) the importance of radiation pressure appears less
  important due to the higher $M/R^2$ and lower\footnote{Vertical
    radiation pressure support in optically thick nuclear disks imply
    $\kappa T_{\mathrm{eff}}^4\propto \Sigma_g$ \cite[e.g.][]{tho05}, where
    $T_{\mathrm{eff}}$ is the effective dust temperature and $\Sigma_g$ is the
    gas column. Comparison of NGC~4418 (\ccore) and 
    Arp~220 (\cwest) indicates, however, lower $T_{\mathrm{eff}}$ and
      higher $\Sigma_g$ in Arp~220.} $T_d$. Nevertheless, the disk in the
  western nucleus may be inclined such that our view of the warmer dust is
  obscured in the far-IR and we have a ``rotationally-supported'' edge-on view
  of the structure.

\subsection{The extended component of Arp~220}

The extended and halo components (\cext\ and \chalo) in Arp~220 reveal a
chemistry very different from that found in the nuclear region. 
The \hdo\ abundance drops to the relatively low value of $\sim5\times10^{-8}$,
typical of Galactic PDR regions, and the OH/\hdo\ ratio reverses
the nuclear value becoming $>1$, as in Galactic translucent clouds
\citep[e.g.][]{neu02}. Our estimated OH abundances of 
$\mathrm{several}\,\times10^{-7}$ are in general agreement with those derived
by \cite{pol05} in the intervining translucent clouds toward Sgr B2. 
On the other hand, the outflow signatures observed in the OH 163 and 79
$\mu$m doublet, in the [O {\sc i}] 63 and 145 $\mu$m lines
(Fig.~\ref{inoutflow}), 
and in the \hdo\ \t303212\ line (Fig.~\ref{h2o-b}p), appear to indicate that
the outflow is spatially extended on the back side of the nuclei.

The extended component of Arp~220 resembles the translucent
Galactic medium, though with a higher ionization rate as derived from the
absorption in lines of molecular ions, as we will report in a forthcoming
paper.

\subsection{Inflowing gas in NGC~4418?}
\label{sec:disc:inflow}

The intringuing redshifted absorption in the ground-state lines of OH and
O {\sc i}, and the blueshifted emission in the [O {\sc i}] 63 $\mu$m line,
caused us to consider in \S\ref{sec:n4418low} the possibility that an
extended envelope around the nucleus of NGC~4418 is inflowing onto the nuclear
region, thus feeding the nucleus and providing an external ram pressure that
may further confine the nuclear gas. \cite{kru10} have reported the 
  detection of inflows in Seyferts based on absorption measurements of
the Na I doublet at optical wavelengths. 

The morphology of the region as seen at near-IR wavelengths \citep{eva03}
and the absorption depths indeed suggest that the redshifted absorbing
component is more than a single cloud in the galaxy disk with an anomalous
velocity; the redshifted absorption affects the whole column of gas 
external to the nuclear region ($\sim10^{23}$ cm$^{-2}$), 
as otherwise the ground-state OH absorption at the systemic velocity
would be deeper than observed. 
Thus the OH and O {\sc i} absorptions suggest bulk inward gas motions.

If spherical symmetry is assumed for this component, a mass inflow rate of
$12$ \Msun\ yr$^{-1}$ is obtained at the surface of the nucleus, based on
the model shown in Fig.~\ref{inoutflow}. 
Spherical symmetry is consistent with the 
OH line profiles, in particular with the emission in the OH 163 $\mu$m
doublet. However, the spherical approach overpredicts the [O {\sc i}] 63
$\mu$m blueshifted emission (Fig.~\ref{inoutflow}a). Furthermore, pure
radiative transfer cannot rule out that a significant fraction of 
the [O {\sc i}] 63 $\mu$m 
emission feature is formed in the nucleus, rather than in the envelope, if the
high density component (several $\times10^6$ \cmt, as derived from OH, \hdo,
and HCN) is sufficiently rich in atomic oxygen (a few
$\times10^{-5}$). Observational surveys of the [O {\sc i}] 63 $\mu$m line 
in Galactic hot cores may help further evaluate the latter possibility,
but pure hot core chemistry predicts 
O {\sc i} to have an abundance of only $\lesssim10^{-6}$ \citep{cha97}. Guided
by these models, we do not expect much contribution from 
dense hot cores to the [O {\sc i}] 63 $\mu$m emission feature. However, the
high-lying OH lines are tracing a region where photodissociation of \hdo\ is
probably taking place (\S\ref{sec:xcrays}), so OH photodissociation
could produce significant columns of dense O {\sc i}. Still, the C-shock
models by \cite{war99} with enhanced \hdo\ photodissociation due to X- or
cosmic rays produce little O {\sc i}, with $\mathrm{OH/O} {\mathsc i}>100$,
but models for XDRs and CRs by \cite{mei05} and  \cite{mei11} predict large
amounts of oxygen in atomic form, though they use densities significantly
lower than $3\times10^6$ \cmt. 

In summary, our observations and analysis favor an inflow in NGC~4418, though
the geometry and inflow rate of the inflowing gas are rather
uncertain.  
If the gas were inflowing only from the front side of the nucleus with
an opening angle of, say, $90^{\circ}$, the mass inflow rate would be
of only $\approx2$ \Msun\ yr$^{-1}$. \cite{kaw90} first suggested a
possible inflow in NGC~4418 based on the narrow CO $J=1-0$ line. 
The interaction with the nearby galaxy VV 655 could
have trigered the inflow via a close (perigalactic), non-merging encounter
$\sim10^8$ yr ago for an average velocity difference of 
$\sim200$ \kms\ in the plane of the sky, and models \citep{nog88,ion04}
predict that the inflow triggered by interacting galaxies may last that 
long. Alternatively, a non-axisymmetric gravitational potential as in
barred galaxies can fuel the nuclear region with large amounts of gas
\citep{fri93}, which is supported by the peculiar morphology of NGC~4418.

\subsection{Evolutionary effects}

In spite of the very different global parameters (mass, luminosity, size, and
merger versus non-merger class) of Arp~220 and NGC~4418, their nuclear regions
share common properties.  They contain warm dust, very excited molecular
absorption, and high surface column densities of \hdo, OH, HCN, and
NH$_3$. There are, however, significant differences in the properties
revealed by their far-infrared spectra that probably stem from and further
probe their individual evolutionary histories.  The smaller, less massive, and
less luminous NGC~4418 shows warmer dust and higher molecular excitation than
Arp~220. Perhaps the most suggestive spectroscopic differences between the two
sources are their inferred kinematics  
and $^{16}$O to $^{18}$O ratios:  while NGC~4418 appears to have a ratio
similar (to within a factor of 2) to that seen
in the local solar neighbourhood, the ratio in Arp~220 is significantly
lower.  

The outflow signatures in Arp~220, which range from 800 \kms\ in its OH
megamaser radio emission to 100 \kms\ in submillimeter transitions of HCO$^+$
and CO and up to 500 \kms\ in some moderate excitation submillimeter and
  far-IR emission lines, suggest 
that injection of enriched gas from current and previous
stellar processing has occurred. On the other hand, mechanical feedback has
not yet occurred in NGC~4418, which may still be in a stage
of inflow.  Analysis of the H$_2^{18}$O and the $^{18}$OH lines indicate
$^{16}$O-to-$^{18}$O ratios of $70-130$ in Arp~220, reminiscent of the low
ratio of $\sim40$ derived for Mrk 231 \citep{fis10}. 
The strong absorption in the red $\Lambda-$component of the ground-state
$^{18}$OH 120 $\mu$m doublet (Fig.~\ref{18oh}c) indicates that the
$^{18}$O enhancement is 
not restricted to the nuclear region, but also applies to the \cext\ and
\chalo\ components. At millimeter wavelengths, \cite{mar11} derive a
$^{16}$O-to-$^{18}$O ratio of $>80$ in Arp~220, while \cite{har99}
estimated a ratio of $\sim150$ in NGC 253. In NGC~4418, however, we find that
the $^{16}$O-to-$^{18}$O ratio is $\gtrsim250$, with a best fit value of
$\sim500$. While $^{16}$O is a primary product of stellar nucleosynthesis,
$^{18}$O is a secondary one, and this ratio decreases with increasing star
forming generations \citep{pra96}.  Since galactic outflows are dominated by
entrained ISM gas \citep{vei05}, the low ratios found in Arp~220 and Mrk 231
probably trace stellar processing throughout the merger lifetime.  The
apparent ISM inflow, lack of outflowing gas, and higher $^{16}$O-to-$^{18}$O
ratios in NGC~4418 appear to indicate an earlier evolutionary phase in the
lifetime of the central AGN or nuclear starburst.

We tentatively suggest that the nuclei of Arp~220 and NGC~4418 are in an
earlier evolutionary phase than the QSO driven ULIRG Mrk~231
\citep[e.g.][]{vei09}. In this source, high velocity ($\sim1000$ \kms)
feedback \citep{fis10,stu11,rup11} in the form of a massive 
molecular/atomic outflow 
has more strongly disrupted its nuclear interstellar medium, as evidenced by the
lower dust opacity of its warm component \citep{gon10}
compared with the nuclear regions of NGC~4418 and Arp~220.  The lack of
measurable HCN emission/absorption in its Herschel/SPIRE spectrum
\citep{vdw10}, suggests that exposure to UV and X-ray emission from its 
AGN appears to have at least partially shut down the hot core nitrogen
chemistry\footnote{But not completely: during the final revision of this
  manuscript, \cite{aal11} reported the detection of bright HC$_3$N
  (10-9) in the nuclear region of Mrk 231.}. 
Further, analysis of Herschel/SPIRE observations of \hdo\ emission 
in Mrk 231 \citep{gon10} suggest an \hdo\ abundance of $\sim10^{-6}$,
significantly lower than in NGC~4418 and Arp~220, but observations of the
same lines would help verify this issue.  Lastly, the stronger stellar
processing suggested by the increased enhancement of $^{18}$OH in Mrk 231 may
indicate an older age of its merger induced star-formation than in
Arp~220. This evolutionary scenario is consistent with the fact that in the
mid-IR Mrk~231 looks AGN-like, while Arp~220 and NGC~4418 are still dominated
by deep silicate absorption indicative of a buried, earlier evolutionary 
stage \citep{spo07}.

\section{Conclusions}
\label{sec:conclusions}

The salient observational results presented here on NGC~4418 and Arp~220
  are: 
\begin{itemize}
\item Full range PACS spectroscopy of NGC~4418 and Arp~220 reveals high
  excitation in \hdo\, OH, HCN, and NH$_3$. In NGC~4418, absorption lines were
  detected with $E_{\mathrm{lower}}>800$ K (\hdo), 600 K (OH), 1075
  K (HCN), and 600 K (NH$_3$), while in Arp~220 the excitation is somewhat
  lower with $E_{\mathrm{lower}}>600$ K (\hdo), 600 K (OH), 650 K (HCN), and
  400 K (NH$_3$).   
\item More specifically, NGC~4418 shows 38 \hdo\ lines in absorption, and with
  marginal detections of even higher-lying lines, while Arp~220 shows 
  absorption in 28 \hdo\ lines. All strong OH
  lines within the PACS domain are 
  detected in both sources. In NGC~4418, at least 6 HCN rotational lines, from
  $J=18\rightarrow17$ to $J=23\rightarrow22$ with $E_{\mathrm{lower}}$
  from 650 to 1075 K are detected in absorption, while the
  $J=18\rightarrow17$ and possibly the $J=21\rightarrow20$ lines are detected
  in Arp~220. Rotational NH$_3$ lines arising from $E_{\mathrm{lower}}$ 
    up to $\sim600$ K are detected in NGC~4418, while Arp~220 shows NH$_3$
    lines up to $\sim400$ K. Molecular absorption equivalent widths from
    high-lying levels are higher in NGC~4418 than in Arp~220.
\item Up to 10 lines of H$_2^{18}$O and 3 $^{18}$OH $\Lambda-$doublets 
are detected in Arp~220, while 6 H$_2^{18}$O lines are detected in NGC~4418,
  where only a marginal detection of $^{18}$OH is found. The lines 
    of the rare isotopologues are thus significantly stronger in Arp~220 than
  in NGC~4418. 
\item Lines with moderate excitation in Arp~220 show indications of
  outflowing gas, specifically in the redshifted emission features
    observed in lines of \hdo, OH, and O {\sc i}. The absorption in more
    excited molecular lines peak at central velocities relative to
  our adopted redshift of $0.0181$, though a contribution to the absorption
    by outflowing gas with velocities up to $\sim150$ \kms\ is not ruled out. 
  In NGC~4418, however, the molecular lines do not show 
  any indication of outflowing gas, and the absorption in the low-lying OH
  doublets as well as in the [O {\sc i}] 63 $\mu$m line are redshifted.
\end{itemize}

We have analyzed the observed absorption-dominated line profiles
together with the far-IR continuum,  
using multicomponent radiative transfer modeling with a
minimum number of parameterized components. The conclusions stemming from the
analysis are: 
\begin{itemize}
\item Extremely high \hdo\ column densities per unit of continuum optical
    depth at 50 $\mu$m ($\tau_{50}$), 
  $\gtrsim3\times10^{18}$ \cmd, and OH/\hdo\ and HCN/\hdo\ ratios of
  $\approx0.5$ and $0.1-0.3$, respectively, are derived in the nuclear
  region of NGC~4418, along with dust 
  temperatures $\sim150$ K in the very inner nuclear core. In the nuclear
  region of Arp~220, $T_{d}$ is in the range $90-130$ K, implying
  \hdo\ columns of $(6-0.8)\times10^{18}$ \cmd\ per $\tau_{50}$ and
  OH/\hdo\ and HCN/\hdo\ 
  ratios similar to those of NGC~4418. The HCN/NH$_3$ ratios are $2-5$ in
    both sources. The continuum emission in these nuclear regions is optically
    thick at far-IR wavelengths in both galaxies.
\item The nuclear molecular abundances relative to H are estimated to
  be high, with $\chi_{\mathrm{H_2O}}\sim10^{-5}$ in NGC~4418 and
  $(0.2-1.5)\times10^{-5}$ in Arp~220.
\item The absorption by the $^{18}$O isotopologues indicate a strong
  enhancement of $^{18}$O in Arp~220, with a $^{16}$O-to-$^{18}$O ratio of
  $70-130$, while the ratio in NGC~4418 is significantly higher,
    $\gtrsim250$ with a best fit value of $\sim500$. The
    apparent $^{18}$O enhancement in Arp~220 probably traces stellar
    processing throughout the merger lifetime.
\item We also require an extended component to produce the absorption in
  the low-lying lines of OH and \hdo; the fractional \hdo\ abundance in
  Arp~220 drops in this region to $\sim5\times10^{-8}$ and the OH/\hdo\ is
  reversed relative to the nuclear region to $2.5-10$. In the envelope of
  NGC~4418, we find $\chi_{\mathrm{OH}}=\mathrm{a \, few}\times10^{-7}$.  
\item The extreme \hdo, HCN, and NH$_3$ columns and abundances in the
  nuclear region of both galaxies, together with the high dust temperatures,
  are interpreted 
  in terms of a hot core chemistry, where grain mantles are evaporated. The
  high OH columns and abundances require in addition photoprocesses due to
  X-rays and/or enhancement by high cosmic ray fluxes. \hdo\
  and OH may also be boosted in low-velocity shocks with enhanced
  \hdo\ photodissociation.  
\item Very thick ($N_{\mathrm{H}}\gtrsim10^{25}$ cm$^{-2}$) nuclear
  media are responsible for the frequency degradation of the intrinsic
  luminosity in both sources, as high energy photons are absorbed and
    re-emitted at far-IR wavelengths sampling
  progressively colder media.  Such conditions, and not necessarily a
  starburst-dominated luminosity, may also be responsible for the hot core
  chemistry in the nuclear regions. The high surface brightnesses,
  $\approx6\times10^{13}$ and $(0.9-4)\times10^{13}$ \Lsun/kpc$^2$ on spatial
  scales of $\sim20$ and $150-80$ pc in NGC~4418 and Arp~220, respectively,
  indicate either strongly buried AGNs or hyper star clusters. In NGC~4418,
  radiation pressure is probably responsible for the support of the nuclear
  structure against self-gravity.
\item Both the outflowing gas and the low $^{16}$O-to-$^{18}$O ratio in Arp
  220 suggest a stage more evolved than in NGC~4418, but less evolved
    than the QSO driven ULIRG Mrk 231.
\item In NGC~4418, the redshifted absorption observed in the low-lying OH and
  [O {\sc i}] lines is best explained in terms of an envelope inflowing onto
  the nuclear region, but the geometry of this inflow is rather uncertain; we
  estimate the mass inflow rate to be $\lesssim12$ \Msun\ yr$^{-1}$.
\end{itemize}

\begin{acknowledgements}
We thank David S. N. Rupke for deriving the redshift of NGC~4418 from SDSS,
Kazushi Sakamoto for providing us with the spectra of the HCO$^+$ (3-2) 
and (4-3) lines in the nuclei of Arp~220, and the referee Christian Henkel for
many useful indications and comments that much improved the manuscript. 
PACS has been developed by a consortium of institutes
led by MPE (Germany) and including UVIE (Austria); KU Leuven, CSL, IMEC
(Belgium); CEA, LAM (France); MPIA (Germany); 
INAFIFSI/OAA/OAP/OAT, LENS, SISSA (Italy); IAC (Spain). This development
has been supported by the funding agencies BMVIT (Austria), ESA-PRODEX
(Belgium), CEA/CNES (France), DLR (Germany), ASI/INAF (Italy), and
CICYT/MCYT (Spain). E.G-A thanks the support by the Spanish 
Ministerio de Ciencia e Innovaci\'on under project AYA2010-21697-C05-01, and
is a Research Associate at the Harvard-Smithsonian
Center for Astrophysics. Basic research in IR astronomy at NRL is funded by
the US ONR; J.F. also acknowledges support from the NHSC. S.V. thanks NASA for
partial support of this research via Research Support Agreement RSA
1427277. He also acknowledges support from a Senior NPP Award from NASA and
thanks his host institution, the Goddard Space Flight Center.
\end{acknowledgements}

   \begin{table*}
      \caption[]{\hdo\ line equivalent widths, fluxes, and velocity
        shifts.} 
         \label{tab:fluxesh2o}
          \begin{tabular}{lc|ccc|ccc}   
            \hline
            \noalign{\smallskip}
& & \multicolumn{3}{c}{NGC 4418} & \multicolumn{3}{c}{Arp 220} \\
            Line  &  $\lambda$ & $W_{\mathrm{eq}}^{\mathrm{a}}$ & 
            Flux$^{\mathrm{a}}$ &
            $V_{\mathrm{shift}}^{\mathrm{a,b}}$ & 
            $W_{\mathrm{eq}}^{\mathrm{a}}$ & Flux$^{\mathrm{a}}$ & 
            $V_{\mathrm{shift}}^{\mathrm{a,b}}$ \\  
                & ($\mu$m)  & ($\mathrm{km\,\,s^{-1}}$) & 
                ($10^{-21}$ W cm$^{-2}$) &
                 ($\mathrm{km\,\,s^{-1}}$) & ($\mathrm{km\,\,s^{-1}}$) & 
                 ($10^{-21}$ W cm$^{-2}$) & ($\mathrm{km\,\,s^{-1}}$) \\
            \noalign{\smallskip}
            \hline
            \noalign{\smallskip}
  \hdo\ \t533422\  & $ 53.138$ & $24.9( 6.5)$ & $-20.0( 5.2)$ & $  35(  29)$ & $  1.9(  2.6)$ & $  -4.5( 6.3)$ &     \\
  \hdo\ \t827716$^{\mathrm{c}}$  & $ 55.131$ & $ 6.2( 2.3)$ & $ -4.8( 1.8)$ & $  -3(  16)$ &      &      &      \\
  \hdo\ \t431322\  & $ 56.325$ & $35.6( 1.8)$ & $-26.6( 1.4)$ & $  -8(  33)$ & $ 30.0(  1.2)$ & $ -65.3( 2.6)$ & $ -29(  49)$ \\
  \hdo\ \t422313\  & $ 57.636$ & $37.1( 2.1)$ & $-28.0( 1.5)$ & $ -15(  35)$ & $ 22.8(  1.3)$ & $ -49.3( 2.8)$ & $  14(  43)$ \\
  \hdo\ \t432321\  & $ 58.699$ & $50.7( 0.9)$ & $-37.8( 0.7)$ & $  16(  34)$ & $ 34.7(  1.0)$ & $ -74.9( 2.1)$ & $ -87(  47)$ \\
  \hdo\ \t726615$^{\mathrm{c}}$  & $ 59.987$ & $ 5.6( 1.3)$ & $ -4.1( 1.0)$ & $  -6(  22)$ &      &      &      \\
  \hdo\ \t431404\  & $ 61.809$ & $11.1( 2.6)$ & $ -7.3( 1.7)$ & $  37(  24)$ & $  2.3(  1.4)$ & $  -4.9( 3.0)$ & \\
  \hdo\ \t818707\  & $ 63.324$ & $13.3( 1.0)$ & $ -9.1( 0.7)$ & $  33(  18)$ & $  2.4(  0.6)$ & $  -5.1( 1.3)$ & \\
  \hdo\ \t716625\  & $ 66.093$ & $11.8( 0.9)$ & $ -7.5( 0.6)$ & $  -6(  24)$ & $  5.7(  1.0)$ & $ -12.3( 2.1)$ & \\
  \hdo\ \t330221\  & $ 66.438$ & $64.7( 1.3)$ & $-41.8( 0.9)$ & $ -18(  34)$ & $ 40.8(  0.8)$ & $ -87.8( 1.8)$ & $ -55(  38)$ \\
  \hdo\ \t331220\  & $ 67.089$ & $43.6( 2.2)$ & $-28.1( 1.4)$ & $  12(  31)$ & $ 41.6(  0.9)$ & $ -90.4( 2.1)$ & $ -30(  49)$ \\
  \hdo\ \t330303\  & $ 67.269$ & $13.2( 1.3)$ & $ -8.5( 0.8)$ & $  -9(  29)$ & $  7.7(  2.3)$ & $ -16.7( 4.9)$ & $ -24(  51)$ \\
  \hdo\ \t524413\  & $ 71.067$ & $23.9( 1.7)$ & $-13.6( 1.0)$ & $ -45(  24)$ & $ 16.7(  1.0)$ & $ -33.2( 2.0)$ & $ 35(  44)$ \\
  \hdo\ \t717606\  & $ 71.540$ & $17.6( 1.8)$ & $ -9.7( 1.0)$ & $  27(  34)$ & $  9.5(  0.8)$ & $ -19.0( 1.6)$ & \\
  \hdo\ \t707616\  & $ 71.947$ & $25.5( 2.3)$ & $-13.8( 1.2)$ & $ -11(  32)$ & $ 10.0(  0.8)$ & $ -20.1( 1.7)$ & $ 35(  38)$ \\
  \hdo\ \t725634$^{\mathrm{c}}$  & $ 74.945$ & $ 5.0( 1.5)$ & $ -2.6( 0.8)$ & $ -15(  23)$ &      &      &      \\
  \hdo\ \t321212\  & $ 75.381$ & $73.9( 1.1)$ & $-38.6( 0.6)$ & $ -23(  32)$ & $ 79.7(  0.8)$ & $-153.4( 1.6)$ & $ 4(  41)$ \\
  \hdo\ \t423312\  & $ 78.742$ & $40.3( 1.9)$ & $-19.1( 0.9)$ & $ -36(  27)$ & $ 34.4(  0.9)$ & $ -62.9( 1.6)$ & $ -42(  43)$ \\
  \hdo\ \t615524\  & $ 78.928$ & $19.4( 1.8)$ & $ -9.3( 0.9)$ & $  -4(  27)$ & $  4.9(  0.8)$ & $  -8.8( 1.4)$ & $-70(  31)$ \\
  \hdo\ \t616505\  & $ 82.031$ & $31.4( 1.4)$ & $-14.1( 0.6)$ & $   2(  25)$ & $ 20.3(  0.6)$ & $ -35.6( 1.1)$ & $ -21(  37)$ \\
  \hdo\ \t606515\  & $ 83.284$ & $25.1( 1.7)$ & $-11.3( 0.7)$ & $ -16(  27)$ & $ 16.1(  1.2)$ & $ -27.5( 2.0)$ & $ -4(  41)$ \\
  \hdo\ \t322211\  & $ 89.988$ & $60.4( 1.7)$ & $-23.3( 0.7)$ & $   5(  27)$ & $ 39.8(  1.1)$ & $ -60.6( 1.6)$ & $ -32(  35)$ \\
  \hdo\ \t643634$^{\mathrm{c}}$  & $ 92.811$ & $ 8.0( 1.6)$ & $ -3.0( 0.6)$ & $ -35(  19)$ &      &      &      \\
  \hdo\ \t625616\  & $ 94.644$ & $10.1( 1.7)$ & $ -3.6( 0.6)$ & $ -37(  17)$ &      &      &      \\
  \hdo\ \t441432\  & $ 94.705$ & $14.3( 1.8)$ & $ -5.1( 0.6)$ & $ -33(  18)$ &      &      &      \\
  \hdo\ \t515404\  & $ 95.627$ & $42.9( 2.4)$ & $-15.2( 0.8)$ & $ -11(  30)$ & $ 22.8(  2.1)$ & $ -31.6( 2.9)$ & $ 9(  41)$ \\
  \hdo\ \t440431\  & $ 95.885$ & $12.5( 2.0)$ & $ -4.3( 0.7)$ & $  -2(  20)$ & $  3.0(  1.7)$ & $  -4.1( 2.3)$ & \\
  \hdo\ \t615606$^{\mathrm{d}}$  & $103.940$ & $10.0( 1.4)$ & $ -2.9( 0.4)$ & $  68(  29)$ & $  7.6(  1.4)$ & $  -9.5( 1.7)$ & $ -34(  42)$ \\
  \hdo\ \t221110\  & $108.073$ & $67.4( 2.6)$ & $-17.5( 0.7)$ & $ -23(  46)$ & $ 72.2(  2.1)$ & $ -82.9( 2.4)$ & $  40(  47)$ \\
  \hdo\ \t524515\  & $111.628$ & $24.8( 2.2)$ & $ -6.1( 0.5)$ & $  42(  55)$ & $  4.2(  0.7)$ & $  -4.5( 0.8)$ & \\
  \hdo\ \t414303\  & $113.537$ & $36.7( 2.1)$ & $ -8.5( 0.5)$ & $  14(  39)$ & $ 26.0(  2.9)$ & $ -27.0( 3.1)$ & $ -32(  44)$ \\
  \hdo\ \t404313\  & $125.354$ & $25.4( 1.8)$ & $ -4.6( 0.3)$ & $ -19(  33)$ & $ 27.7(  1.2)$ & $ -23.7( 1.0)$ & $  -6(  54)$ \\
  \hdo\ \t331322$^{\mathrm{d}}$  & $126.714$ & $12.0( 1.8)$ & $ -2.1( 0.3)$ & $  55(  35)$ & $  8.8(  5.3)$ & $  -7.4( 4.4)$ & $  14( 119)$ \\
  \hdo\ \t423414\  & $132.408$ & $28.1( 2.5)$ & $ -4.6( 0.4)$ & $   5(  68)$ & $ 12.6(  0.8)$ & $  -9.7( 0.6)$ & $ -29(  46)$ \\
  \hdo\ \t514505\  & $134.935$ & $26.1( 1.9)$ & $ -4.1( 0.3)$ & $   0(  49)$ & $  8.6(  0.8)$ & $  -6.3( 0.6)$ & $  18(  40)$ \\
  \hdo\ \t330321\  & $136.496$ & $36.9( 1.9)$ & $ -5.5( 0.3)$ & $   0(  46)$ & $ 19.8(  0.8)$ & $ -14.3( 0.6)$ & $ -37(  48)$ \\
  \hdo\ \t313202\  & $138.528$ & $30.7( 1.4)$ & $ -4.5( 0.2)$ & $  30(  39)$ & $ 29.0(  1.2)$ & $ -20.2( 0.8)$ & $ -40(  45)$ \\
  \hdo\ \t413322\  & $144.518$ & $11.9( 1.7)$ & $ -1.5( 0.2)$ & $ -32(  28)$ & $ 11.6(  0.9)$ & $  -7.3( 0.6)$ & $ -24(  40)$ \\
  \hdo\ \t431422\  & $146.923$ & $12.2( 2.1)$ & $ -1.5( 0.3)$ & $ -17(  33)$ & $  3.3(  0.6)$ & $  -2.0( 0.3)$ & $ -23(  27)$ \\
  \hdo\ \t322313\  & $156.194$ & $ 9.2( 2.3)$ & $ -1.0( 0.2)$ & $  30(  26)$ & $ 14.5(  0.9)$ & $  -7.5( 0.5)$ & $ -29(  38)$ \\
  \hdo\ \t303212\  & $174.626$ & $13.2( 2.5)$ & $ -1.0( 0.2)$ & $  -7(  15)$ & $ 20.3(  3.7)$ & $  -7.3( 1.3)$ & $ -33(  26)$ \\
  \hdo\ \t212101$^{\mathrm{d}}$  & $179.527$ & $81.9( 4.9)$ & $ -5.5( 0.3)$ & $ 118(  35)$ & $167.3(  5.1)$ & $ -51.5( 1.6)$ & $  34(  40)$ \\
               \noalign{\smallskip}
            \hline
         \end{tabular} 
\begin{list}{}{}
\item[$^{\mathrm{a}}$] Numbers in parenthesis indicate 1 $\sigma$
  uncertainties from Gaussian fits to the lines. 
\item[$^{\mathrm{b}}$] Velocity shifts are relative to $z=0.00705$ for NGC~4418
  and $z=0.0181$ for Arp~220.
\item[$^{\mathrm{c}}$] Marginal detection.
\item[$^{\mathrm{d}}$] Probably contaminated.
\end{list}
   \end{table*}

   \begin{table*}
      \caption[]{[O {\sc i}] and OH line equivalent widths, fluxes, and
        velocity shifts.} 
         \label{tab:fluxesoh}
          \begin{tabular}{lc|ccc|ccc}   
            \hline
            \noalign{\smallskip}
& & \multicolumn{3}{c}{NGC 4418} & \multicolumn{3}{c}{Arp 220} \\
            Line  &  $\lambda$ & $W_{\mathrm{eq}}^{\mathrm{a}}$ & 
            Flux$^{\mathrm{a}}$ &
            $V_{\mathrm{shift}}^{\mathrm{a,b}}$ & 
            $W_{\mathrm{eq}}^{\mathrm{a}}$ & Flux$^{\mathrm{a}}$ & 
            $V_{\mathrm{shift}}^{\mathrm{a,b}}$ \\  
                & ($\mu$m)  & ($\mathrm{km\,\,s^{-1}}$) & 
                ($10^{-21}$ W cm$^{-2}$) &
                 ($\mathrm{km\,\,s^{-1}}$) & ($\mathrm{km\,\,s^{-1}}$) & 
                 ($10^{-21}$ W cm$^{-2}$) & ($\mathrm{km\,\,s^{-1}}$) \\
            \noalign{\smallskip}
            \hline
            \noalign{\smallskip}
{\rm [O {\sc i}]} $1-2$ &  $ 63.184$  &  $ 25.2( 1.0)$ & $-17.2( 0.7)$ & $ 115(  16)$  & $ 45.4(  2.0)$ & $ -96.3( 4.3)$ & $  20(  30)$  \\
{\rm [O {\sc i}]} $0-1$ &  $145.525$  &  $ -7.3( 1.6)$ & $  0.9( 0.2)$ & $  35(  22)$  & $-25.4(  1.1)$ & $  15.6( 0.7)$ & $  75(  77)$  \\
 OH    $\Pi_{3/2}-\Pi_{3/2}\, \frac{5}{2}^--\frac{3}{2}^+$ & $119.233$ & $ 58.4( 2.0)$ & $-12.0( 0.4)$ & $ 125(  36)$ & $212.5(  2.2)$ & $-204.5( 2.1)$ & $  41(  52)$ \\
 OH    $\Pi_{3/2}-\Pi_{3/2}\, \frac{5}{2}^+-\frac{3}{2}^-$ & $119.441$ & $ 54.9( 1.9)$ & $-11.2( 0.4)$ & $ 114(  32)$ & $174.9(  2.0)$ & $-168.1( 1.9)$ & $  39(  44)$ \\
 OH    $\Pi_{1/2}-\Pi_{3/2}\, \frac{1}{2}^--\frac{3}{2}^+$ & $ 79.118$ & $ 17.8( 1.4)$ & $ -8.7( 0.7)$ & $  97(  24)$ & $ 63.5(  1.3)$ & $-115.0( 2.3)$ & $  90(  35)$ \\
 OH    $\Pi_{1/2}-\Pi_{3/2}\, \frac{1}{2}^+-\frac{3}{2}^-$ & $ 79.181$ & $ 12.5( 1.3)$ & $ -6.1( 0.6)$ & $  72(  19)$ & $ 63.5(  1.3)$ & $-115.0( 2.3)$ & $  90(  35)$ \\
 OH    $\Pi_{1/2}-\Pi_{3/2}\, \frac{3}{2}^+-\frac{3}{2}^-$ & $ 53.261$ & $ 38.8( 4.5)$ & $-31.3( 3.6)$ & $  76(  35)$ & $ 95.7(  3.8)$ & $-230.5( 9.1)$ & $  66(  44)$ \\
 OH    $\Pi_{1/2}-\Pi_{3/2}\, \frac{3}{2}^--\frac{3}{2}^+$ & $ 53.351$ & $ 59.2( 5.6)$ & $-47.7( 4.5)$ & $  56(  52)$ & $ 94.4(  3.7)$ & $-227.3( 8.8)$ & $  31(  42)$ \\
 OH    $\Pi_{1/2}-\Pi_{3/2}\, \frac{3}{2}-\frac{5}{2}^{\mathrm{c}}$ & $ 96.340$ & $  9.1( 3.3)$ & $ -3.1( 1.1)$ & $  38(  34)$ &      &      \\
 OH    $\Pi_{3/2}-\Pi_{3/2}\, \frac{7}{2}^+-\frac{5}{2}^-$ & $ 84.420$ & $ 77.8( 1.9)$ & $-34.2( 0.8)$ & $  10(  30)$ & $ 88.6(  1.4)$ & $-150.8( 2.3)$ & $   2(  38)$ \\
 OH    $\Pi_{3/2}-\Pi_{3/2}\, \frac{7}{2}^--\frac{5}{2}^+$ & $ 84.597$ & $ 89.6( 2.0)$ & $-39.2( 0.9)$ & $   1(  34)$ & $ 95.4(  1.4)$ & $-161.7( 2.3)$ & $ -16(  39)$ \\
 OH    $\Pi_{3/2}-\Pi_{3/2}\, \frac{9}{2}^--\frac{7}{2}^+$ & $ 65.132$ & $ 74.1( 1.7)$ & $-48.7( 1.1)$ & $  16(  39)$ & $ 49.8(  1.2)$ & $-107.7( 2.6)$ & $  -6(  50)$ \\
 OH    $\Pi_{3/2}-\Pi_{3/2}\, \frac{9}{2}^+-\frac{7}{2}^-$ & $ 65.279$ & $ 62.7( 1.8)$ & $-40.8( 1.1)$ & $ -13(  39)$ & $ 45.8(  1.2)$ & $ -98.8( 2.5)$ & $ -57(  49)$ \\
 OH    $\Pi_{1/2}-\Pi_{1/2}\, \frac{7}{2}-\frac{5}{2}^{\mathrm{c}}$ & $ 71.197$ & $ 72.1( 2.8)$ & $-40.5( 1.6)$ & $ -16(  45)$ & $ 42.4(  1.0)$ & $ -84.3( 2.0)$ & $ -17(  54)$ \\
 OH    $\Pi_{3/2}-\Pi_{3/2}\, \frac{11}{2}^+-\frac{9}{2}^-$ & $ 52.934$ & $ 24.1( 5.0)$ & $-18.8( 3.9)$ & $ -57(  22)$ & $  3.0(  2.2)$ & $  -7.4( 5.5)$ & $ -67(  40)$ \\
 OH    $\Pi_{3/2}-\Pi_{3/2}\, \frac{11}{2}^--\frac{9}{2}^+$ & $ 53.057$ & $ 23.5( 5.4)$ & $-18.5( 4.3)$ & $  91(  25)$ & $  3.6(  3.8)$ & $  -8.8( 9.3)$ & $   4(  33)$ \\
 OH    $\Pi_{1/2}-\Pi_{1/2}\, \frac{9}{2}-\frac{7}{2}^{\mathrm{c}}$ & $ 55.920$ & $ 54.1( 4.4)$ & $-41.0( 3.3)$ & $  40(  91)$ & $ 12.5(  2.5)$ & $ -26.9( 5.4)$ & $ -52(  58)$ \\
 OH    $\Pi_{1/2}-\Pi_{1/2}\, \frac{3}{2}^+-\frac{1}{2}^-$ & $163.124$ & $-10.0( 2.2)$ & $  1.0( 0.2)$ & $ -78(  24)$ & $-71.9(  2.7)$ & $  31.9( 1.2)$ & $ 150(  77)$ \\
 OH    $\Pi_{1/2}-\Pi_{1/2}\, \frac{3}{2}^--\frac{1}{2}^+$ & $163.397$ & $ -9.6( 2.2)$ & $  0.9( 0.2)$ & $ -16(  23)$ & $-41.8(  5.0)$ & $  18.5( 2.2)$ & $ 161(  57)$ \\
               \noalign{\smallskip}
            \hline
         \end{tabular} 
\begin{list}{}{}
\item[$^{\mathrm{a}}$] Numbers in parenthesis indicate 1 $\sigma$
  uncertainties from Gaussian fits to the lines.
\item[$^{\mathrm{b}}$] Velocity shifts are relative to $z=0.00705$ for NGC~4418
  and $z=0.0181$ for Arp~220.
\item[$^{\mathrm{c}}$] The two $\Lambda-$components are (nearly) blended into
  a single spectral feature.
\end{list}
   \end{table*}

   \begin{table*}
      \caption[]{HCN line equivalent widths, fluxes, and velocity shifts.}
         \label{tab:fluxeshcn}
          \begin{tabular}{lc|ccc|ccc}   
            \hline
            \noalign{\smallskip}
& & \multicolumn{3}{c}{NGC 4418} & \multicolumn{3}{c}{Arp 220} \\
            Line  &  $\lambda$ & $W_{\mathrm{eq}}^{\mathrm{a}}$ & 
            Flux$^{\mathrm{a}}$ &
            $V_{\mathrm{shift}}^{\mathrm{a,b}}$ & 
            $W_{\mathrm{eq}}^{\mathrm{a}}$ & Flux$^{\mathrm{a}}$ & 
            $V_{\mathrm{shift}}^{\mathrm{a,b}}$ \\  
                & ($\mu$m)  & ($\mathrm{km\,\,s^{-1}}$) & 
                ($10^{-21}$ W cm$^{-2}$) &
                 ($\mathrm{km\,\,s^{-1}}$) & ($\mathrm{km\,\,s^{-1}}$) & 
                 ($10^{-21}$ W cm$^{-2}$) & ($\mathrm{km\,\,s^{-1}}$) \\
            \noalign{\smallskip}
            \hline
            \noalign{\smallskip}
 HCN $25\rightarrow24$ & $135.626$ & $ 5.8( 1.8)$ & $ -0.9( 0.3)$ & $ -20(  43)$ &      &      &      \\
 HCN $23\rightarrow22$ & $147.365$ & $ 9.4( 1.9)$ & $ -1.1( 0.2)$ & $  69(  23)$ &    &  &  \\
 HCN $22\rightarrow21$ & $154.037$ & $12.3( 1.6)$ & $ -1.3( 0.2)$ & $  19(  35)$ & $  4.3(  2.0)$ & $ -2.3( 1.1)$ & $ -35(  58)$ \\
 HCN $21\rightarrow20^{\mathrm{c}}$ & $161.346$ & $14.7( 2.7)$ & $ -1.5( 0.3)$ & $   7(  21)$ & $  9.6(  1.5)$ & $ -4.5( 0.7)$ & $ -53(  45)$ \\
 HCN $20\rightarrow19$ & $169.386$ & $12.7( 2.7)$ & $ -1.0( 0.2)$ & $  -6(  20)$ &      &      &      \\
 HCN $19\rightarrow18$ & $178.274$ & $11.0( 3.8)$ & $ -0.8( 0.3)$ & $  21(  27)$ & $  3.8(  2.0)$ & $ -1.2( 0.6)$ & $  24(  72)$ \\
 HCN $18\rightarrow17$ & $188.151$ & $21.0( 5.4)$ & $ -0.9( 0.2)$ & $ -16(  20)$ & $ 15.2(  3.2)$ & $ -3.1( 0.6)$ & $-100(  37)$ \\
               \noalign{\smallskip}
            \hline
         \end{tabular} 
\begin{list}{}{}
\item[$^{\mathrm{a}}$] Numbers in parenthesis indicate 1 $\sigma$
  uncertainties from Gaussian fits to the lines.
\item[$^{\mathrm{b}}$] Velocity shifts are relative to $z=0.00705$ for NGC~4418
  and $z=0.0181$ for Arp~220.
\item[$^{\mathrm{c}}$] Doubtful identification in Arp~220.
\end{list}
   \end{table*}

   \begin{table*}
      \caption{NH$_3$ line equivalent widths and fluxes for relatively isolated spectral features.}
         \label{tab:fluxesnh3}
          \begin{tabular}{lc|cc|cc}   
            \hline
            \noalign{\smallskip}
& & \multicolumn{2}{c}{NGC 4418} & \multicolumn{2}{c}{Arp 220} \\
            Line  &  $\lambda$ & $W_{\mathrm{eq}}^{\mathrm{a}}$ & 
            Flux$^{\mathrm{a}}$ &
            $W_{\mathrm{eq}}^{\mathrm{a}}$ & Flux$^{\mathrm{a}}$ \\  
                & ($\mu$m)  & ($\mathrm{km\,\,s^{-1}}$) & 
                ($10^{-21}$ W cm$^{-2}$) & ($\mathrm{km\,\,s^{-1}}$) & 
                 ($10^{-21}$ W cm$^{-2}$)  \\
            \noalign{\smallskip}
            \hline
            \noalign{\smallskip}
 NH$_3$ $(8,6)s\rightarrow(7,6)a$ & $ 62.888$ & $ 3.9( 1.7)$ & $ -2.7( 1.1)$ & &       \\
 NH$_3$ $(7,6)a\rightarrow(6,6)s$ & $ 72.439$ & $ 6.7( 0.6)$ & $ -3.6( 0.3)$ & $  4.6(  0.5)$ & $ -9.2( 1.0)$ \\
 NH$_3$ $(6,5)s\rightarrow(5,5)a$ & $ 83.432$ & $10.1( 1.2)$ & $ -4.5( 0.5)$ & $  4.5(  0.8)$ & $ -7.6( 1.4)$ \\
 NH$_3$ $(6,4)s\rightarrow(5,4)a$ & $ 83.590$ & $13.6( 1.4)$ & $ -6.1( 0.6)$ &$^{\mathrm{b}}$ &  $^{\mathrm{b}}$   \\
 NH$_3$ $(6,3)s\rightarrow(5,3)a$ & $ 83.709$ & $10.1( 1.0)$ & $ -4.5( 0.5)$ & $  3.0(  0.7)$ & $ -5.1( 1.1)$ \\
 NH$_3$ $(6,2)s\rightarrow(5,2)a$ & $ 83.794$ & $ 4.5( 1.0)$ & $ -2.0( 0.4)$ &
 $^{\mathrm{b}}$ &  $^{\mathrm{b}}$   \\
 NH$_3$ $(6,1-0)s\rightarrow(5,1-0)a$$^{\mathrm{c}}$ & $83.844-83.861$ & $11.0( 1.1)$ & $ -4.9( 0.5)$ & $  7.5(  1.2)$ & $-12.9( 2.1)$ \\
 NH$_3$ $(4,3)s\rightarrow(3,3)a$ & $124.648$ & $17.9( 1.7)$ & $ -3.3( 0.3)$ & $ 25.1(  1.1)$ & $-21.6( 0.9)$ \\
 NH$_3$ $(4,2-1-0)s\rightarrow(3,2-1-0)a$$^{\mathrm{c}}$ & $124.796-124.913$ & $38.3( 1.6)$ & $ -7.0( 0.3)$ & $ 21.5(  1.2)$ & $-18.5( 1.1)$ \\
 NH$_3$ $(4,3-2-1)a\rightarrow(3,3-2-1)s$$^{\mathrm{c}}$ & $127.108-127.181$ & $38.9( 1.3)$ & $ -6.9( 0.2)$ & $ 28.9(  0.9)$ & $-24.1( 0.7)$ \\
 NH$_3$ $(3,2)s\rightarrow(2,2)a$$^{\mathrm{d}}$ & $165.597$ & $30.9( 2.3)$ & $ -2.8( 0.2)$ & $ 27.6(  2.0)$ & $-11.8( 0.9)$ \\
 NH$_3$ $(3,1)s\rightarrow(2,1)a$$^{\mathrm{d}}$ & $165.729$ & $13.6( 2.4)$ &
 $ -1.2( 0.2)$ & $^{\mathrm{b}}$ & $^{\mathrm{b}}$   \\
 NH$_3$ $(3,2-1-0)a\rightarrow(2,2-1-0)s$$^{\mathrm{c}}$ & $169.968-169.996$ & $26.9( 3.4)$ & $ -2.2( 0.3)$ & $ 33.7(  1.5)$ & $-13.2( 0.6)$ \\
  \noalign{\smallskip}
            \hline
         \end{tabular} 
\begin{list}{}{}
\item[$^{\mathrm{a}}$] Numbers in parenthesis indicate 1 $\sigma$
  uncertainties from Gaussian fits to the lines. 
\item[$^{\mathrm{b}}$] Uncertain due to baseline subtraction or blending.
\item[$^{\mathrm{c}}$] Joint emission of several $K-$lines
\item[$^{\mathrm{d}}$] Probably contaminated by H$_3$O$^+$.
\end{list}
   \end{table*}

   \begin{table*}
      \caption[]{H$_2^{18}$O line equivalent widths and fluxes.}
         \label{tab:fluxesh218o}
          \begin{tabular}{lc|cc|cc}   
            \hline
            \noalign{\smallskip}
& & \multicolumn{2}{c}{NGC 4418} & \multicolumn{2}{c}{Arp 220} \\
            Line  &  $\lambda$ & $W_{\mathrm{eq}}^{\mathrm{a}}$ & 
            Flux$^{\mathrm{a}}$ &
            $W_{\mathrm{eq}}^{\mathrm{a}}$ & Flux$^{\mathrm{a}}$ \\  
                & ($\mu$m)  & ($\mathrm{km\,\,s^{-1}}$) & 
                ($10^{-21}$ W cm$^{-2}$) & ($\mathrm{km\,\,s^{-1}}$) & 
                 ($10^{-21}$ W cm$^{-2}$)  \\
            \noalign{\smallskip}
            \hline
            \noalign{\smallskip}
  H$_2^{18}$O \t432321\  & $ 59.350$ & $ 1.5( 0.9)$ & $ -1.0( 0.6)$ & $  7.9(  1.1)$ & $ -17.2( 2.5)$ \\
  H$_2^{18}$O \t331220\  & $ 67.883$ & $ 2.0( 0.9)$ & $ -1.2( 0.6)$ & $  8.3(  0.9)$ & $ -18.0( 1.9)$ \\
  H$_2^{18}$O \t321212\  & $ 75.867$ & $ 7.0( 1.6)$ & $ -3.5( 0.8)$ & $ 17.9(  0.8)$ & $ -34.7( 1.5)$ \\
  H$_2^{18}$O \t423312\  & $ 79.531$ & $ 6.0( 1.1)$ & $ -2.9( 0.5)$ & $  5.4(  0.8)$ & $  -9.6( 1.4)$ \\
  H$_2^{18}$O \t322211\  & $ 90.936$ & $ 3.5( 1.5)$ & $ -1.3( 0.6)$ & $  4.2(  0.7)$ & $  -6.4( 1.1)$ \\
  H$_2^{18}$O \t221110\  & $109.347$ & $ 5.7( 1.6)$ & $ -1.5( 0.4)$ & $ 13.2(  1.8)$ & $ -14.8( 2.0)$ \\
  H$_2^{18}$O \t414303\  & $114.296$ & $^{\mathrm{b}}$& $^{\mathrm{b}}$& $  7.0( 1.5)$ & $  -7.2( 1.5)$ \\
  H$_2^{18}$O \t313202\  & $139.586$ & & & $ 10.0( 0.8)$ & $  -6.8( 0.5)$ \\
  H$_2^{18}$O \t303212\  & $174.374$ & & & $  7.9( 1.5)$ & $  -2.9( 0.6)$ \\
  H$_2^{18}$O \t221212\  & $183.530$ & & & $ 14.8( 2.5)$ & $  -3.8( 0.7)$ \\
  \noalign{\smallskip}
            \hline
         \end{tabular} 
\begin{list}{}{}
\item[$^{\mathrm{a}}$] Numbers in parenthesis indicate 1 $\sigma$
  uncertainties from Gaussian fits to the lines. 
\item[$^{\mathrm{b}}$] Uncertain due to baseline subtraction.
\end{list}
   \end{table*}

   \begin{table*}
      \caption[]{$^{18}$OH line equivalent widths and fluxes.}
         \label{tab:fluxes18oh}
          \begin{tabular}{lc|cc|cc}   
            \hline
            \noalign{\smallskip}
& & \multicolumn{2}{c}{NGC 4418} & \multicolumn{2}{c}{Arp 220} \\
            Line  &  $\lambda$ & $W_{\mathrm{eq}}^{\mathrm{a}}$ & 
            Flux$^{\mathrm{a}}$ &
            $W_{\mathrm{eq}}^{\mathrm{a}}$ & Flux$^{\mathrm{a}}$ \\  
                & ($\mu$m)  & ($\mathrm{km\,\,s^{-1}}$) & 
                ($10^{-21}$ W cm$^{-2}$) & ($\mathrm{km\,\,s^{-1}}$) & 
                 ($10^{-21}$ W cm$^{-2}$)  \\
            \noalign{\smallskip}
            \hline
            \noalign{\smallskip}
  $^{18}$OH $\Pi_{3/2}-\Pi_{3/2} \, \frac{9}{2}^+-\frac{7}{2}^-$ & $ 65.690$ & $ 1.7( 0.9)$ & $ -1.1( 0.6)$ & $  4.9(  0.9)$ & $ -10.6( 1.9)$ \\
  $^{18}$OH $\Pi_{3/2}-\Pi_{3/2} \, \frac{7}{2}^+-\frac{5}{2}^-$ & $ 84.947$ & $ 2.1( 0.9)$ & $ -0.9( 0.4)$ & $  7.0(  0.8)$ & $ -11.8( 1.3)$ \\
  $^{18}$OH $\Pi_{3/2}-\Pi_{3/2} \, \frac{7}{2}^--\frac{5}{2}^+$ & $ 85.123$ & $ 0.7( 0.7)$ & $ -0.3( 0.3)$ & $ 12.9(  0.8)$ & $ -21.5( 1.4)$ \\
  $^{18}$OH $\Pi_{3/2}-\Pi_{3/2} \, \frac{5}{2}^+-\frac{3}{2}^-$ & $120.171$ & $ 0.4( 0.8)$ & $ -0.1( 0.2)$ & $ 38.7(  2.2)$ & $ -36.7( 2.0)$ \\
  \noalign{\smallskip}
            \hline
         \end{tabular} 
\begin{list}{}{}
\item[$^{\mathrm{a}}$] Numbers in parenthesis indicate 1 $\sigma$
  uncertainties from Gaussian fits to the lines. 
\end{list}
   \end{table*}

\end{document}